\documentclass[%
 reprint,
nofootinbib,
 amsmath,amssymb,
 aps,
]{revtex4-2}

\usepackage{graphicx}
\usepackage{dcolumn}
\usepackage{bm}
\usepackage{tikz}
\usetikzlibrary{quantikz}
\usepackage{adjustbox}
\usepackage{mathdots}
\usepackage{braket}
\usepackage {graphicx}
\usepackage{caption}
\RequirePackage{subfigure}
\usepackage{amsmath}
\usepackage{multirow}

\begin{document}

\preprint{APS/123-QED}

\title{Efficient Quantum State Preparation with Walsh Series}

\author{Julien Zylberman}

\author{Fabrice Debbasch}%
\affiliation{%
 Sorbonne Université, Observatoire de Paris, Université PSL, CNRS, LERMA, F-75005 Paris, France }%



\date{\today}

\begin{abstract}

A new approximate Quantum State Preparation (QSP) method is introduced, called the Walsh Series Loader (WSL). The WSL approximates quantum states defined by real-valued functions of single real variables with a depth independent of the number $n$ of qubits. Two approaches are presented: the first one approximates the target quantum state by a Walsh Series truncated at order $O(1/\sqrt{\epsilon})$, where $\epsilon$ is the precision of the approximation in terms of infidelity. The circuit depth is also $O(1/\sqrt{\epsilon})$, the size is $O(n+1/\sqrt{\epsilon})$ and only one ancilla qubit is needed. The second method represents accurately quantum states with sparse Walsh series.
The WSL loads $s$-sparse Walsh Series into $n$-qubits with a depth doubly-sparse in $s$ and $k$, the maximum number of bits with value $1$ in the binary decomposition of the Walsh function indices. The associated quantum circuit approximates the sparse Walsh Series up to an error $\epsilon$ with a depth $O(sk)$, a size $O(n+sk)$ and one ancilla qubit.
In both cases, the protocol is a Repeat-Until-Success (RUS) procedure with a probability of success $P=\Theta(\epsilon)$, giving an averaged total time of $O(1/\epsilon^{3/2})$ for the WSL ({\sl resp.} $O(sk/\epsilon)$ for the sparse WSL). Amplitude amplification can be used to reach a probability of success $P=\Theta(1)$, modifying the quantum circuit size to $O((n+1/\sqrt{\epsilon})/\sqrt{\epsilon})$ ({\sl resp.} $O((n+sk)/\sqrt{\epsilon})$) and the depth to $O((n+1/\sqrt{\epsilon})/\sqrt{\epsilon})$ ({\sl resp.} $O((n+sk)/\sqrt{\epsilon})$). Amplitude amplification reduces by a factor $O(1/\sqrt{\epsilon})$ the total time dependency with $\epsilon$ but increases the size and depth of the associated quantum circuits, making them linearly dependent on $n$. These protocols give overall efficient algorithms with no exponential scaling in any parameter.
They can be generalized to any complex-valued, multi-variate, almost-everywhere-differentiable function.
The Repeat-Until-Success Walsh Series Loader is so far the only method which prepares a quantum
state with a circuit depth and an averaged total time independent of the number of qubits.

\end{abstract}

\maketitle

The second quantum revolution relies on the manipulation of individual quantum systems. One of the key technologies promised by this revolution is quantum computing, which is made possible by the manipulation of individual quantum bits (qubits). Because they can use quantum superposition and entanglement, Quantum Computers (QCs) will perform some computations faster than classical computers and mapping computationally demanding problems into a form tractable by a QC has become an active area of research. 

In particular, the problem of solving Partial Differential Equations (PDEs) on QCs has recently attracted a lot of attention,
with publications discussing digital quantum algorithms \cite{berry2014high,berry2017quantum,leyton2008quantum,liu2021efficient,lloyd2020quantum,an2022efficient,costa2019quantum,cao2013quantum,wang2020quantum,scherer2017concrete,childs2020quantum,engel2019quantum,oz2023efficient,zylberman2022quantum}, hybrid and variational quantum-classical methods \cite{ zylberman2022quantum,shukla2023hybrid,lubasch2020variational,liu2021variational,liu2022application, kubo2021variational,demirdjian2022variational,garcia2022quantum}, and adiabatic and annealing quantum algorithms \cite{zanger2021quantum, costa2022optimal,goes2023qboost, chandra2014quadratic, PhysRevA.104.032426, criado2022qade,greer2020approach}. To solve the Cauchy problems for differential equations on any digital computer, be it classical or quantum, one needs (i) to discretize space and time (ii) to load the initial condition onto the computer.  
An initial condition for a PDE can always be represented by a function $f$ of a certain variable $x$, where both $x$ and $f$ are possibly multi-dimensional. 

On a classical computer, the cost of loading the initial condition is negligible when compared to the cost of the integration steps. This is not so on a QC. Indeed, encoding classical data into an $n$-qubit state may cost an exponential amount of primitive operations because the space of all $n$-qubit states has dimension $2^n$. 
Thus, exact methods for Quantum State Preparation  (QSP) have an exponential scaling with $n$ either in depth, size or number of ancilla qubits \cite{grover2002creating, mottonen2004transformation, plesch2011quantum, sun2023asymptotically,PhysRevResearch.3.043200,zhang2022quantum,araujo2021divide}. It has been suggested that these issues can be overcome by using quantum generative adversarial networks and variational methods with low depth and size trained-quantum circuits \cite{nakaji2022approximate, zoufal2019quantum}. However, these methods suffer from usual optimization problems such as Barren plateaus, local minima and scalability \cite{cerezo2021variational,mcclean2018barren}. This has prompted the introduction of approximate methods with efficient complexities scaling at most as $O(\text{poly}(n,1/\epsilon))$
and, in particular, no exponential scaling \cite{marin2021quantum,moosa2023linear,rattew2022preparing}. 

In this article, we present a new simple quantum algorithm for QSP based on Walsh functions : the Walsh Series Loader (WSL). Consider for example initial conditions corresponding to real valued functions of a single real variable. Given an error $\epsilon>0$, one can then implement a quantum state $\epsilon$-close to the target quantum state using only one ancilla qubit, with a quantum circuit of depth $O(1/\sqrt{\epsilon})$ independent of the number of qubits $n$ and of size $O(n+1/\sqrt{\epsilon})$. The efficiency of the algorithm is guaranteed for any function  with bounded first derivative. The algorithm also applies to complex-valued functions and/or functions of $d$ real variables loaded into $nd$ qubits. More generally, the WSL can load any Walsh Series of $s$ terms up to an error $\epsilon>0$ with a quantum circuit of depth $O(sk)$, and size $O(n+sk)$  where $k$ is the maximum number of bits with value $1$ in the binary decomposition of the Walsh function indices (maximum Hamming weight of the Walsh function indices). The protocol is presented as a Repeat-Until-Success (RUS) procedure with a probability of success $P=\Theta(\epsilon)$, and an averaged time for success given by $T= \text{depth}/P$. Amplitude Amplification (AA) can be performed to increase the probability of success to $P=\Theta(1)$ and decrease the averaged total time by a factor $O(1/\sqrt{\epsilon})$, all at the cost of increasing  the quantum circuit size and depth. All complexity scalings are summarized in Table \ref{Table1}.

Assume for the time being that $f$ is a real-valued function of the single real variable $x \in [0, 1]$.
To solve the PDE numerically, be it on a classical or a quantum computer, the variable $x$ of the function $f$ must be discretised and we take that step for granted in what follows. Initialising a quantum algorithm solving the PDE means loading onto a digital quantum computer the state $\ket{f} = \frac{1}{||f||_2}\sum_{x} f(x) \ket{x}$ where the kets $\ket{x}$ are eigenstates of the operator representing the classical variable $x$. Consider now, for any given $f$, the operator $\hat{f} = \sum_{x} f(x) \ket{x} \bra{x}$ and the unitary operator $\hat{U}_{f, \epsilon_0}=e^{-i\hat{f} \epsilon_0}$ where $\epsilon_0$ is an arbitrary strictly positive real number. Both operators are diagonal in the $x$-basis. At given $\epsilon_0$, the operator $\hat{U}_{f, \epsilon_0}$ contains all the information present in the state $\hat{f}$. So, encoding $\hat{U}_{f, \epsilon_0}$ in an efficient way is tantamount to encoding the information present in $\ket{f}$ in an efficient way. 

The new quantum algorithm for QSP that we propose is thus based on two key ingredients. The first one is an efficient implementation of diagonal unitary operators through their actions on the set of orthogonal functions called Walsh functions $w_j:[0,1]\rightarrow \{-1,1\}$ \cite{welch2014efficient}. This set of functions was first introduced by Walsh in 1923  \cite{walsh1923closed} who showed that every continuous function of bounded variations defined on $[0,1]$ can be expanded into a series of Walsh functions.\footnote{In other words, Walsh functions can be used to perform spectral analysis.}

\begin{table*}
    \centering
\begin{tabular}{ |p{3.7 cm}|p{1.7 cm}|p{3cm}|p{3cm}|p{1.7cm}|p{3cm}|  }

\hline
\multicolumn{6}{|c|}{Walsh Series Loader complexity} 
\\
 \hline
  \multicolumn{2}{|c|}{Method} & Depth & Size & Probability of success & Averaged total time \\
 \hline
 \multirow{2}{*}{Repeat Until Success} & WSL    & $O(1/\sqrt{\epsilon})$ & $O(n+1/\sqrt{\epsilon})$   & $\Theta(\epsilon)$ &$O(1/\epsilon^{3/2})$ \\
 \cline{2-6}
    & Sparse WSL & $O(sk)$ & $O(n+sk)$   & $\Theta(\epsilon)$ &$O(sk/\epsilon)$\\
    \hline
  \multirow{2}{*}{Amplitude Amplification}& WSL    & $O((n+1/\sqrt{\epsilon})/\sqrt{\epsilon})$ & $O((n+1/\sqrt{\epsilon})/\sqrt{\epsilon})$   & $\Theta(1)$ &$O((n+1/\sqrt{\epsilon})/\sqrt{\epsilon})$ \\
 \cline{2-6}
     &Sparse WSL & $O((n+sk)/\sqrt{\epsilon})$ & $O((n+sk)/\sqrt{\epsilon})$   & $\Theta(1)$ &$O((n+sk)/\sqrt{\epsilon})$\\
 \hline
\end{tabular}
\caption{
Scaling laws of  the depth, size, probability of success and averaged total time for the WSL and for the sparse WSL with Repeat-Until-Success and Amplitude Amplification protocols. The number of qubits is $n$, the sparsity of the Walsh Series is $s$ and $k\leq n$ is the maximum Hamming weight of the indices of the Walsh functions appearing in the sparse Walsh Series. For the WSL (resp. the sparse WSL), $\epsilon$ is the error in terms of infidelity between the implemented state and a target quantum state $\ket{f}$ associated to a differentiable function $f$ (resp. $\ket{f_s}$ associated to a sparse Walsh series $f_s$).
}
\label{Table1}
\end{table*}

Walsh functions are ideal in the general context of binary logic and binary arithmetic and, in particular, in quantum information. Indeed, the operator $ \hat{w}_j$ associated to the Walsh functions $w_j$ can be written as a tensor product of $Z$-Pauli gates $\hat{w}_j=(Z_1)^{j_1}\otimes...\otimes (Z_n)^{j_n}$, where $j_i$ is the $i$-th coefficient in the binary expansion of $j=\sum_{i=1}^nj_i2^{i-1}$. 
Given a function $f$ of the variable $x \in [0, 1]$, one can expand it in terms of Walsh functions,  $f=\sum_{j=0}^\infty a_j w_j$. On a finite set of $M$ points, one can expand the restricted function $f$ as a series of $M$ Walsh functions which approximate the function $f$ on $[0,1]$ up to an error $\epsilon_1$ (see Lemma 1.1 of Appendix B). 

Since all Walsh operators $\hat{w}_j$ commute with each other, implementing (approximately) $\hat{U}_{f, \epsilon_0}$ comes down
to implementing the $M$ operators $\hat{W}_{j, \epsilon_0}=e^{-ia_j\hat{w}_j\epsilon_0}$, $j = 0, ..., M-1$, and the efficiency of the method is ensured by the simplicity of the quantum circuits implementing each 
$\hat{W}_{j, \epsilon_0}$ \cite{welch2014efficient}.

The second ingredient in the algorithm is a repeat-until-success method which transforms the unitary $\hat{U}_{f, \epsilon_0}=e^{-i\hat{f}\epsilon_0}$ into an operator proportional to $\hat{f}$ and, ultimately, into the 
desired quantum state $\ket{f}$. This is achieved by an interference scheme where an ancilla qubit is manipulated to generate the operator $\hat{I}-e^{-i\hat{f}\epsilon_0}$ which, for small enough $\epsilon_0$, coincides with $i \hat{f} \epsilon_0$. It turns out that measuring the ancilla qubit delivers, at leading order in $\epsilon_0$, the desired state $\ket{f}$. This is so because measurement introduces an extra normalisation factor $\mathcal{N}\simeq ||\hat{f}\epsilon_0||_2$ which, at leading order in $\epsilon_0$, cancels the $\epsilon_0$ dependence 
present in $\hat{I}-e^{-i\hat{f}\epsilon_0}$. 

Let us now give some details about the way the ancilla qubit is used. 

Suppose that the $n$-qubit register for the position $x$ is initially in the state $\ket{0,...,0}$. We apply to the register a Hadamard tower to get from that state the uniform superposition:
\begin{equation}
    \ket{s}=\hat{H}^{\otimes n}\ket{0,...,0}=\frac{1}{\sqrt{N}}\sum_{x}\ket{x}.
\label{all state}
\end{equation}

We then add an ancillary qubit in state $\ket{q_A}=\hat{H}\ket{0}=\frac{1}{\sqrt{2}}(\ket{0}+\ket{1})$, so the state of the total system is $\ket{\psi_1} = \frac{1}{\sqrt{2}}(\ket{s}\ket{0}+\ket{s}\ket{1})$. We now let the ancilla control the action of ${\hat U}_{f, \epsilon_0}$ by introducing a new operator controlled$-\hat{U}_{f, \epsilon_0}$ whose action on $\ket{\psi_1}$ gives
\begin{equation}
\ket{\psi_2}= \frac{1}{\sqrt{2}}(\ket{s}\ket{0}+e^{-i\hat{f}\epsilon_0}\ket{s}\ket{1}).
\end{equation}
Technically, a quantum circuit for controlled$-\hat{U}_{f, \epsilon_0}$ can be obtained from a quantum circuit for $\hat{U}_{f, \epsilon_0}$ by letting every gate be controlled by the ancilla qubit, changing CNOT gates into Toffoli gates and single-qubits-rotations into controlled-rotations. 
\begin{figure}[t]
    \centering

        \includegraphics[width=0.47\textwidth]{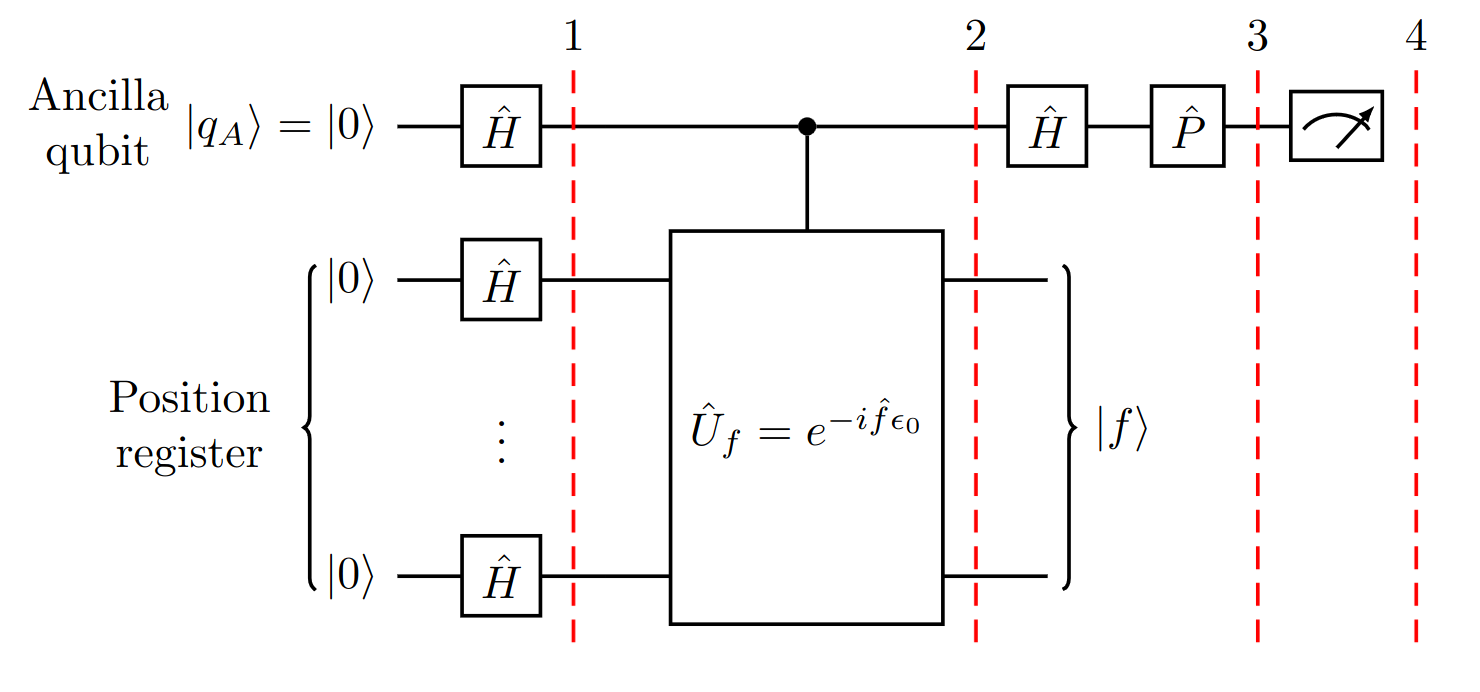}
        
\caption{Quantum circuit for the preparation of an initial quantum state $\ket{f}=\frac{1}{||f||_{2}}\sum_{x}f(x)\ket{x}$ associated to a real-valued function $f$. At each red line, the quantum state  corresponds respectively to equation (\ref{all state}), (2), (3) and (4). }
\label{quantum circuit scheme}
\end{figure}
 The Hadamard gate $\hat{H}$ and the gate $\hat{P}=\begin{pmatrix} 1 &0 \\ 0 & -i\end{pmatrix}$ can then be used to mix components and get the state
 \begin{equation}
\ket{\psi_3}  =  \frac{\hat{I}+e^{-i\hat{f}\epsilon_0}}{2}\ket{s}\ket{0}-i\frac{\hat{I}-e^{-i\hat{f}\epsilon_0}}{2}\ket{s}\ket{1}.
\label{Eq3}
 \end{equation}
One then measures the ancilla qubit  (in the computational basis). If $\ket{q_A}=\ket{0}$, the protocol starts again. If $\ket{q_A}=\ket{1}$, the output state is 
\begin{equation}
\begin{split}
  \ket{\psi_4}=-i\frac{\hat{I}-e^{-i\hat{f}\epsilon_0}}{2||\frac{\hat{I}-e^{-i\hat{f}\epsilon_0}}{2}\ket{s} ||_2}\ket{s}  \simeq \ket{f} + O(\epsilon_0),
\end{split}
\end{equation}
which, at leading order in $\epsilon_0$, is identical to the desired state. Note the very act of measuring the ancilla introduces the correct renormalization which makes it possible to obtain the desired state.

The part of the algorithm that we have just described, which involves the ancilla qubit is represented  in Fig. \ref{quantum circuit scheme}.

The probability of success of the protocol, {\sl i.e}, the probability $P(1)$ to measure $\ket{q_A}=\ket{1}$, scales as 
\begin{equation}
    P(1)=||\frac{\hat{I}-e^{-i\hat{f}\epsilon_0}}{2}\ket{s}||_{2}^2 \simeq \frac{\epsilon_0^2}{4}||f||_2^2.
\end{equation}

The repeat-until-success procedure does not increase the size nor the depth of the quantum circuit. On average, the time T required to achieve success equals the depth of the quantum circuit divided by the probability of success. Once success is reached, the initialization has been performed and one can implement the evolution of the initial quantum state without initializing again. Futhermore, one could perform amplitude amplification to reach $P(1)=\Theta(1)$ and to reduce the total time by a factor $1/\epsilon_0$ but at the cost of increasing the size and the depth of the WSL (more details in Appendix C).

This procedures works for real-valued functions $f$, but it obviously fails if $f$ is complex-valued, because a complex-valued $f$ makes the operator ${\hat U}_{f, \epsilon_0}$ non unitary.
The way to handle complex-valued functions is to add a layer to the algorithm. One introduces the modulus $\mid f\mid$ and the phase $\phi_f$. One carries out the above procedure
for $\mid f \mid$ (instead of $f$) and then implements efficiently the unitary operator $\exp(i \phi_f)$ separately using again Walsh functions as developed in \cite{welch2014efficient} adding an additional $O(1/\sqrt{\epsilon})$ in terms of size and depth (more details in Appendix A.2).

\paragraph*{Error analysis.} 

The discrepancy between the target quantum state and the implemented quantum state has two distinct origins. The first one is the error $\epsilon_1$ introduced
by computing the finite Walsh series of $f$ on a set of $M(\epsilon_1)$ points.  The second one is the error $\epsilon_0$ introduced by the interference scheme.

Let us be a bit more specific about the first source of error. The diagonal unitary operator $\hat{U}_{f}$ is implemented efficiently using the scheme introduced by Welsh et al. in \cite{welch2014efficient}. The differentiable real function $f$ defined on $[0,1]$ is expanded into a Walsh series $f^{\epsilon_1}$. The Walsh series of $f$ corresponds to a piece-wise constant function which coincides with $f$ on a finite number of points and the error associated to the Walsh series can be bounded by the maximum value of the first derivative of $f$ on $[0,1]$: $||f(x)-f^{\epsilon_1}(x)||_{\infty}\leq \epsilon_1||f'||_\infty$, where $||f||_\infty=\sup_{x\in[0,1]}|f(x)|$.
These two errors result in an infidelity $1-F=O((\epsilon_0+\epsilon_1||f'||_\infty)^2)$ emphasizing the fact that the method is efficient for slowly varying functions, ie when the space-step of the discretization of the continuous problem is small compared to the characteristic length of variations of the PDE problem. 

The results of this article can be summarized in two theorems. First, consider a state defined by a real valued function $f$ defined on $[0,1]^d$ and suppose one wants to load that state unto $n = \sum_{i = 1}^d n_i$ qubits with errors $\vec{\epsilon} = (\epsilon_1, ..., \epsilon_d)$. Then:

\paragraph*{Theorem 1 (informal version). } 
There is an efficient quantum circuit of size $O(n_1+...+n_d+1/(\epsilon_1...\epsilon_d))$ and depth $O(1/(\epsilon_1...\epsilon_d))$, which, using one ancillary qubit, implements the quantum state $\ket{f}$ with a  probability of success $P(1)=\Theta(\epsilon_0^2)$ and infidelity $1-F=O((\epsilon_0+\sum_{i=1}^d\epsilon_i ||\partial_i f||_{\infty,[0,1]^d})^2)$.

The proof of this theorem can be found in the Appendix B. A direct corollary of this theorem in the one dimensional case is: there is a quantum circuit of size $O(n+1/\sqrt{\epsilon})$, depth $O(1/\sqrt{\epsilon})$ which uses only one ancillary qubit and implements the quantum state $\ket{f}$ with a probability of success $P(1)=\Theta(\epsilon)$ and infidelity $1-F\leq \epsilon$. Also, note that the size is affine in $n_1+...+n_d$ (or $n$) because of the Hadamard gates applied on each qubits at the first step of the QSP algorithm. 

Additionnally, real-valued functions which are accurately represented by a sparse Walsh Series of $s$ terms can be efficiently loaded unto $n$ qubits. Let us consider
$f\simeq f_s:=\sum_{j\in S} a_jw_j$ with $S\subset \{0,1,...,2^n-1\}$ where the $a_j$'s can be choosen to minimize the difference between $f_s$ and $f$. The problem of finding the best set $S$ and the best coefficients $\{a_j , j \in S\}$ to approximate a function $f$ is called the Minimax Series problem \cite{yuen1975function}. 
A simple but efficient way to find a sparse Walsh series approximating a given function $f$ is to keep in the Walsh series of $f$ the terms with the largest $|a_j|$. 
The complexity of implementing a given $s$-sparse Walsh series depends directly on $s$ and $k$, the maximum Hamming weight of the binary decomposition of the Walsh coefficient indices: $k=\max_{j\in S}(\sum_{i=0}^{l_j}j_i)$ with $j=\sum_{i=0}^{l_j}j_i2^i$.

\paragraph*{Theorem 2 (informal version).} 
For a given set $S\subset \{0,1,...,2^n-1\} $, and real Walsh coefficients $\{a_j , j \in S\}$, there is an efficient quantum circuit of size $O(n+sk)$ and depth $O(sk)$ which, using one ancillary qubit, implements the quantum state $\ket{f_s}$ with a  probability of success $P(1)=\Theta(\epsilon)$ and infidelity $1-F\leq \epsilon$.

A corollary of theorem 2 concerns the case of a function $f$ approximated by a sparse Walsh series $f_s$  such that $||f-f_s||_{\infty}\leq \sqrt{\epsilon}$, then there is an efficient quantum circuit of size $O(n+sk)$ and depth $O(sk)$ which, using one ancillary qubit, implements the quantum state $\ket{f}$ with a  probability of success $P(1)=\Theta(\epsilon)$ and infidelity $1-F \leq \epsilon$.
On $n$ qubits, the parameter $k$ is necessary smaller or equal to $n$. So, 
in the worst case scenario, the sparse WSL method has depth $O(sn)$ and size $O(sn)$. The proof of this theorem and corollary can be found in the Appendix B.7. 


\begin{figure}[b]

     \begin{subfigure}{(a)}  
        \centering 
        \includegraphics[width=0.4\textwidth]{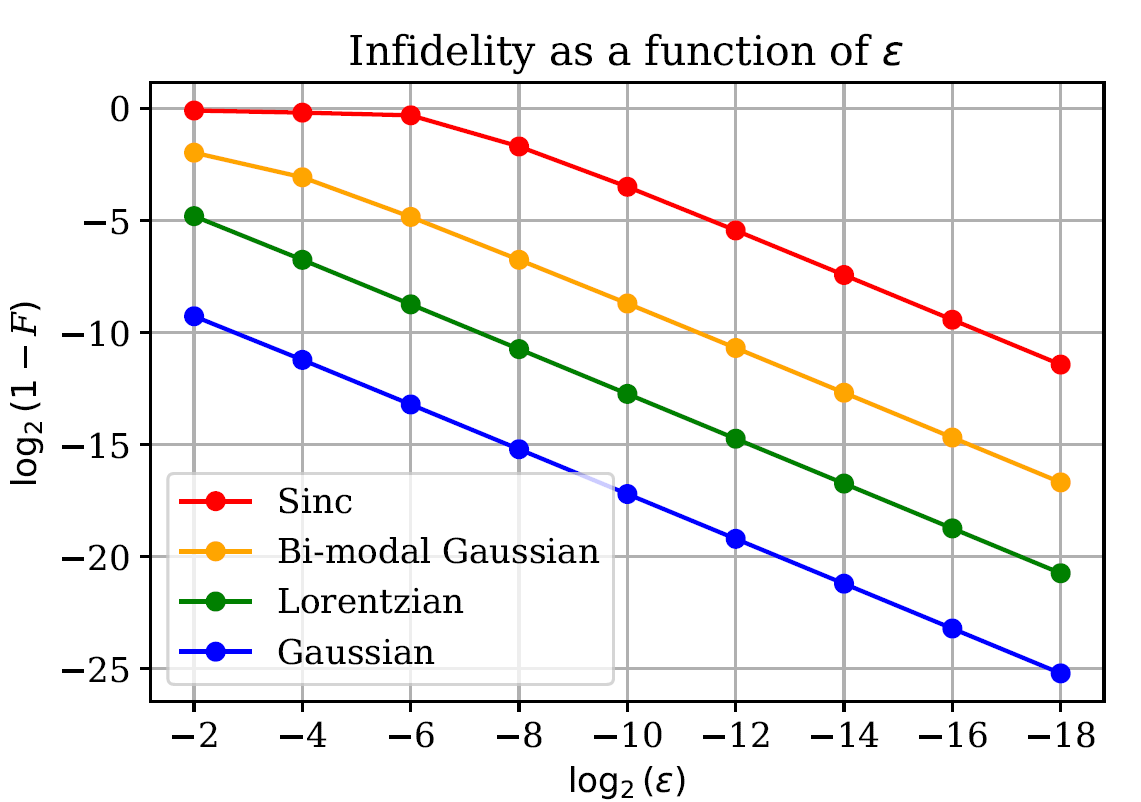}
        
     \end{subfigure}

     \begin{subfigure}{(b)}
         \centering 
         \includegraphics[width=0.4\textwidth]{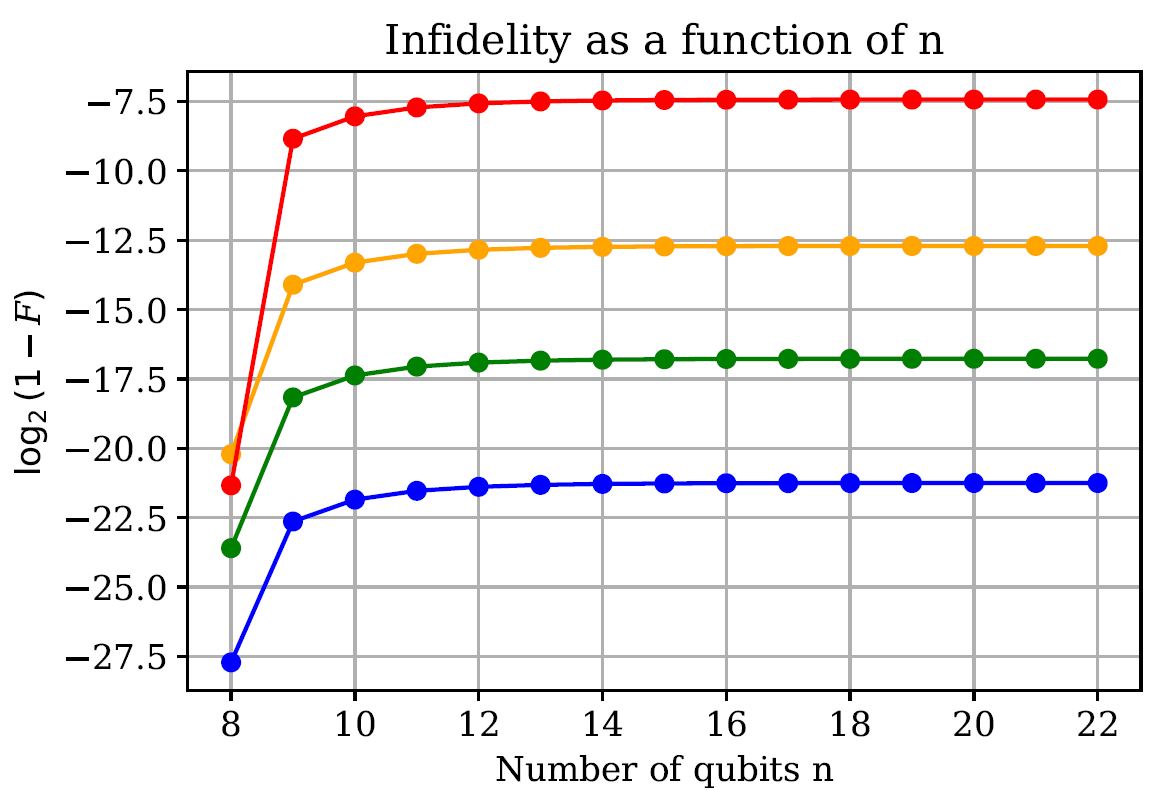}
         
     \end{subfigure}
     
\caption{Infidelity $1-F$ as a function of $\epsilon=\epsilon_0^2=\epsilon_1^2$ for $n=20$ qubits (top) and infidelity with $n$ for $\epsilon_0=\epsilon_1=1/2^7$ (bottom) for different probability distributions: Gaussian $g_{\mu,\sigma}(x)=\text{exp}(-(x-\mu)^2/(2\sigma^2))/\sigma$ with $\mu=0.5, \sigma=1$, Bi-modal Gaussian $g_{\mu_1,\sigma_1,\mu_2,\sigma_2,s}(x)$ with $\mu_1=0.25$, $ \mu_2=0.75$, $ \sigma_1=0.3$, $ \sigma_2=0.04$, $s=0.1$, Lorentzian $L_{\mu,\Gamma}(x)=(\Gamma+4(x-\mu)^2/\Gamma)^{-1}$ with $\mu=0.5$, $ \Gamma=1$ and $\text{sinc}(x)=\sin(6\pi x)/(6\pi x)$.}
\label{Infidelity scalings}
\end{figure}

\paragraph*{Numerical results.}

The scaling laws stated in theorem 1 can be illustrated by numerical examples. Fig. \ref{Infidelity scalings} displays how the infidelity $1 - F$ scales with $\epsilon = \epsilon_0^2 = \epsilon_1^2$ (Fig. 2.a) and with $n$ (Fig. 2b) for various functions. Fig. 2a confirms the linear scaling with $\epsilon$ while Fig. 2b clearly illustrates the fact that, for a given target state, the infidelity admits an $n$-independent (but $\epsilon$-dependent) upper-bound. 

\begin{figure}[h]
    \centering
    
    \includegraphics[width=0.4\textwidth]{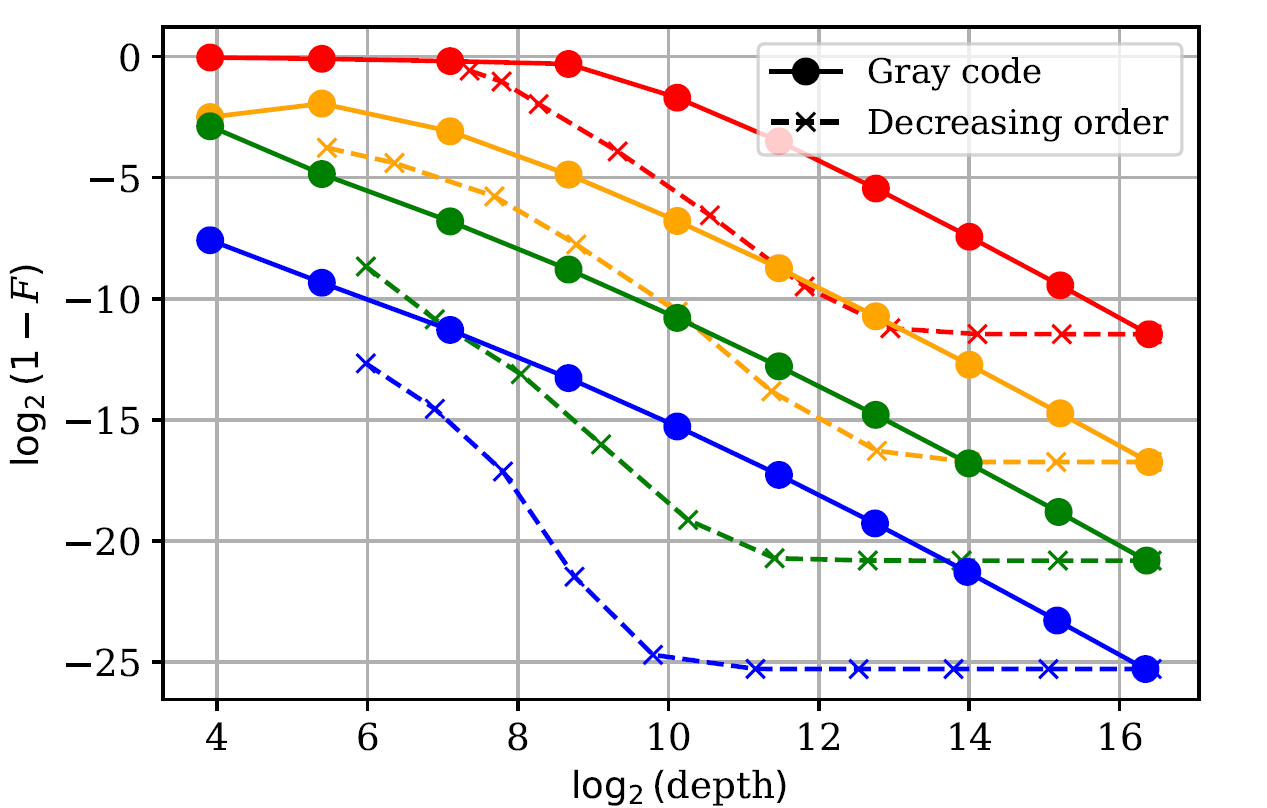} 
    \caption{Infidelity $1-F$ as a function of the depth of the quantum circuits associated to the Gray code order (full lines) and the decreasing order (dashed lines) for the functions defined in Fig. 2 and parameters $n=16$, $\epsilon_0=10^{-3}$. Each dot corresponds to a number of Walsh operators $2^m$ going from $2^1$ to $2^{10}$. For the Gray code order, the $2^m$-Walsh Series is computed for each points. For the decreasing order method, $2^{10}$ Walsh coefficients are computed and only the $2^m$ largest are implemented.}
\label{Infidelity for Gray code & decreasing order.}
\end{figure}

Furthermore, the WSL offers two ways of arranging the Walsh operators. The first one is to use a Gray code which cancels a maximum number of CNOT gates: out of two CNOT stairs, only one CNOT remains, reaching optimality in terms of size \cite{welch2014efficient,bullock2004asymptotically}.
The second method consists in implementing a sparse Walsh Series by listing the $M$-Walsh coefficients of $f$ in decreasing order, keeping then only the first, dominant coefficients. One can thus obtain surprisingly accurate approximations of the targeted state with a very small numbers of Walsh operators. 
Numerical results show that, at given infidelity, the second method has a depth smaller than first method (see Fig \ref{Infidelity for Gray code & decreasing order.}). The dominant Walsh coefficients actually do not depend on the total number of qubits $n$. The procedure thus delivers another QSP method with depth independent of $n$. The number of classical computations needed to implement the Gray code or the decreasing
order method depends only on $M$, and not $n$. 
More details on the WSL for complex and non-differentiable functions can be found in Appendix A.

\paragraph*{Discussion/Comparison with other methods.}
 Our method needs $(n+1)$ Hadamard gates to initialise the state into a full superposition of all possible ket vectors. 
 The control-diagonal unitary which is applied afterwards can be implemented with $M$ controlled-Z-rotations ($\widehat{CR}_Z$) and $M-1$ Toffoli gates, where $M$ depends on $\epsilon_1$, which is the error made in representing the function $f$ by its Walsh series $f^{\epsilon_1}$. To be precise, $M=2^m$ with $m=\lfloor \log_2(1/\epsilon_1) \rfloor+1$. Now, each Toffoli gate can be decomposed into $6$ CNOT gates, $2$ Hadamard gates and $7$ $\hat{T}$ and $\hat{T}^\dagger$ gates, without using ancilla qubits \cite{shende2008cnot}, and each two-qubit gate $\widehat{CR}_Z(\theta)$  can be decomposed into $2$ CNOT gates and $3$ $\hat{R}_Z$ gates with the formula $\widehat{CR}_Z(\theta)=(\hat{I}_2\otimes \hat{R}_Z(\theta))CNOT(\hat{I}_2\otimes \hat{R}_Z(-\theta/2))CNOT(\hat{I}_2\otimes \hat{R}_Z(-\theta/2))$ \cite{nielsen2002quantum}. Finally, an Hadamard gates and an optional Phase gate  $\hat{P}=\begin{pmatrix} 1 &0 \\ 0 & -i\end{pmatrix}=\hat{S}\hat{Z}$, with $\hat{S}=\begin{pmatrix} 1 &0 \\ 0 & i\end{pmatrix}$, are performed on the ancilla qubit, giving a total count of $8M-6$ CNOT gates, $n+2M$ Hadamard gates, $3M$ $\hat{R}_Z$ gates, $7M-7$ $\hat{T}$ and $\hat{T}^\dagger$ gates and an optional $\hat{Z}$ and $\hat{S}$ gates. This leads to a size scaling as $O(n+1/\epsilon_1)$ and a depth $O(1/\epsilon_1)$. Let us remark that the complexity scalings are given in function of $\epsilon_1$ because $\epsilon_1$ is linked to the target error which is an input parameter of the problem while $M$ is not an input parameter of the problem. Nevertheless, the scalings with $m$ or $M$ are equivalent to the one stated here.
 
 In the case of loading an $s$-sparse Walsh Series of parameter $k$, the controlled-$\hat{W}_{j, \epsilon_0}$ is implementable for each of the $s$ Walsh operators with a quantum circuit composed of two Toffoli stairs of $H(j)$ Toffoli gates, with $H(j)$ the Hamming weight of the binary decomposition of $j$ such that $k=\max_{j\in S} H(j) \leq n$, and one $\widehat{CR}_Z$ rotation. The number of Toffoli gates is bounded by $2ks$ and the number of $\widehat{CR}_Z$ is $s$. In the worst case scenario, this gives a total number of $12k+2s$ CNOTs, $4ks+n+2$ Hadamard gates, $14ks$ $\hat{T}$ and $\hat{T}^\dagger$ gates, $3s$ $\hat{R}_Z$ gates and an optional $\hat{S}$ and $\hat{Z}$ gates. In many particular cases, the number of Toffoli gates can be reduced by choosing the order of implementation of the Walsh operators which cancel a maximum number of Toffoli gates in two successive Toffoli stairs. Several algorithms minimize the number of gates in a quantum circuit composed only of CNOT, Toffoli and Rz gates using Phase polynomials synthesis for fully-connected hardware and hardware with constrained connectivity \cite{vandaele2022phase}. As expected, the Gray code ordering appears as the optimal solution in the case of a dense Walsh Series with $s=2^m$ and $k=m$ and it reduces the gate complexity up to a factor $2m$. These results can be compared to other approximate QSP algorithms preparing quantum states associated to continuous functions.


The recent Fourier Series Loader (FSL) \cite{moosa2023linear} makes it possible to prepare continuous functions with a depth linear in the number of Fourier components and in the number of qubits.  The idea behind this method is to first load the $2^m$ Fourier components of the target $f$ on the quantum computer, and then to apply an inverse Quantum Fourier Transform to get the function $f$ in `real space'. 
This result can be compared to ours since the number of Fourier components in the Fourier series of a function can be directly related to the error one makes in the truncation, leading to a gate complexity scaling at most as $O(1/\epsilon^{1/p})$ for $p$-differentiable functions. Nevertheless, the inverse-QFT leads to a final quantum circuit of size $O(n^2+2^m)$ and depth $O(n+2^m)$
while the Walsh Series Loader has only size $O(n+2^m)$ and depth $O(2^m)$\footnote{and a probability of success $P(1)=\Theta(\epsilon_0^2)$.}. This difference mainly comes from the fact that Walsh series can be loaded directly in real space.

In \cite{rattew2022preparing}, quantum state preparation for continuous real functions $f_1$ is achieved going adiabatically from Hamiltonian $H_0=\ket{f}\bra{f}$ with $\ket{f}=H^{\otimes n}\ket{0}$ to the target Hamitonian $H_1=\ket{f_1}\bra{f_1}$. The adiabatic evolution is implemented via `small' Trotterization steps. 
To thus prepare the target quantum state with error $\epsilon$, the query complexity $O(\mathcal{F}^p/\epsilon^2)$, where $\mathcal{F}$ is a constant depending on $f_1$ and  the number of necessary ancilla qubits scales as $O(n+d)$, where $d$ is the number of digits used in the discretised encoding of $f_1$. 
Even if the WSL is a repeat-until-success procedure, it offers a quadratic advantage in terms of size and depth from the fact that the complexity scales with the L2 error $\epsilon$ ($\frac{1}{\epsilon}$ instead of  $\frac{1}{\epsilon^2}$) and necessitates only one ancilla qubit.

Another method \cite{holmes2020efficient} suggests to approximate quantum states associated to smooth, differentiable, real (SDR) valued functions using Matrix Product States methods. Approximating SDR functions as polynomials admitting MPS representation, one can use MPS compressions and mappings
from MPS representations to quantum circuit.  The presented quantum circuits are linear in $n$ (depth and size) and are obtained with a linear number of classical computations.
However,  \cite{holmes2020efficient} offers only empirical arguments in favour of the method's efficiency and does not produce analytically proven scaling laws involving the error $\epsilon$. 

Another approximate QSP method \cite{marin2021quantum} makes use of a modified version of the Grover-Rudolph algorithm \cite{grover2002creating}. 
To load a real valued, positive and twice differentiable function on $n$ qubits with infidelity less than $\epsilon$, Sanchez et al. implement only $2^{k(\epsilon,n)}-1$ multi-controlled rotations (instead of $2^n$) with $k(\epsilon,n)$ asymptotically independent of $n$. For other functions, Sanchez et al. use a variational generalisation of the original algorithm. Even if the Walsh Series Loader presented above is a repeat-until-success procedure, it does not involve variational steps and it can be used for any once (as opposed to twice) differentiable functions, including real-valued but non-positive functions, or complex functions, or even multivariate ones. Also, the depth of the WSL is exactly, and not only asymptotically independent of the number of qubits $n$.

\paragraph*{Conclusion.}
The WSL is the first in a new family of quantum algorithms. These approximate quantum states efficiently with a depth independent of the number of qubits. This remarkable property brings us one step closer to quantum supremacy for all algorithms needing a QSP step. This work should be extended by investigating other, alternative methods to compute finite Walsh Series approximations. Possible candidates include threshold sampling, data compression \cite{yuen1975function}  or efficient estimation of the number $M$ of best Walsh coefficients \cite{kushilevitz1991learning}.

\begin{acknowledgments}

The authors thank N.F.Loureiro, A.B.Grilo, T.Fredon, U.Remond, U.Nzongani, B.Claudon and C.Feniou for their usefull feedbacks on our research. The quantum circuit diagrams in this manuscript were prepared using quantikz package \cite{kay2018tutorial} and the plots were prepared using Matplotlib library \cite{hunter2007matplotlib}. 

\end{acknowledgments}

\bibliography{apssamp}

\providecommand{\noopsort}[1]{}\providecommand{\singleletter}[1]{#1}%
\begin{thebibliography}{59}%
\makeatletter
\providecommand \@ifxundefined [1]{%
 \@ifx{#1\undefined}
}%
\providecommand \@ifnum [1]{%
 \ifnum #1\expandafter \@firstoftwo
 \else \expandafter \@secondoftwo
 \fi
}%
\providecommand \@ifx [1]{%
 \ifx #1\expandafter \@firstoftwo
 \else \expandafter \@secondoftwo
 \fi
}%
\providecommand \natexlab [1]{#1}%
\providecommand \enquote  [1]{``#1''}%
\providecommand \bibnamefont  [1]{#1}%
\providecommand \bibfnamefont [1]{#1}%
\providecommand \citenamefont [1]{#1}%
\providecommand \href@noop [0]{\@secondoftwo}%
\providecommand \href [0]{\begingroup \@sanitize@url \@href}%
\providecommand \@href[1]{\@@startlink{#1}\@@href}%
\providecommand \@@href[1]{\endgroup#1\@@endlink}%
\providecommand \@sanitize@url [0]{\catcode `\\12\catcode `\$12\catcode `\&12\catcode `\#12\catcode `\^12\catcode `\_12\catcode `\%12\relax}%
\providecommand \@@startlink[1]{}%
\providecommand \@@endlink[0]{}%
\providecommand \url  [0]{\begingroup\@sanitize@url \@url }%
\providecommand \@url [1]{\endgroup\@href {#1}{\urlprefix }}%
\providecommand \urlprefix  [0]{URL }%
\providecommand \Eprint [0]{\href }%
\providecommand \doibase [0]{https://doi.org/}%
\providecommand \selectlanguage [0]{\@gobble}%
\providecommand \bibinfo  [0]{\@secondoftwo}%
\providecommand \bibfield  [0]{\@secondoftwo}%
\providecommand \translation [1]{[#1]}%
\providecommand \BibitemOpen [0]{}%
\providecommand \bibitemStop [0]{}%
\providecommand \bibitemNoStop [0]{.\EOS\space}%
\providecommand \EOS [0]{\spacefactor3000\relax}%
\providecommand \BibitemShut  [1]{\csname bibitem#1\endcsname}%
\let\auto@bib@innerbib\@empty
\bibitem [{\citenamefont {Berry}(2014)}]{berry2014high}%
  \BibitemOpen
  \bibfield  {author} {\bibinfo {author} {\bibfnamefont {D.~W.}\ \bibnamefont {Berry}},\ }\bibfield  {title} {\bibinfo {title} {High-order quantum algorithm for solving linear differential equations},\ }\href@noop {} {\bibfield  {journal} {\bibinfo  {journal} {Journal of Physics A: Mathematical and Theoretical}\ }\textbf {\bibinfo {volume} {47}},\ \bibinfo {pages} {105301} (\bibinfo {year} {2014})}\BibitemShut {NoStop}%
\bibitem [{\citenamefont {Berry}\ \emph {et~al.}(2017)\citenamefont {Berry}, \citenamefont {Childs}, \citenamefont {Ostrander},\ and\ \citenamefont {Wang}}]{berry2017quantum}%
  \BibitemOpen
  \bibfield  {author} {\bibinfo {author} {\bibfnamefont {D.~W.}\ \bibnamefont {Berry}}, \bibinfo {author} {\bibfnamefont {A.~M.}\ \bibnamefont {Childs}}, \bibinfo {author} {\bibfnamefont {A.}~\bibnamefont {Ostrander}},\ and\ \bibinfo {author} {\bibfnamefont {G.}~\bibnamefont {Wang}},\ }\bibfield  {title} {\bibinfo {title} {Quantum algorithm for linear differential equations with exponentially improved dependence on precision},\ }\href@noop {} {\bibfield  {journal} {\bibinfo  {journal} {Communications in Mathematical Physics}\ }\textbf {\bibinfo {volume} {356}},\ \bibinfo {pages} {1057} (\bibinfo {year} {2017})}\BibitemShut {NoStop}%
\bibitem [{\citenamefont {Leyton}\ and\ \citenamefont {Osborne}(2008)}]{leyton2008quantum}%
  \BibitemOpen
  \bibfield  {author} {\bibinfo {author} {\bibfnamefont {S.~K.}\ \bibnamefont {Leyton}}\ and\ \bibinfo {author} {\bibfnamefont {T.~J.}\ \bibnamefont {Osborne}},\ }\bibfield  {title} {\bibinfo {title} {A quantum algorithm to solve nonlinear differential equations},\ }\href@noop {} {\bibfield  {journal} {\bibinfo  {journal} {arXiv preprint arXiv:0812.4423}\ } (\bibinfo {year} {2008})}\BibitemShut {NoStop}%
\bibitem [{\citenamefont {Liu}\ \emph {et~al.}(2021{\natexlab{a}})\citenamefont {Liu}, \citenamefont {Kolden}, \citenamefont {Krovi}, \citenamefont {Loureiro}, \citenamefont {Trivisa},\ and\ \citenamefont {Childs}}]{liu2021efficient}%
  \BibitemOpen
  \bibfield  {author} {\bibinfo {author} {\bibfnamefont {J.-P.}\ \bibnamefont {Liu}}, \bibinfo {author} {\bibfnamefont {H.~{\O}.}\ \bibnamefont {Kolden}}, \bibinfo {author} {\bibfnamefont {H.~K.}\ \bibnamefont {Krovi}}, \bibinfo {author} {\bibfnamefont {N.~F.}\ \bibnamefont {Loureiro}}, \bibinfo {author} {\bibfnamefont {K.}~\bibnamefont {Trivisa}},\ and\ \bibinfo {author} {\bibfnamefont {A.~M.}\ \bibnamefont {Childs}},\ }\bibfield  {title} {\bibinfo {title} {Efficient quantum algorithm for dissipative nonlinear differential equations},\ }\href@noop {} {\bibfield  {journal} {\bibinfo  {journal} {Proceedings of the National Academy of Sciences}\ }\textbf {\bibinfo {volume} {118}},\ \bibinfo {pages} {e2026805118} (\bibinfo {year} {2021}{\natexlab{a}})}\BibitemShut {NoStop}%
\bibitem [{\citenamefont {Lloyd}\ \emph {et~al.}(2020)\citenamefont {Lloyd}, \citenamefont {De~Palma}, \citenamefont {Gokler}, \citenamefont {Kiani}, \citenamefont {Liu}, \citenamefont {Marvian}, \citenamefont {Tennie},\ and\ \citenamefont {Palmer}}]{lloyd2020quantum}%
  \BibitemOpen
  \bibfield  {author} {\bibinfo {author} {\bibfnamefont {S.}~\bibnamefont {Lloyd}}, \bibinfo {author} {\bibfnamefont {G.}~\bibnamefont {De~Palma}}, \bibinfo {author} {\bibfnamefont {C.}~\bibnamefont {Gokler}}, \bibinfo {author} {\bibfnamefont {B.}~\bibnamefont {Kiani}}, \bibinfo {author} {\bibfnamefont {Z.-W.}\ \bibnamefont {Liu}}, \bibinfo {author} {\bibfnamefont {M.}~\bibnamefont {Marvian}}, \bibinfo {author} {\bibfnamefont {F.}~\bibnamefont {Tennie}},\ and\ \bibinfo {author} {\bibfnamefont {T.}~\bibnamefont {Palmer}},\ }\bibfield  {title} {\bibinfo {title} {Quantum algorithm for nonlinear differential equations},\ }\href@noop {} {\bibfield  {journal} {\bibinfo  {journal} {arXiv preprint arXiv:2011.06571}\ } (\bibinfo {year} {2020})}\BibitemShut {NoStop}%
\bibitem [{\citenamefont {An}\ \emph {et~al.}(2022)\citenamefont {An}, \citenamefont {Fang}, \citenamefont {Jordan}, \citenamefont {Liu}, \citenamefont {Low},\ and\ \citenamefont {Wang}}]{an2022efficient}%
  \BibitemOpen
  \bibfield  {author} {\bibinfo {author} {\bibfnamefont {D.}~\bibnamefont {An}}, \bibinfo {author} {\bibfnamefont {D.}~\bibnamefont {Fang}}, \bibinfo {author} {\bibfnamefont {S.}~\bibnamefont {Jordan}}, \bibinfo {author} {\bibfnamefont {J.-P.}\ \bibnamefont {Liu}}, \bibinfo {author} {\bibfnamefont {G.~H.}\ \bibnamefont {Low}},\ and\ \bibinfo {author} {\bibfnamefont {J.}~\bibnamefont {Wang}},\ }\bibfield  {title} {\bibinfo {title} {Efficient quantum algorithm for nonlinear reaction-diffusion equations and energy estimation},\ }\href@noop {} {\bibfield  {journal} {\bibinfo  {journal} {arXiv preprint arXiv:2205.01141}\ } (\bibinfo {year} {2022})}\BibitemShut {NoStop}%
\bibitem [{\citenamefont {Costa}\ \emph {et~al.}(2019)\citenamefont {Costa}, \citenamefont {Jordan},\ and\ \citenamefont {Ostrander}}]{costa2019quantum}%
  \BibitemOpen
  \bibfield  {author} {\bibinfo {author} {\bibfnamefont {P.~C.}\ \bibnamefont {Costa}}, \bibinfo {author} {\bibfnamefont {S.}~\bibnamefont {Jordan}},\ and\ \bibinfo {author} {\bibfnamefont {A.}~\bibnamefont {Ostrander}},\ }\bibfield  {title} {\bibinfo {title} {Quantum algorithm for simulating the wave equation},\ }\href@noop {} {\bibfield  {journal} {\bibinfo  {journal} {Physical Review A}\ }\textbf {\bibinfo {volume} {99}},\ \bibinfo {pages} {012323} (\bibinfo {year} {2019})}\BibitemShut {NoStop}%
\bibitem [{\citenamefont {Cao}\ \emph {et~al.}(2013)\citenamefont {Cao}, \citenamefont {Papageorgiou}, \citenamefont {Petras}, \citenamefont {Traub},\ and\ \citenamefont {Kais}}]{cao2013quantum}%
  \BibitemOpen
  \bibfield  {author} {\bibinfo {author} {\bibfnamefont {Y.}~\bibnamefont {Cao}}, \bibinfo {author} {\bibfnamefont {A.}~\bibnamefont {Papageorgiou}}, \bibinfo {author} {\bibfnamefont {I.}~\bibnamefont {Petras}}, \bibinfo {author} {\bibfnamefont {J.}~\bibnamefont {Traub}},\ and\ \bibinfo {author} {\bibfnamefont {S.}~\bibnamefont {Kais}},\ }\bibfield  {title} {\bibinfo {title} {Quantum algorithm and circuit design solving the poisson equation},\ }\href@noop {} {\bibfield  {journal} {\bibinfo  {journal} {New Journal of Physics}\ }\textbf {\bibinfo {volume} {15}},\ \bibinfo {pages} {013021} (\bibinfo {year} {2013})}\BibitemShut {NoStop}%
\bibitem [{\citenamefont {Wang}\ \emph {et~al.}(2020)\citenamefont {Wang}, \citenamefont {Wang}, \citenamefont {Li}, \citenamefont {Fan}, \citenamefont {Wei},\ and\ \citenamefont {Gu}}]{wang2020quantum}%
  \BibitemOpen
  \bibfield  {author} {\bibinfo {author} {\bibfnamefont {S.}~\bibnamefont {Wang}}, \bibinfo {author} {\bibfnamefont {Z.}~\bibnamefont {Wang}}, \bibinfo {author} {\bibfnamefont {W.}~\bibnamefont {Li}}, \bibinfo {author} {\bibfnamefont {L.}~\bibnamefont {Fan}}, \bibinfo {author} {\bibfnamefont {Z.}~\bibnamefont {Wei}},\ and\ \bibinfo {author} {\bibfnamefont {Y.}~\bibnamefont {Gu}},\ }\bibfield  {title} {\bibinfo {title} {Quantum fast poisson solver: the algorithm and complete and modular circuit design},\ }\href@noop {} {\bibfield  {journal} {\bibinfo  {journal} {Quantum Information Processing}\ }\textbf {\bibinfo {volume} {19}},\ \bibinfo {pages} {1} (\bibinfo {year} {2020})}\BibitemShut {NoStop}%
\bibitem [{\citenamefont {Scherer}\ \emph {et~al.}(2017)\citenamefont {Scherer}, \citenamefont {Valiron}, \citenamefont {Mau}, \citenamefont {Alexander}, \citenamefont {Van~den Berg},\ and\ \citenamefont {Chapuran}}]{scherer2017concrete}%
  \BibitemOpen
  \bibfield  {author} {\bibinfo {author} {\bibfnamefont {A.}~\bibnamefont {Scherer}}, \bibinfo {author} {\bibfnamefont {B.}~\bibnamefont {Valiron}}, \bibinfo {author} {\bibfnamefont {S.-C.}\ \bibnamefont {Mau}}, \bibinfo {author} {\bibfnamefont {S.}~\bibnamefont {Alexander}}, \bibinfo {author} {\bibfnamefont {E.}~\bibnamefont {Van~den Berg}},\ and\ \bibinfo {author} {\bibfnamefont {T.~E.}\ \bibnamefont {Chapuran}},\ }\bibfield  {title} {\bibinfo {title} {Concrete resource analysis of the quantum linear-system algorithm used to compute the electromagnetic scattering cross section of a 2d target},\ }\href@noop {} {\bibfield  {journal} {\bibinfo  {journal} {Quantum Information Processing}\ }\textbf {\bibinfo {volume} {16}},\ \bibinfo {pages} {1} (\bibinfo {year} {2017})}\BibitemShut {NoStop}%
\bibitem [{\citenamefont {Childs}\ and\ \citenamefont {Liu}(2020)}]{childs2020quantum}%
  \BibitemOpen
  \bibfield  {author} {\bibinfo {author} {\bibfnamefont {A.~M.}\ \bibnamefont {Childs}}\ and\ \bibinfo {author} {\bibfnamefont {J.-P.}\ \bibnamefont {Liu}},\ }\bibfield  {title} {\bibinfo {title} {Quantum spectral methods for differential equations},\ }\href@noop {} {\bibfield  {journal} {\bibinfo  {journal} {Communications in Mathematical Physics}\ }\textbf {\bibinfo {volume} {375}},\ \bibinfo {pages} {1427} (\bibinfo {year} {2020})}\BibitemShut {NoStop}%
\bibitem [{\citenamefont {Engel}\ \emph {et~al.}(2019)\citenamefont {Engel}, \citenamefont {Smith},\ and\ \citenamefont {Parker}}]{engel2019quantum}%
  \BibitemOpen
  \bibfield  {author} {\bibinfo {author} {\bibfnamefont {A.}~\bibnamefont {Engel}}, \bibinfo {author} {\bibfnamefont {G.}~\bibnamefont {Smith}},\ and\ \bibinfo {author} {\bibfnamefont {S.~E.}\ \bibnamefont {Parker}},\ }\bibfield  {title} {\bibinfo {title} {Quantum algorithm for the vlasov equation},\ }\href@noop {} {\bibfield  {journal} {\bibinfo  {journal} {Physical Review A}\ }\textbf {\bibinfo {volume} {100}},\ \bibinfo {pages} {062315} (\bibinfo {year} {2019})}\BibitemShut {NoStop}%
\bibitem [{\citenamefont {Oz}\ \emph {et~al.}(2023)\citenamefont {Oz}, \citenamefont {San},\ and\ \citenamefont {Kara}}]{oz2023efficient}%
  \BibitemOpen
  \bibfield  {author} {\bibinfo {author} {\bibfnamefont {F.}~\bibnamefont {Oz}}, \bibinfo {author} {\bibfnamefont {O.}~\bibnamefont {San}},\ and\ \bibinfo {author} {\bibfnamefont {K.}~\bibnamefont {Kara}},\ }\bibfield  {title} {\bibinfo {title} {An efficient quantum partial differential equation solver with chebyshev points},\ }\href@noop {} {\bibfield  {journal} {\bibinfo  {journal} {Scientific Reports}\ }\textbf {\bibinfo {volume} {13}},\ \bibinfo {pages} {7767} (\bibinfo {year} {2023})}\BibitemShut {NoStop}%
\bibitem [{\citenamefont {Zylberman}\ \emph {et~al.}(2022)\citenamefont {Zylberman}, \citenamefont {Di~Molfetta}, \citenamefont {Brachet}, \citenamefont {Loureiro},\ and\ \citenamefont {Debbasch}}]{zylberman2022quantum}%
  \BibitemOpen
  \bibfield  {author} {\bibinfo {author} {\bibfnamefont {J.}~\bibnamefont {Zylberman}}, \bibinfo {author} {\bibfnamefont {G.}~\bibnamefont {Di~Molfetta}}, \bibinfo {author} {\bibfnamefont {M.}~\bibnamefont {Brachet}}, \bibinfo {author} {\bibfnamefont {N.~F.}\ \bibnamefont {Loureiro}},\ and\ \bibinfo {author} {\bibfnamefont {F.}~\bibnamefont {Debbasch}},\ }\bibfield  {title} {\bibinfo {title} {Quantum simulations of hydrodynamics via the madelung transformation},\ }\href@noop {} {\bibfield  {journal} {\bibinfo  {journal} {Physical Review A}\ }\textbf {\bibinfo {volume} {106}},\ \bibinfo {pages} {032408} (\bibinfo {year} {2022})}\BibitemShut {NoStop}%
\bibitem [{\citenamefont {Shukla}\ and\ \citenamefont {Vedula}(2023)}]{shukla2023hybrid}%
  \BibitemOpen
  \bibfield  {author} {\bibinfo {author} {\bibfnamefont {A.}~\bibnamefont {Shukla}}\ and\ \bibinfo {author} {\bibfnamefont {P.}~\bibnamefont {Vedula}},\ }\bibfield  {title} {\bibinfo {title} {A hybrid classical-quantum algorithm for solution of nonlinear ordinary differential equations},\ }\href@noop {} {\bibfield  {journal} {\bibinfo  {journal} {Applied Mathematics and Computation}\ }\textbf {\bibinfo {volume} {442}},\ \bibinfo {pages} {127708} (\bibinfo {year} {2023})}\BibitemShut {NoStop}%
\bibitem [{\citenamefont {Lubasch}\ \emph {et~al.}(2020)\citenamefont {Lubasch}, \citenamefont {Joo}, \citenamefont {Moinier}, \citenamefont {Kiffner},\ and\ \citenamefont {Jaksch}}]{lubasch2020variational}%
  \BibitemOpen
  \bibfield  {author} {\bibinfo {author} {\bibfnamefont {M.}~\bibnamefont {Lubasch}}, \bibinfo {author} {\bibfnamefont {J.}~\bibnamefont {Joo}}, \bibinfo {author} {\bibfnamefont {P.}~\bibnamefont {Moinier}}, \bibinfo {author} {\bibfnamefont {M.}~\bibnamefont {Kiffner}},\ and\ \bibinfo {author} {\bibfnamefont {D.}~\bibnamefont {Jaksch}},\ }\bibfield  {title} {\bibinfo {title} {Variational quantum algorithms for nonlinear problems},\ }\href@noop {} {\bibfield  {journal} {\bibinfo  {journal} {Physical Review A}\ }\textbf {\bibinfo {volume} {101}},\ \bibinfo {pages} {010301} (\bibinfo {year} {2020})}\BibitemShut {NoStop}%
\bibitem [{\citenamefont {Liu}\ \emph {et~al.}(2021{\natexlab{b}})\citenamefont {Liu}, \citenamefont {Wu}, \citenamefont {Wan}, \citenamefont {Pan}, \citenamefont {Qin}, \citenamefont {Gao},\ and\ \citenamefont {Wen}}]{liu2021variational}%
  \BibitemOpen
  \bibfield  {author} {\bibinfo {author} {\bibfnamefont {H.-L.}\ \bibnamefont {Liu}}, \bibinfo {author} {\bibfnamefont {Y.-S.}\ \bibnamefont {Wu}}, \bibinfo {author} {\bibfnamefont {L.-C.}\ \bibnamefont {Wan}}, \bibinfo {author} {\bibfnamefont {S.-J.}\ \bibnamefont {Pan}}, \bibinfo {author} {\bibfnamefont {S.-J.}\ \bibnamefont {Qin}}, \bibinfo {author} {\bibfnamefont {F.}~\bibnamefont {Gao}},\ and\ \bibinfo {author} {\bibfnamefont {Q.-Y.}\ \bibnamefont {Wen}},\ }\bibfield  {title} {\bibinfo {title} {Variational quantum algorithm for the poisson equation},\ }\href@noop {} {\bibfield  {journal} {\bibinfo  {journal} {Physical Review A}\ }\textbf {\bibinfo {volume} {104}},\ \bibinfo {pages} {022418} (\bibinfo {year} {2021}{\natexlab{b}})}\BibitemShut {NoStop}%
\bibitem [{\citenamefont {Liu}\ \emph {et~al.}(2022)\citenamefont {Liu}, \citenamefont {Chen}, \citenamefont {Shu}, \citenamefont {Chew}, \citenamefont {Khoo}, \citenamefont {Zhao},\ and\ \citenamefont {Cui}}]{liu2022application}%
  \BibitemOpen
  \bibfield  {author} {\bibinfo {author} {\bibfnamefont {Y.}~\bibnamefont {Liu}}, \bibinfo {author} {\bibfnamefont {Z.}~\bibnamefont {Chen}}, \bibinfo {author} {\bibfnamefont {C.}~\bibnamefont {Shu}}, \bibinfo {author} {\bibfnamefont {S.-C.}\ \bibnamefont {Chew}}, \bibinfo {author} {\bibfnamefont {B.~C.}\ \bibnamefont {Khoo}}, \bibinfo {author} {\bibfnamefont {X.}~\bibnamefont {Zhao}},\ and\ \bibinfo {author} {\bibfnamefont {Y.}~\bibnamefont {Cui}},\ }\bibfield  {title} {\bibinfo {title} {Application of a variational hybrid quantum-classical algorithm to heat conduction equation and analysis of time complexity},\ }\href@noop {} {\bibfield  {journal} {\bibinfo  {journal} {Physics of Fluids}\ }\textbf {\bibinfo {volume} {34}},\ \bibinfo {pages} {117121} (\bibinfo {year} {2022})}\BibitemShut {NoStop}%
\bibitem [{\citenamefont {Kubo}\ \emph {et~al.}(2021)\citenamefont {Kubo}, \citenamefont {Nakagawa}, \citenamefont {Endo},\ and\ \citenamefont {Nagayama}}]{kubo2021variational}%
  \BibitemOpen
  \bibfield  {author} {\bibinfo {author} {\bibfnamefont {K.}~\bibnamefont {Kubo}}, \bibinfo {author} {\bibfnamefont {Y.~O.}\ \bibnamefont {Nakagawa}}, \bibinfo {author} {\bibfnamefont {S.}~\bibnamefont {Endo}},\ and\ \bibinfo {author} {\bibfnamefont {S.}~\bibnamefont {Nagayama}},\ }\bibfield  {title} {\bibinfo {title} {Variational quantum simulations of stochastic differential equations},\ }\href@noop {} {\bibfield  {journal} {\bibinfo  {journal} {Physical Review A}\ }\textbf {\bibinfo {volume} {103}},\ \bibinfo {pages} {052425} (\bibinfo {year} {2021})}\BibitemShut {NoStop}%
\bibitem [{\citenamefont {Demirdjian}\ \emph {et~al.}(2022)\citenamefont {Demirdjian}, \citenamefont {Gunlycke}, \citenamefont {Reynolds}, \citenamefont {Doyle},\ and\ \citenamefont {Tafur}}]{demirdjian2022variational}%
  \BibitemOpen
  \bibfield  {author} {\bibinfo {author} {\bibfnamefont {R.}~\bibnamefont {Demirdjian}}, \bibinfo {author} {\bibfnamefont {D.}~\bibnamefont {Gunlycke}}, \bibinfo {author} {\bibfnamefont {C.~A.}\ \bibnamefont {Reynolds}}, \bibinfo {author} {\bibfnamefont {J.~D.}\ \bibnamefont {Doyle}},\ and\ \bibinfo {author} {\bibfnamefont {S.}~\bibnamefont {Tafur}},\ }\bibfield  {title} {\bibinfo {title} {Variational quantum solutions to the advection--diffusion equation for applications in fluid dynamics},\ }\href@noop {} {\bibfield  {journal} {\bibinfo  {journal} {Quantum Information Processing}\ }\textbf {\bibinfo {volume} {21}},\ \bibinfo {pages} {322} (\bibinfo {year} {2022})}\BibitemShut {NoStop}%
\bibitem [{\citenamefont {Garc{\'\i}a-Molina}\ \emph {et~al.}(2022)\citenamefont {Garc{\'\i}a-Molina}, \citenamefont {Rodr{\'\i}guez-Mediavilla},\ and\ \citenamefont {Garc{\'\i}a-Ripoll}}]{garcia2022quantum}%
  \BibitemOpen
  \bibfield  {author} {\bibinfo {author} {\bibfnamefont {P.}~\bibnamefont {Garc{\'\i}a-Molina}}, \bibinfo {author} {\bibfnamefont {J.}~\bibnamefont {Rodr{\'\i}guez-Mediavilla}},\ and\ \bibinfo {author} {\bibfnamefont {J.~J.}\ \bibnamefont {Garc{\'\i}a-Ripoll}},\ }\bibfield  {title} {\bibinfo {title} {Quantum fourier analysis for multivariate functions and applications to a class of schr{\"o}dinger-type partial differential equations},\ }\href@noop {} {\bibfield  {journal} {\bibinfo  {journal} {Physical Review A}\ }\textbf {\bibinfo {volume} {105}},\ \bibinfo {pages} {012433} (\bibinfo {year} {2022})}\BibitemShut {NoStop}%
\bibitem [{\citenamefont {Zanger}\ \emph {et~al.}(2021)\citenamefont {Zanger}, \citenamefont {Mendl}, \citenamefont {Schulz},\ and\ \citenamefont {Schreiber}}]{zanger2021quantum}%
  \BibitemOpen
  \bibfield  {author} {\bibinfo {author} {\bibfnamefont {B.}~\bibnamefont {Zanger}}, \bibinfo {author} {\bibfnamefont {C.~B.}\ \bibnamefont {Mendl}}, \bibinfo {author} {\bibfnamefont {M.}~\bibnamefont {Schulz}},\ and\ \bibinfo {author} {\bibfnamefont {M.}~\bibnamefont {Schreiber}},\ }\bibfield  {title} {\bibinfo {title} {Quantum algorithms for solving ordinary differential equations via classical integration methods},\ }\href@noop {} {\bibfield  {journal} {\bibinfo  {journal} {Quantum}\ }\textbf {\bibinfo {volume} {5}},\ \bibinfo {pages} {502} (\bibinfo {year} {2021})}\BibitemShut {NoStop}%
\bibitem [{\citenamefont {Costa}\ \emph {et~al.}(2022)\citenamefont {Costa}, \citenamefont {An}, \citenamefont {Sanders}, \citenamefont {Su}, \citenamefont {Babbush},\ and\ \citenamefont {Berry}}]{costa2022optimal}%
  \BibitemOpen
  \bibfield  {author} {\bibinfo {author} {\bibfnamefont {P.~C.}\ \bibnamefont {Costa}}, \bibinfo {author} {\bibfnamefont {D.}~\bibnamefont {An}}, \bibinfo {author} {\bibfnamefont {Y.~R.}\ \bibnamefont {Sanders}}, \bibinfo {author} {\bibfnamefont {Y.}~\bibnamefont {Su}}, \bibinfo {author} {\bibfnamefont {R.}~\bibnamefont {Babbush}},\ and\ \bibinfo {author} {\bibfnamefont {D.~W.}\ \bibnamefont {Berry}},\ }\bibfield  {title} {\bibinfo {title} {Optimal scaling quantum linear-systems solver via discrete adiabatic theorem},\ }\href@noop {} {\bibfield  {journal} {\bibinfo  {journal} {PRX Quantum}\ }\textbf {\bibinfo {volume} {3}},\ \bibinfo {pages} {040303} (\bibinfo {year} {2022})}\BibitemShut {NoStop}%
\bibitem [{\citenamefont {G{\'o}es}\ \emph {et~al.}(2023)\citenamefont {G{\'o}es}, \citenamefont {Maciel}, \citenamefont {Pollachini}, \citenamefont {Salazar}, \citenamefont {Cuenca},\ and\ \citenamefont {Duzzioni}}]{goes2023qboost}%
  \BibitemOpen
  \bibfield  {author} {\bibinfo {author} {\bibfnamefont {C.~B.}\ \bibnamefont {G{\'o}es}}, \bibinfo {author} {\bibfnamefont {T.~O.}\ \bibnamefont {Maciel}}, \bibinfo {author} {\bibfnamefont {G.~G.}\ \bibnamefont {Pollachini}}, \bibinfo {author} {\bibfnamefont {J.~P.}\ \bibnamefont {Salazar}}, \bibinfo {author} {\bibfnamefont {R.~G.}\ \bibnamefont {Cuenca}},\ and\ \bibinfo {author} {\bibfnamefont {E.~I.}\ \bibnamefont {Duzzioni}},\ }\bibfield  {title} {\bibinfo {title} {Qboost for regression problems: solving partial differential equations},\ }\href@noop {} {\bibfield  {journal} {\bibinfo  {journal} {Quantum Information Processing}\ }\textbf {\bibinfo {volume} {22}},\ \bibinfo {pages} {129} (\bibinfo {year} {2023})}\BibitemShut {NoStop}%
\bibitem [{\citenamefont {Chandra}\ \emph {et~al.}(2014)\citenamefont {Chandra}, \citenamefont {Jacobson}, \citenamefont {Moussa}, \citenamefont {Frankel},\ and\ \citenamefont {Kais}}]{chandra2014quadratic}%
  \BibitemOpen
  \bibfield  {author} {\bibinfo {author} {\bibfnamefont {R.}~\bibnamefont {Chandra}}, \bibinfo {author} {\bibfnamefont {N.~T.}\ \bibnamefont {Jacobson}}, \bibinfo {author} {\bibfnamefont {J.~E.}\ \bibnamefont {Moussa}}, \bibinfo {author} {\bibfnamefont {S.~H.}\ \bibnamefont {Frankel}},\ and\ \bibinfo {author} {\bibfnamefont {S.}~\bibnamefont {Kais}},\ }\bibfield  {title} {\bibinfo {title} {Quadratic constrained mixed discrete optimization with an adiabatic quantum optimizer},\ }\href@noop {} {\bibfield  {journal} {\bibinfo  {journal} {Physical Review A}\ }\textbf {\bibinfo {volume} {90}},\ \bibinfo {pages} {012308} (\bibinfo {year} {2014})}\BibitemShut {NoStop}%
\bibitem [{\citenamefont {Pollachini}\ \emph {et~al.}(2021)\citenamefont {Pollachini}, \citenamefont {Salazar}, \citenamefont {G\'oes}, \citenamefont {Maciel},\ and\ \citenamefont {Duzzioni}}]{PhysRevA.104.032426}%
  \BibitemOpen
  \bibfield  {author} {\bibinfo {author} {\bibfnamefont {G.~G.}\ \bibnamefont {Pollachini}}, \bibinfo {author} {\bibfnamefont {J.~P. L.~C.}\ \bibnamefont {Salazar}}, \bibinfo {author} {\bibfnamefont {C.~B.~D.}\ \bibnamefont {G\'oes}}, \bibinfo {author} {\bibfnamefont {T.~O.}\ \bibnamefont {Maciel}},\ and\ \bibinfo {author} {\bibfnamefont {E.~I.}\ \bibnamefont {Duzzioni}},\ }\bibfield  {title} {\bibinfo {title} {Hybrid classical-quantum approach to solve the heat equation using quantum annealers},\ }\href {https://doi.org/10.1103/PhysRevA.104.032426} {\bibfield  {journal} {\bibinfo  {journal} {Phys. Rev. A}\ }\textbf {\bibinfo {volume} {104}},\ \bibinfo {pages} {032426} (\bibinfo {year} {2021})}\BibitemShut {NoStop}%
\bibitem [{\citenamefont {Criado}\ and\ \citenamefont {Spannowsky}(2022)}]{criado2022qade}%
  \BibitemOpen
  \bibfield  {author} {\bibinfo {author} {\bibfnamefont {J.~C.}\ \bibnamefont {Criado}}\ and\ \bibinfo {author} {\bibfnamefont {M.}~\bibnamefont {Spannowsky}},\ }\bibfield  {title} {\bibinfo {title} {Qade: solving differential equations on quantum annealers},\ }\href@noop {} {\bibfield  {journal} {\bibinfo  {journal} {Quantum Science and Technology}\ }\textbf {\bibinfo {volume} {8}},\ \bibinfo {pages} {015021} (\bibinfo {year} {2022})}\BibitemShut {NoStop}%
\bibitem [{\citenamefont {Greer}\ and\ \citenamefont {O’Malley}(2020)}]{greer2020approach}%
  \BibitemOpen
  \bibfield  {author} {\bibinfo {author} {\bibfnamefont {S.}~\bibnamefont {Greer}}\ and\ \bibinfo {author} {\bibfnamefont {D.}~\bibnamefont {O’Malley}},\ }\bibfield  {title} {\bibinfo {title} {An approach to seismic inversion with quantum annealing},\ }in\ \href@noop {} {\emph {\bibinfo {booktitle} {SEG Technical Program Expanded Abstracts 2020}}}\ (\bibinfo  {publisher} {Society of Exploration Geophysicists},\ \bibinfo {year} {2020})\ pp.\ \bibinfo {pages} {2845--2849}\BibitemShut {NoStop}%
\bibitem [{\citenamefont {Grover}\ and\ \citenamefont {Rudolph}(2002)}]{grover2002creating}%
  \BibitemOpen
  \bibfield  {author} {\bibinfo {author} {\bibfnamefont {L.}~\bibnamefont {Grover}}\ and\ \bibinfo {author} {\bibfnamefont {T.}~\bibnamefont {Rudolph}},\ }\bibfield  {title} {\bibinfo {title} {Creating superpositions that correspond to efficiently integrable probability distributions},\ }\href@noop {} {\bibfield  {journal} {\bibinfo  {journal} {arXiv preprint quant-ph/0208112}\ } (\bibinfo {year} {2002})}\BibitemShut {NoStop}%
\bibitem [{\citenamefont {Mottonen}\ \emph {et~al.}(2004)\citenamefont {Mottonen}, \citenamefont {Vartiainen}, \citenamefont {Bergholm},\ and\ \citenamefont {Salomaa}}]{mottonen2004transformation}%
  \BibitemOpen
  \bibfield  {author} {\bibinfo {author} {\bibfnamefont {M.}~\bibnamefont {Mottonen}}, \bibinfo {author} {\bibfnamefont {J.~J.}\ \bibnamefont {Vartiainen}}, \bibinfo {author} {\bibfnamefont {V.}~\bibnamefont {Bergholm}},\ and\ \bibinfo {author} {\bibfnamefont {M.~M.}\ \bibnamefont {Salomaa}},\ }\bibfield  {title} {\bibinfo {title} {Transformation of quantum states using uniformly controlled rotations},\ }\href@noop {} {\bibfield  {journal} {\bibinfo  {journal} {arXiv preprint quant-ph/0407010}\ } (\bibinfo {year} {2004})}\BibitemShut {NoStop}%
\bibitem [{\citenamefont {Plesch}\ and\ \citenamefont {Brukner}(2011)}]{plesch2011quantum}%
  \BibitemOpen
  \bibfield  {author} {\bibinfo {author} {\bibfnamefont {M.}~\bibnamefont {Plesch}}\ and\ \bibinfo {author} {\bibfnamefont {{\v{C}}.}~\bibnamefont {Brukner}},\ }\bibfield  {title} {\bibinfo {title} {Quantum-state preparation with universal gate decompositions},\ }\href@noop {} {\bibfield  {journal} {\bibinfo  {journal} {Physical Review A}\ }\textbf {\bibinfo {volume} {83}},\ \bibinfo {pages} {032302} (\bibinfo {year} {2011})}\BibitemShut {NoStop}%
\bibitem [{\citenamefont {Sun}\ \emph {et~al.}(2023)\citenamefont {Sun}, \citenamefont {Tian}, \citenamefont {Yang}, \citenamefont {Yuan},\ and\ \citenamefont {Zhang}}]{sun2023asymptotically}%
  \BibitemOpen
  \bibfield  {author} {\bibinfo {author} {\bibfnamefont {X.}~\bibnamefont {Sun}}, \bibinfo {author} {\bibfnamefont {G.}~\bibnamefont {Tian}}, \bibinfo {author} {\bibfnamefont {S.}~\bibnamefont {Yang}}, \bibinfo {author} {\bibfnamefont {P.}~\bibnamefont {Yuan}},\ and\ \bibinfo {author} {\bibfnamefont {S.}~\bibnamefont {Zhang}},\ }\bibfield  {title} {\bibinfo {title} {Asymptotically optimal circuit depth for quantum state preparation and general unitary synthesis},\ }\href@noop {} {\bibfield  {journal} {\bibinfo  {journal} {IEEE Transactions on Computer-Aided Design of Integrated Circuits and Systems}\ } (\bibinfo {year} {2023})}\BibitemShut {NoStop}%
\bibitem [{\citenamefont {Zhang}\ \emph {et~al.}(2021)\citenamefont {Zhang}, \citenamefont {Yung},\ and\ \citenamefont {Yuan}}]{PhysRevResearch.3.043200}%
  \BibitemOpen
  \bibfield  {author} {\bibinfo {author} {\bibfnamefont {X.-M.}\ \bibnamefont {Zhang}}, \bibinfo {author} {\bibfnamefont {M.-H.}\ \bibnamefont {Yung}},\ and\ \bibinfo {author} {\bibfnamefont {X.}~\bibnamefont {Yuan}},\ }\bibfield  {title} {\bibinfo {title} {Low-depth quantum state preparation},\ }\href {https://doi.org/10.1103/PhysRevResearch.3.043200} {\bibfield  {journal} {\bibinfo  {journal} {Phys. Rev. Res.}\ }\textbf {\bibinfo {volume} {3}},\ \bibinfo {pages} {043200} (\bibinfo {year} {2021})}\BibitemShut {NoStop}%
\bibitem [{\citenamefont {Zhang}\ \emph {et~al.}(2022)\citenamefont {Zhang}, \citenamefont {Li},\ and\ \citenamefont {Yuan}}]{zhang2022quantum}%
  \BibitemOpen
  \bibfield  {author} {\bibinfo {author} {\bibfnamefont {X.-M.}\ \bibnamefont {Zhang}}, \bibinfo {author} {\bibfnamefont {T.}~\bibnamefont {Li}},\ and\ \bibinfo {author} {\bibfnamefont {X.}~\bibnamefont {Yuan}},\ }\bibfield  {title} {\bibinfo {title} {Quantum state preparation with optimal circuit depth: Implementations and applications},\ }\href@noop {} {\bibfield  {journal} {\bibinfo  {journal} {Physical Review Letters}\ }\textbf {\bibinfo {volume} {129}},\ \bibinfo {pages} {230504} (\bibinfo {year} {2022})}\BibitemShut {NoStop}%
\bibitem [{\citenamefont {Araujo}\ \emph {et~al.}(2021)\citenamefont {Araujo}, \citenamefont {Park}, \citenamefont {Petruccione},\ and\ \citenamefont {da~Silva}}]{araujo2021divide}%
  \BibitemOpen
  \bibfield  {author} {\bibinfo {author} {\bibfnamefont {I.~F.}\ \bibnamefont {Araujo}}, \bibinfo {author} {\bibfnamefont {D.~K.}\ \bibnamefont {Park}}, \bibinfo {author} {\bibfnamefont {F.}~\bibnamefont {Petruccione}},\ and\ \bibinfo {author} {\bibfnamefont {A.~J.}\ \bibnamefont {da~Silva}},\ }\bibfield  {title} {\bibinfo {title} {A divide-and-conquer algorithm for quantum state preparation},\ }\href@noop {} {\bibfield  {journal} {\bibinfo  {journal} {Scientific reports}\ }\textbf {\bibinfo {volume} {11}},\ \bibinfo {pages} {1} (\bibinfo {year} {2021})}\BibitemShut {NoStop}%
\bibitem [{\citenamefont {Nakaji}\ \emph {et~al.}(2022)\citenamefont {Nakaji}, \citenamefont {Uno}, \citenamefont {Suzuki}, \citenamefont {Raymond}, \citenamefont {Onodera}, \citenamefont {Tanaka}, \citenamefont {Tezuka}, \citenamefont {Mitsuda},\ and\ \citenamefont {Yamamoto}}]{nakaji2022approximate}%
  \BibitemOpen
  \bibfield  {author} {\bibinfo {author} {\bibfnamefont {K.}~\bibnamefont {Nakaji}}, \bibinfo {author} {\bibfnamefont {S.}~\bibnamefont {Uno}}, \bibinfo {author} {\bibfnamefont {Y.}~\bibnamefont {Suzuki}}, \bibinfo {author} {\bibfnamefont {R.}~\bibnamefont {Raymond}}, \bibinfo {author} {\bibfnamefont {T.}~\bibnamefont {Onodera}}, \bibinfo {author} {\bibfnamefont {T.}~\bibnamefont {Tanaka}}, \bibinfo {author} {\bibfnamefont {H.}~\bibnamefont {Tezuka}}, \bibinfo {author} {\bibfnamefont {N.}~\bibnamefont {Mitsuda}},\ and\ \bibinfo {author} {\bibfnamefont {N.}~\bibnamefont {Yamamoto}},\ }\bibfield  {title} {\bibinfo {title} {Approximate amplitude encoding in shallow parameterized quantum circuits and its application to financial market indicators},\ }\href@noop {} {\bibfield  {journal} {\bibinfo  {journal} {Physical Review Research}\ }\textbf {\bibinfo {volume} {4}},\ \bibinfo {pages} {023136} (\bibinfo {year} {2022})}\BibitemShut {NoStop}%
\bibitem [{\citenamefont {Zoufal}\ \emph {et~al.}(2019)\citenamefont {Zoufal}, \citenamefont {Lucchi},\ and\ \citenamefont {Woerner}}]{zoufal2019quantum}%
  \BibitemOpen
  \bibfield  {author} {\bibinfo {author} {\bibfnamefont {C.}~\bibnamefont {Zoufal}}, \bibinfo {author} {\bibfnamefont {A.}~\bibnamefont {Lucchi}},\ and\ \bibinfo {author} {\bibfnamefont {S.}~\bibnamefont {Woerner}},\ }\bibfield  {title} {\bibinfo {title} {Quantum generative adversarial networks for learning and loading random distributions},\ }\href@noop {} {\bibfield  {journal} {\bibinfo  {journal} {npj Quantum Information}\ }\textbf {\bibinfo {volume} {5}},\ \bibinfo {pages} {103} (\bibinfo {year} {2019})}\BibitemShut {NoStop}%
\bibitem [{\citenamefont {Cerezo}\ \emph {et~al.}(2021)\citenamefont {Cerezo}, \citenamefont {Arrasmith}, \citenamefont {Babbush}, \citenamefont {Benjamin}, \citenamefont {Endo}, \citenamefont {Fujii}, \citenamefont {McClean}, \citenamefont {Mitarai}, \citenamefont {Yuan}, \citenamefont {Cincio} \emph {et~al.}}]{cerezo2021variational}%
  \BibitemOpen
  \bibfield  {author} {\bibinfo {author} {\bibfnamefont {M.}~\bibnamefont {Cerezo}}, \bibinfo {author} {\bibfnamefont {A.}~\bibnamefont {Arrasmith}}, \bibinfo {author} {\bibfnamefont {R.}~\bibnamefont {Babbush}}, \bibinfo {author} {\bibfnamefont {S.~C.}\ \bibnamefont {Benjamin}}, \bibinfo {author} {\bibfnamefont {S.}~\bibnamefont {Endo}}, \bibinfo {author} {\bibfnamefont {K.}~\bibnamefont {Fujii}}, \bibinfo {author} {\bibfnamefont {J.~R.}\ \bibnamefont {McClean}}, \bibinfo {author} {\bibfnamefont {K.}~\bibnamefont {Mitarai}}, \bibinfo {author} {\bibfnamefont {X.}~\bibnamefont {Yuan}}, \bibinfo {author} {\bibfnamefont {L.}~\bibnamefont {Cincio}}, \emph {et~al.},\ }\bibfield  {title} {\bibinfo {title} {Variational quantum algorithms},\ }\href@noop {} {\bibfield  {journal} {\bibinfo  {journal} {Nature Reviews Physics}\ }\textbf {\bibinfo {volume} {3}},\ \bibinfo {pages} {625} (\bibinfo {year} {2021})}\BibitemShut {NoStop}%
\bibitem [{\citenamefont {McClean}\ \emph {et~al.}(2018)\citenamefont {McClean}, \citenamefont {Boixo}, \citenamefont {Smelyanskiy}, \citenamefont {Babbush},\ and\ \citenamefont {Neven}}]{mcclean2018barren}%
  \BibitemOpen
  \bibfield  {author} {\bibinfo {author} {\bibfnamefont {J.~R.}\ \bibnamefont {McClean}}, \bibinfo {author} {\bibfnamefont {S.}~\bibnamefont {Boixo}}, \bibinfo {author} {\bibfnamefont {V.~N.}\ \bibnamefont {Smelyanskiy}}, \bibinfo {author} {\bibfnamefont {R.}~\bibnamefont {Babbush}},\ and\ \bibinfo {author} {\bibfnamefont {H.}~\bibnamefont {Neven}},\ }\bibfield  {title} {\bibinfo {title} {Barren plateaus in quantum neural network training landscapes},\ }\href@noop {} {\bibfield  {journal} {\bibinfo  {journal} {Nature communications}\ }\textbf {\bibinfo {volume} {9}},\ \bibinfo {pages} {4812} (\bibinfo {year} {2018})}\BibitemShut {NoStop}%
\bibitem [{\citenamefont {Marin-Sanchez}\ \emph {et~al.}(2021)\citenamefont {Marin-Sanchez}, \citenamefont {Gonzalez-Conde},\ and\ \citenamefont {Sanz}}]{marin2021quantum}%
  \BibitemOpen
  \bibfield  {author} {\bibinfo {author} {\bibfnamefont {G.}~\bibnamefont {Marin-Sanchez}}, \bibinfo {author} {\bibfnamefont {J.}~\bibnamefont {Gonzalez-Conde}},\ and\ \bibinfo {author} {\bibfnamefont {M.}~\bibnamefont {Sanz}},\ }\bibfield  {title} {\bibinfo {title} {Quantum algorithms for approximate function loading},\ }\href@noop {} {\bibfield  {journal} {\bibinfo  {journal} {arXiv preprint arXiv:2111.07933}\ } (\bibinfo {year} {2021})}\BibitemShut {NoStop}%
\bibitem [{\citenamefont {Moosa}\ \emph {et~al.}(2023)\citenamefont {Moosa}, \citenamefont {Watts}, \citenamefont {Chen}, \citenamefont {Sarma},\ and\ \citenamefont {McMahon}}]{moosa2023linear}%
  \BibitemOpen
  \bibfield  {author} {\bibinfo {author} {\bibfnamefont {M.}~\bibnamefont {Moosa}}, \bibinfo {author} {\bibfnamefont {T.~W.}\ \bibnamefont {Watts}}, \bibinfo {author} {\bibfnamefont {Y.}~\bibnamefont {Chen}}, \bibinfo {author} {\bibfnamefont {A.}~\bibnamefont {Sarma}},\ and\ \bibinfo {author} {\bibfnamefont {P.~L.}\ \bibnamefont {McMahon}},\ }\bibfield  {title} {\bibinfo {title} {Linear-depth quantum circuits for loading fourier approximations of arbitrary functions},\ }\href@noop {} {\bibfield  {journal} {\bibinfo  {journal} {arXiv preprint arXiv:2302.03888}\ } (\bibinfo {year} {2023})}\BibitemShut {NoStop}%
\bibitem [{\citenamefont {Rattew}\ and\ \citenamefont {Koczor}(2022)}]{rattew2022preparing}%
  \BibitemOpen
  \bibfield  {author} {\bibinfo {author} {\bibfnamefont {A.~G.}\ \bibnamefont {Rattew}}\ and\ \bibinfo {author} {\bibfnamefont {B.}~\bibnamefont {Koczor}},\ }\bibfield  {title} {\bibinfo {title} {Preparing arbitrary continuous functions in quantum registers with logarithmic complexity},\ }\href@noop {} {\bibfield  {journal} {\bibinfo  {journal} {arXiv preprint arXiv:2205.00519}\ } (\bibinfo {year} {2022})}\BibitemShut {NoStop}%
\bibitem [{\citenamefont {Welch}\ \emph {et~al.}(2014)\citenamefont {Welch}, \citenamefont {Greenbaum}, \citenamefont {Mostame},\ and\ \citenamefont {Aspuru-Guzik}}]{welch2014efficient}%
  \BibitemOpen
  \bibfield  {author} {\bibinfo {author} {\bibfnamefont {J.}~\bibnamefont {Welch}}, \bibinfo {author} {\bibfnamefont {D.}~\bibnamefont {Greenbaum}}, \bibinfo {author} {\bibfnamefont {S.}~\bibnamefont {Mostame}},\ and\ \bibinfo {author} {\bibfnamefont {A.}~\bibnamefont {Aspuru-Guzik}},\ }\bibfield  {title} {\bibinfo {title} {Efficient quantum circuits for diagonal unitaries without ancillas},\ }\href@noop {} {\bibfield  {journal} {\bibinfo  {journal} {New Journal of Physics}\ }\textbf {\bibinfo {volume} {16}},\ \bibinfo {pages} {033040} (\bibinfo {year} {2014})}\BibitemShut {NoStop}%
\bibitem [{\citenamefont {Walsh}(1923)}]{walsh1923closed}%
  \BibitemOpen
  \bibfield  {author} {\bibinfo {author} {\bibfnamefont {J.~L.}\ \bibnamefont {Walsh}},\ }\bibfield  {title} {\bibinfo {title} {A closed set of normal orthogonal functions},\ }\href@noop {} {\bibfield  {journal} {\bibinfo  {journal} {American Journal of Mathematics}\ }\textbf {\bibinfo {volume} {45}},\ \bibinfo {pages} {5} (\bibinfo {year} {1923})}\BibitemShut {NoStop}%
\bibitem [{\citenamefont {Yuen}(1975)}]{yuen1975function}%
  \BibitemOpen
  \bibfield  {author} {\bibinfo {author} {\bibfnamefont {C.-K.}\ \bibnamefont {Yuen}},\ }\bibfield  {title} {\bibinfo {title} {Function approximation by walsh series},\ }\href@noop {} {\bibfield  {journal} {\bibinfo  {journal} {IEEE Transactions on Computers}\ }\textbf {\bibinfo {volume} {100}},\ \bibinfo {pages} {590} (\bibinfo {year} {1975})}\BibitemShut {NoStop}%
\bibitem [{\citenamefont {Bullock}\ and\ \citenamefont {Markov}(2004)}]{bullock2004asymptotically}%
  \BibitemOpen
  \bibfield  {author} {\bibinfo {author} {\bibfnamefont {S.~S.}\ \bibnamefont {Bullock}}\ and\ \bibinfo {author} {\bibfnamefont {I.~L.}\ \bibnamefont {Markov}},\ }\bibfield  {title} {\bibinfo {title} {Asymptotically optimal circuits for arbitrary n-qubit diagonal comutations},\ }\href@noop {} {\bibfield  {journal} {\bibinfo  {journal} {Quantum Information \& Computation}\ }\textbf {\bibinfo {volume} {4}},\ \bibinfo {pages} {27} (\bibinfo {year} {2004})}\BibitemShut {NoStop}%
\bibitem [{\citenamefont {Shende}\ and\ \citenamefont {Markov}(2008)}]{shende2008cnot}%
  \BibitemOpen
  \bibfield  {author} {\bibinfo {author} {\bibfnamefont {V.~V.}\ \bibnamefont {Shende}}\ and\ \bibinfo {author} {\bibfnamefont {I.~L.}\ \bibnamefont {Markov}},\ }\bibfield  {title} {\bibinfo {title} {On the cnot-cost of toffoli gates},\ }\href@noop {} {\bibfield  {journal} {\bibinfo  {journal} {arXiv preprint arXiv:0803.2316}\ } (\bibinfo {year} {2008})}\BibitemShut {NoStop}%
\bibitem [{\citenamefont {Nielsen}\ and\ \citenamefont {Chuang}(2002)}]{nielsen2002quantum}%
  \BibitemOpen
  \bibfield  {author} {\bibinfo {author} {\bibfnamefont {M.~A.}\ \bibnamefont {Nielsen}}\ and\ \bibinfo {author} {\bibfnamefont {I.}~\bibnamefont {Chuang}},\ }\href@noop {} {\bibinfo {title} {Quantum computation and quantum information}} (\bibinfo {year} {2002})\BibitemShut {NoStop}%
\bibitem [{\citenamefont {Vandaele}\ \emph {et~al.}(2022)\citenamefont {Vandaele}, \citenamefont {Martiel},\ and\ \citenamefont {de~Brugi{\`e}re}}]{vandaele2022phase}%
  \BibitemOpen
  \bibfield  {author} {\bibinfo {author} {\bibfnamefont {V.}~\bibnamefont {Vandaele}}, \bibinfo {author} {\bibfnamefont {S.}~\bibnamefont {Martiel}},\ and\ \bibinfo {author} {\bibfnamefont {T.~G.}\ \bibnamefont {de~Brugi{\`e}re}},\ }\bibfield  {title} {\bibinfo {title} {Phase polynomials synthesis algorithms for nisq architectures and beyond},\ }\href@noop {} {\bibfield  {journal} {\bibinfo  {journal} {Quantum Science and Technology}\ }\textbf {\bibinfo {volume} {7}},\ \bibinfo {pages} {045027} (\bibinfo {year} {2022})}\BibitemShut {NoStop}%
\bibitem [{\citenamefont {Holmes}\ and\ \citenamefont {Matsuura}(2020)}]{holmes2020efficient}%
  \BibitemOpen
  \bibfield  {author} {\bibinfo {author} {\bibfnamefont {A.}~\bibnamefont {Holmes}}\ and\ \bibinfo {author} {\bibfnamefont {A.~Y.}\ \bibnamefont {Matsuura}},\ }\bibfield  {title} {\bibinfo {title} {Efficient quantum circuits for accurate state preparation of smooth, differentiable functions},\ }in\ \href@noop {} {\emph {\bibinfo {booktitle} {2020 IEEE International Conference on Quantum Computing and Engineering (QCE)}}}\ (\bibinfo {organization} {IEEE},\ \bibinfo {year} {2020})\ pp.\ \bibinfo {pages} {169--179}\BibitemShut {NoStop}%
\bibitem [{\citenamefont {Kushilevitz}\ and\ \citenamefont {Mansour}(1991)}]{kushilevitz1991learning}%
  \BibitemOpen
  \bibfield  {author} {\bibinfo {author} {\bibfnamefont {E.}~\bibnamefont {Kushilevitz}}\ and\ \bibinfo {author} {\bibfnamefont {Y.}~\bibnamefont {Mansour}},\ }\bibfield  {title} {\bibinfo {title} {Learning decision trees using the fourier spectrum},\ }in\ \href@noop {} {\emph {\bibinfo {booktitle} {Proceedings of the twenty-third annual ACM symposium on Theory of computing}}}\ (\bibinfo {year} {1991})\ pp.\ \bibinfo {pages} {455--464}\BibitemShut {NoStop}%
\bibitem [{\citenamefont {Kay}(2018)}]{kay2018tutorial}%
  \BibitemOpen
  \bibfield  {author} {\bibinfo {author} {\bibfnamefont {A.}~\bibnamefont {Kay}},\ }\bibfield  {title} {\bibinfo {title} {Tutorial on the quantikz package},\ }\href@noop {} {\bibfield  {journal} {\bibinfo  {journal} {arXiv preprint arXiv:1809.03842}\ } (\bibinfo {year} {2018})}\BibitemShut {NoStop}%
\bibitem [{\citenamefont {Hunter}(2007)}]{hunter2007matplotlib}%
  \BibitemOpen
  \bibfield  {author} {\bibinfo {author} {\bibfnamefont {J.~D.}\ \bibnamefont {Hunter}},\ }\bibfield  {title} {\bibinfo {title} {Matplotlib: A 2d graphics environment},\ }\href@noop {} {\bibfield  {journal} {\bibinfo  {journal} {Computing in science \& engineering}\ }\textbf {\bibinfo {volume} {9}},\ \bibinfo {pages} {90} (\bibinfo {year} {2007})}\BibitemShut {NoStop}%
\bibitem [{\citenamefont {Beauchamp}(1984)}]{beauchamp1984applications}%
  \BibitemOpen
  \bibfield  {author} {\bibinfo {author} {\bibfnamefont {K.~G.}\ \bibnamefont {Beauchamp}},\ }\href@noop {} {\emph {\bibinfo {title} {Applications of Walsh and related functions, with an introduction to sequency theory}}},\ Vol.~\bibinfo {volume} {2}\ (\bibinfo  {publisher} {Academic press},\ \bibinfo {year} {1984})\BibitemShut {NoStop}%
\bibitem [{\citenamefont {Debbasch}\ and\ \citenamefont {Brachet}(1995)}]{debbasch1995relativistic}%
  \BibitemOpen
  \bibfield  {author} {\bibinfo {author} {\bibfnamefont {F.}~\bibnamefont {Debbasch}}\ and\ \bibinfo {author} {\bibfnamefont {M.~E.}\ \bibnamefont {Brachet}},\ }\bibfield  {title} {\bibinfo {title} {Relativistic hydrodynamics of semiclassical quantum fluids},\ }\href@noop {} {\bibfield  {journal} {\bibinfo  {journal} {Physica D: Nonlinear Phenomena}\ }\textbf {\bibinfo {volume} {82}},\ \bibinfo {pages} {255} (\bibinfo {year} {1995})}\BibitemShut {NoStop}%
\bibitem [{\citenamefont {Filbet}\ and\ \citenamefont {Sonnendr{\"u}cker}(2003)}]{filbet2003numerical}%
  \BibitemOpen
  \bibfield  {author} {\bibinfo {author} {\bibfnamefont {F.}~\bibnamefont {Filbet}}\ and\ \bibinfo {author} {\bibfnamefont {E.}~\bibnamefont {Sonnendr{\"u}cker}},\ }\bibfield  {title} {\bibinfo {title} {Numerical methods for the vlasov equation},\ }in\ \href@noop {} {\emph {\bibinfo {booktitle} {Numerical Mathematics and Advanced Applications: Proceedings of ENUMATH 2001 the 4th European Conference on Numerical Mathematics and Advanced Applications Ischia, July 2001}}}\ (\bibinfo {organization} {Springer},\ \bibinfo {year} {2003})\ pp.\ \bibinfo {pages} {459--468}\BibitemShut {NoStop}%
\bibitem [{\citenamefont {Brassard}\ \emph {et~al.}(2002)\citenamefont {Brassard}, \citenamefont {Hoyer}, \citenamefont {Mosca},\ and\ \citenamefont {Tapp}}]{brassard2002quantum}%
  \BibitemOpen
  \bibfield  {author} {\bibinfo {author} {\bibfnamefont {G.}~\bibnamefont {Brassard}}, \bibinfo {author} {\bibfnamefont {P.}~\bibnamefont {Hoyer}}, \bibinfo {author} {\bibfnamefont {M.}~\bibnamefont {Mosca}},\ and\ \bibinfo {author} {\bibfnamefont {A.}~\bibnamefont {Tapp}},\ }\bibfield  {title} {\bibinfo {title} {Quantum amplitude amplification and estimation},\ }\href@noop {} {\bibfield  {journal} {\bibinfo  {journal} {Contemporary Mathematics}\ }\textbf {\bibinfo {volume} {305}},\ \bibinfo {pages} {53} (\bibinfo {year} {2002})}\BibitemShut {NoStop}%
\bibitem [{\citenamefont {Saeedi}\ and\ \citenamefont {Pedram}(2013)}]{saeedi2013linear}%
  \BibitemOpen
  \bibfield  {author} {\bibinfo {author} {\bibfnamefont {M.}~\bibnamefont {Saeedi}}\ and\ \bibinfo {author} {\bibfnamefont {M.}~\bibnamefont {Pedram}},\ }\bibfield  {title} {\bibinfo {title} {Linear-depth quantum circuits for n-qubit toffoli gates with no ancilla},\ }\href@noop {} {\bibfield  {journal} {\bibinfo  {journal} {Physical Review A}\ }\textbf {\bibinfo {volume} {87}},\ \bibinfo {pages} {062318} (\bibinfo {year} {2013})}\BibitemShut {NoStop}%
\bibitem [{\citenamefont {Barenco}\ \emph {et~al.}(1995)\citenamefont {Barenco}, \citenamefont {Bennett}, \citenamefont {Cleve}, \citenamefont {DiVincenzo}, \citenamefont {Margolus}, \citenamefont {Shor}, \citenamefont {Sleator}, \citenamefont {Smolin},\ and\ \citenamefont {Weinfurter}}]{barenco1995elementary}%
  \BibitemOpen
  \bibfield  {author} {\bibinfo {author} {\bibfnamefont {A.}~\bibnamefont {Barenco}}, \bibinfo {author} {\bibfnamefont {C.~H.}\ \bibnamefont {Bennett}}, \bibinfo {author} {\bibfnamefont {R.}~\bibnamefont {Cleve}}, \bibinfo {author} {\bibfnamefont {D.~P.}\ \bibnamefont {DiVincenzo}}, \bibinfo {author} {\bibfnamefont {N.}~\bibnamefont {Margolus}}, \bibinfo {author} {\bibfnamefont {P.}~\bibnamefont {Shor}}, \bibinfo {author} {\bibfnamefont {T.}~\bibnamefont {Sleator}}, \bibinfo {author} {\bibfnamefont {J.~A.}\ \bibnamefont {Smolin}},\ and\ \bibinfo {author} {\bibfnamefont {H.}~\bibnamefont {Weinfurter}},\ }\bibfield  {title} {\bibinfo {title} {Elementary gates for quantum computation},\ }\href@noop {} {\bibfield  {journal} {\bibinfo  {journal} {Physical review A}\ }\textbf {\bibinfo {volume} {52}},\ \bibinfo {pages} {3457} (\bibinfo {year} {1995})}\BibitemShut {NoStop}%
\end{thebibliography}%

\newpage

\appendix

\section{Technical Details about the implementation of the Walsh Series Loader}
\label{Technical Details about the Walsh Series Loader}

\subsection{Loading of real-valued functions}
Following the scheme of Welsh et al. \cite{welch2014efficient}, loading a real-valued function $f$ defined on $[0,1]$ unto a $n-$qubit state starts with the discretization of the interval $[0,1]$ into $N=2^n$ discrete points  $\mathcal{X}_n=\{0,1/N,2/N,3/N,...,(N-1)/N\}$.
The second step consists of computing classically the Walsh coefficients of the function $f$. The Walsh function of order $j \in \{0,1,...,N-1\}$ is defined by
\begin{equation}
    w_j(x)=(-1)^{\sum_{i=1}^nj_ix_{i-1}}.
\label{Walsh function definition}
\end{equation}
where $j_i$ is the $i$th bit in the binary expansion $j=\sum_{i=1}^nj_i2^{i-1}$ and $x_i$ is the $i$th bit in the dyadic expansion $x=\sum_{i=0}^{\infty} x_i/ 2^{i+1}$.

The $M-$ Walsh series $f^{\epsilon_1}$ approximating a function $f$ up to an error $\epsilon_1$ is
\begin{equation}
f^{\epsilon_1}=\sum_{j=0}^{M(\epsilon_1)-1}a_j^fw_j,
\label{Walsh serie}
\end{equation}
where $m(\epsilon_1)=\lfloor \log_2(1/\epsilon_1) \rfloor+1$ and $M(\epsilon_1)=2^{m(\epsilon_1)}$ with $\frac{1}{M(\epsilon_1)} < \epsilon_1$,. 

The $j$-th Walsh coefficient $a_j^f$ associated to the function $f$ is defined by:
\begin{equation}
    a_j^f=\frac{1}{M}\sum_{x\in\mathcal{X}_m}^{M-1}f(x)w_{j}(x),
\label{Walsh coeff}
\end{equation}

where $\mathcal{X}_m=\{0,1/M,2/M,3/M,...,(M-1)/M\}$.

The last step consist to perform the WSL quantum circuit by implementing the controlled-diagonal unitary $\hat{U}_{f^{\epsilon_1},\epsilon_0}$ associated to the computed $a_j^f$ coefficients. Let's remark that the decomposition into one-qubit gates and two-qubit gates of the controlled-$\hat{U}_{f,\epsilon_0}$ operator is given by controlling all the gates coming from the decomposition of $\hat{U}_{f,\epsilon_0}$. Thus, in the following, we focus only on the decomposition of  $\hat{U}_{f^{\epsilon_1},\epsilon_0}$:
\begin{equation}
\begin{split}
    \hat{U}_{f^{\epsilon_1},\epsilon_0}=e^{-i\hat{f}^{\epsilon_1}\epsilon_0}&=e^{-i\sum_{j=0}^{M-1}a_j^f\hat{w}_j\epsilon_0}\\&=\prod_{j=0}^{M-1}e^{-i a_j^f\hat{w}_j\epsilon_0}
    \\&=\prod_{j=0}^{M-1}\hat{W}_{j,\epsilon_0},
\end{split}
\end{equation}

where $\hat{W}_{j,\epsilon_0}=e^{-i a_j^f\hat{w}_j\epsilon_0}$ and the Walsh operators are defined as $\hat{w}_j=(Z_1)^{j_1}\otimes...\otimes (Z_n)^{j_n}$, where $j_i$ is the $i$-th coefficient in the binary expansion of $j=\sum_{i=1}^nj_i2^{i-1}$.
Using the fact that a tensor product of $Z$-Pauli gates can be rewritten using CNOT staircases as:
\begin{equation}
\begin{split}    \hat{Z}_{0}\otimes \cdots \otimes \hat{Z}_{p-1} = \hat{A}_p( \hat{I}_{0:p-2}\otimes \hat{Z}_{p-1})\hat{A}_p^{-1},
\label{Z-tower}
\end{split}
\end{equation}
where $\hat{Z}_i$ is the Z-Pauli gate acting on qubit $i$, $\hat{I}_{0:p-2}$ is the identity operator acting on qubits $0,...,p-2$ and $\hat{A}_p=\widehat{CNOT}_{0}^{p-1}\widehat{CNOT}_{1}^{p-1}...\widehat{CNOT}_{p-2}^{p-1}$ and $\widehat{CNOT}_i^j$ is the CNOT quantum gate controlled by qubit $i$ and applied on qubit $j$.

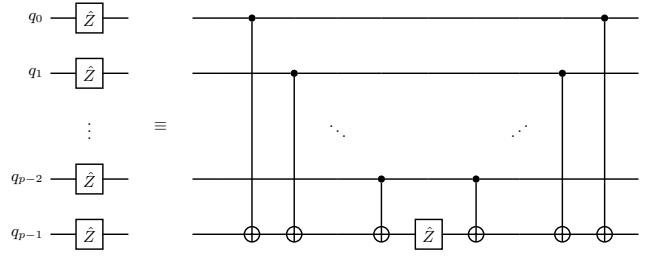
\begin{figure}[h]
    \centering
\begin{adjustbox}{width=0.47\textwidth}
\begin{quantikz}
\lstick[wires=1]{$q_0$} 
&\gate[1]{\hat{Z}} & \qw & && \qw & \ctrl{4} & \qw & \qw& \qw&\qw & \qw&\qw & \qw & \ctrl{4}&\qw
\\ 
 \lstick[wires=1]{$q_1$} 
&\gate[1]{\hat{Z}} & \qw & && \qw & \qw & \ctrl{3}& \qw& \qw &\qw &\qw & \qw &\ctrl{3} & \qw& \qw
\\
& \vdots & & \equiv & &&&& \ddots && && \iddots&
\\  \lstick[wires=1]{$q_{p-2}$} & \gate[1]{\hat{Z}} & \qw &  & & \qw & \qw & \qw& \qw  & \ctrl{1} & \qw & \ctrl{1} & \qw& \qw & \qw & \qw
\\
\lstick[wires=1]{$q_{p-1}$}& \gate[1]{\hat{Z}}& \qw & && \qw & \targ{} &\targ{}& \qw & \targ{} & \gate[1]{\hat{Z}} & \targ{} & \qw&\targ{} &\targ{} & \qw
\\
\end{quantikz}
\end{adjustbox}
\caption{ Equivalent quantum circuit for a tower of $p$ Z-Pauli quantum gates. }
\label{fig:Z tower}
\end{figure}

Therefore, the operator $\hat{W}_{j,\epsilon_0}$ acting on $p$ qubits, can be simply written in term of quantum gates as

\begin{equation}
    \hat{W}_{j,\epsilon_0}=\hat{A}_p(\hat{I}_{0:p-2}\otimes e^{-i a_j^f \epsilon_0 \hat{Z}})\hat{A}_p^{-1}.
\label{def W operator}
\end{equation}

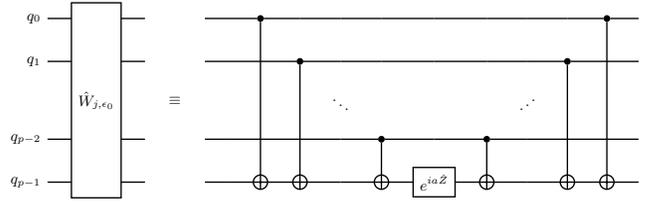
\begin{figure}[h]
    \centering
\begin{adjustbox}{width=0.47\textwidth}
\begin{quantikz}
\lstick[wires=1]{$q_0$} 
&\gate[5,nwires={3}]{\hat{W}_{j,\epsilon_0}} & \qw & && \qw & \ctrl{4} & \qw & \qw& \qw&\qw & \qw&\qw & \qw & \ctrl{4}&\qw
\\ 
 \lstick[wires=1]{$q_1$} 
& & \qw & && \qw & \qw & \ctrl{3}& \qw& \qw &\qw &\qw & \qw &\ctrl{3} & \qw& \qw
\\
& \vdots & & \equiv & &&&& \ddots && && \iddots&
\\  \lstick[wires=1]{$q_{p-2}$} &  & \qw &  & & \qw & \qw & \qw& \qw  & \ctrl{1} & \qw & \ctrl{1} & \qw& \qw & \qw & \qw
\\
\lstick[wires=1]{$q_{p-1}$}& & \qw & && \qw & \targ{} &\targ{}& \qw & \targ{} & \gate[1]{e^{ia\hat{Z}}} & \targ{} & \qw&\targ{} &\targ{} & \qw
\\
\end{quantikz}
\end{adjustbox}
\caption{Quantum circuit for the operator $\hat{W}_j(a)$ acting on $p$ different qubits using CNOT and RZ quantum gates.}
\label{fig:Z tower2}
\end{figure}

Then, the $\hat{W}_{j,\epsilon_0}$ commute with each other, allowing to optimize the order of implementation of the $\hat{W}_{j,\epsilon_0}$.
The first method consist to cancel a maximum number of CNOT gates coming from the operators $\hat{A}_q$ of two consecutive $\hat{W}_{j,\epsilon_0}$ and $\hat{W}_{j',\epsilon_0}$. This is done using a Gray code \cite{beauchamp1984applications} : only one bit changes in the binary decomposition of the index $j$ and $j'$ of two consecutive operators.
The second method consist to sort the $M$ Walsh coefficients $a_j^f$ in order to implement only a finite number $M'<M$ of the operators $\hat{W}_{j,\epsilon_0}$  associated to the largest $|a_j^f|$. 
The two methods are not compared in terms of infidelity scaling with the number of Walsh operators but in terms of infidelity scaling with the depth of the associated quantum circuits (Fig. 3). While the first method has theoretical guarantees, the second one seems numerically more efficient since it does not implement all the smallest coefficients of the $M$-Walsh Series of $f$. 
Other methods exist to compute finite Walsh Series approximating a given function $f$ using threshold sampling, data compression \cite{yuen1975function} or efficient estimation of a number $M'$ of the best Walsh coefficients \cite{kushilevitz1991learning} which could be used for Quantum State Preparation. 

\begin{figure}
     \centering 
    
     \begin{subfigure}  
     
        \includegraphics[width=0.4\textwidth]{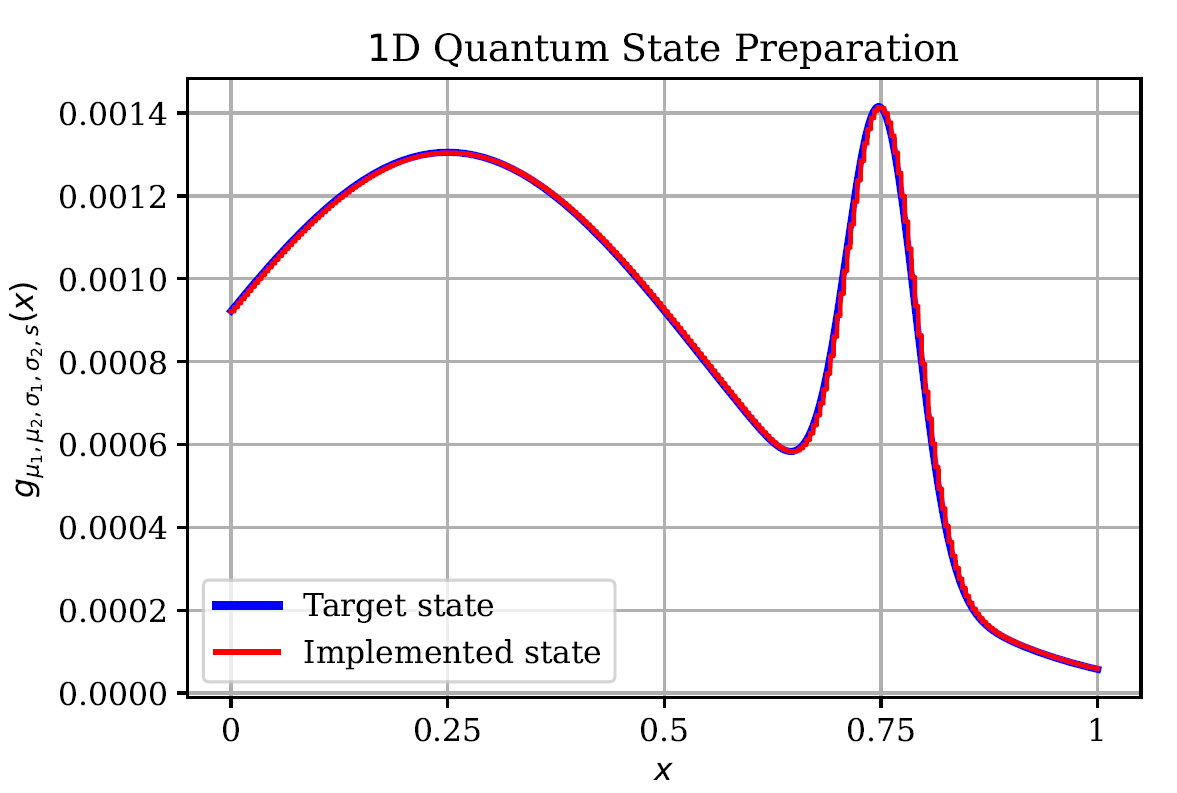}
     \end{subfigure}
     
     \hfill
     \centering
     
     \begin{subfigure}
     
         \includegraphics[width=0.4\textwidth]{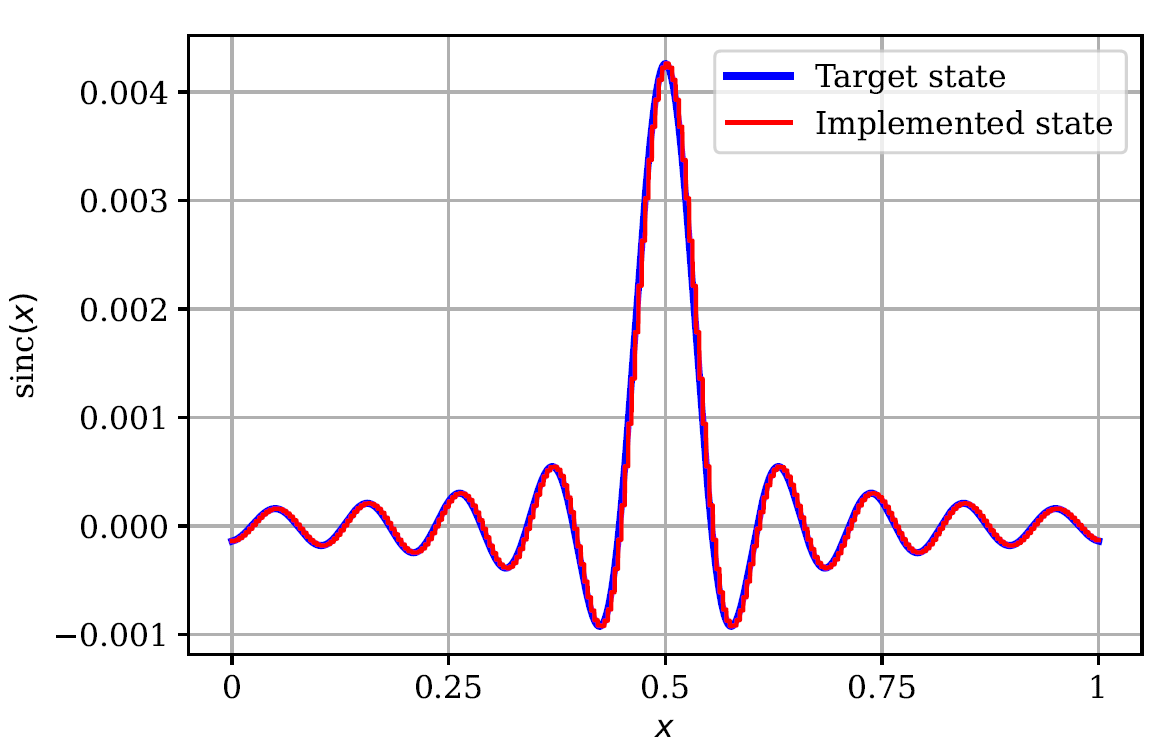}
     \end{subfigure}
     
\caption{Quantum State Preparation on $n=20$ qubits of a Bi-modal Gaussian $g_{\mu_1,\sigma_1,\mu_2,\sigma_2,s}(x)=(1-s)g_{\mu_1,\sigma_1}(x)+sg_{\mu_2,\sigma_2}(x)$ with $g_{\mu,\sigma}(x)=\text{exp}(-(x-\mu)^2/(2\sigma^2))/\sigma$ and $\mu_1=0.25$, $\mu_2=0.75$, $\sigma_1=0.3$, $\sigma_2=0.04$ and $s=0.1$ with infidelity $1-F=1.5\times10^{-4}$ using parameters $\epsilon_0=0.01$ and $\epsilon_1=1/2^{7}$ (top) and Quantum State Preparation of $\text{sinc}(x)=\sin(6\pi x)/(6\pi x)$ with infidelity $1-F=6.0\times10^{-3}$ using parameters $\epsilon_0=0.1$ and $\epsilon_1=1/2^{7}$ (bottom).}
\label{1Q QSP of bimodal gaussian and sinc}
\end{figure}

Fig. \ref{1Q QSP of bimodal gaussian and sinc} illustrates the Gray-ordered WSL method on a Bi-modal Gaussian function and a sinc function on $n=20$ qubits, with fidelity larger than $0.999$ and $0.99$ respectively.

\subsection{Loading of complex-valued functions}
\label{Loading of complex-valued functions}
\begin{figure}[h]
    \centering
\includegraphics[width=0.47\textwidth]{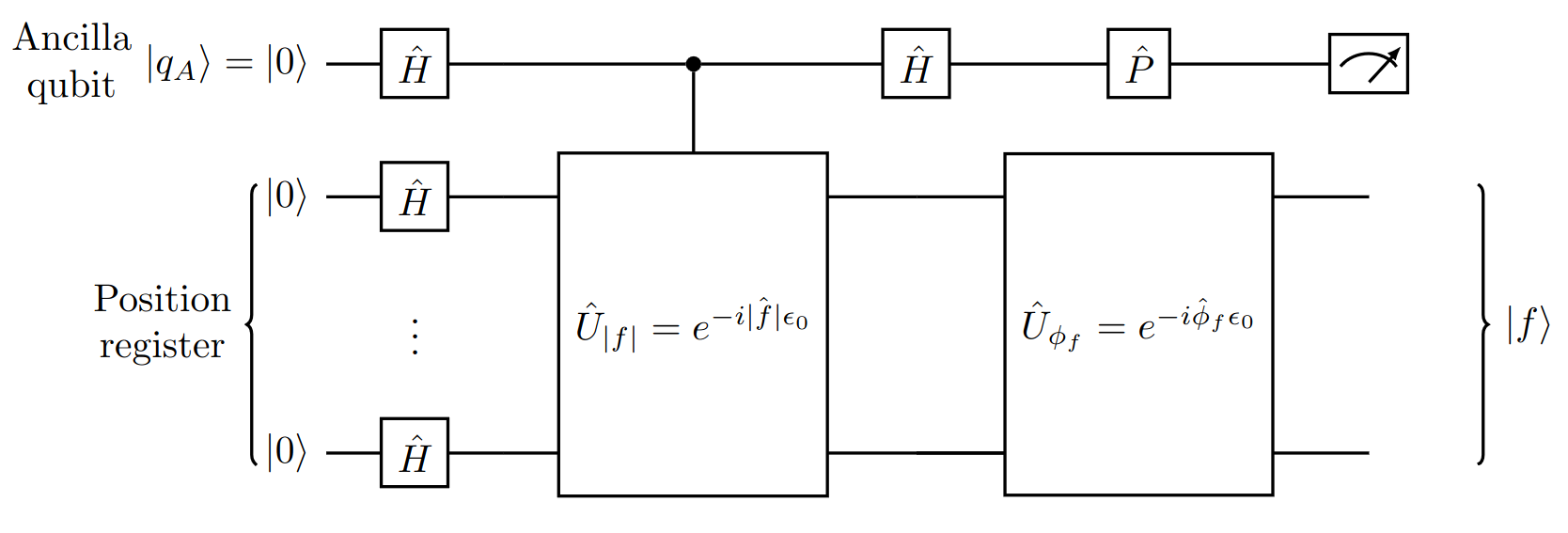}
\caption{Quantum circuit for the preparation of an initial quantum state $\ket{f}=\frac{1}{||f||_{2,N}}\sum_{x\in\mathcal{X}_n}f(x)\ket{x}$ associated to a complex valued function $f=|f|e^{i\phi_f}$.}
\label{quantum circuit scheme complex functions}
\end{figure}

Complex-valued functions are especially useful in contexts involving wave propagation. Associated PDEs include the Maxwell equations and the Klein-Gordon, the Dirac and the Schrödinger equations. Note that complex-valued functions are also useful in studying hydrodynamical potential flows \cite{zylberman2022quantum,debbasch1995relativistic}. The WSL is an efficient method to load the complex-valued initial condition for PDEs.
Loading of a complex-valued function $f$ defined in $[0,1]$ is carried out  by first loading the modulus $|f|$ of $f$ and by then loading the phase $\phi_f$ of $f$. The modulus is loaded with the scheme presented in the Letter. The phase $\phi_f$ is implemented as a diagonal unitary through a Walsh Series of $\phi_f$ using the scheme proposed by Welch et al. \cite{welch2014efficient}. The resulting quantum circuit is illustrated Fig.\ref{quantum circuit scheme complex functions}. 

In term of accuracy, the error one makes on the implementation of $f$ is bounded by the sum of the errors one makes in the implementation of $|f|$ and $e^{i\hat{\phi}_f}$.

\subsection{Loading of non-differentiable functions}
\label{Loading of non-differentiable functions}

Applying the WSL on non-differentiable functions is possible with significant results even if no theoretical guarantees have been proven. In the particular case of real-valued functions defined on $[0,1]$ and differentiable almost everywhere but on a finite set of points with bounded first derivative, theorem 1 generalizes using the fact that in the proof of Lemma 1.1., the difference between the function and its Walsh series can be bounded by the maximum of $|f'|$ on each of the interval where $f'$ is well defined. The WSL is performed for the following real-valued functions defined on $[0,1]$:
\begin{equation}
\begin{split}
    f_1(x)&=\sin{(2\pi(x-1/3))}w_4(x), \\
    f_2(x)&=|x-0.25|-|x-0.5|+|x-0.75| ,\\
    f_3(x)&=\sqrt{|x-0.5|}, \\
    f_4(x)&=\frac{1}{1-x},
\label{"functions test"}
\end{split}
\end{equation}
where $w_4$ is the Walsh function of order $4$.

\begin{figure}[h]
     \centering 
     \begin{subfigure}
     
        \includegraphics[width=0.4\textwidth]{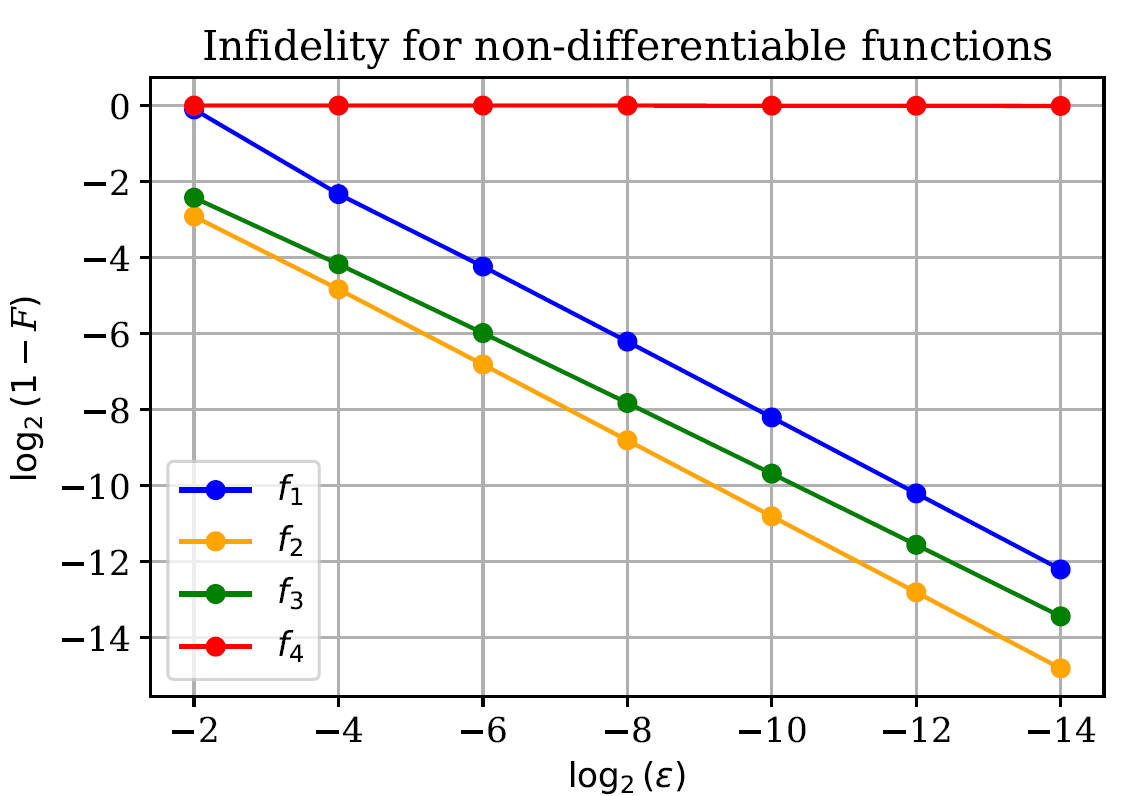}
     \end{subfigure}
     \hfill
     \centering
     \caption{Scaling laws of the infidelity $1-F$ with $\epsilon=\epsilon_0^2=\epsilon_1^2$ for $n=20$ qubits for some non-differentiable functions $f_1,f_2,f_3,f_4$ defined Eq.(\ref{"functions test"}).}
\label{Infidelity scaling of non-differentiable functions}
\end{figure}

\begin{figure}[h]
     \centering 
   
        \includegraphics[width=0.4\textwidth]{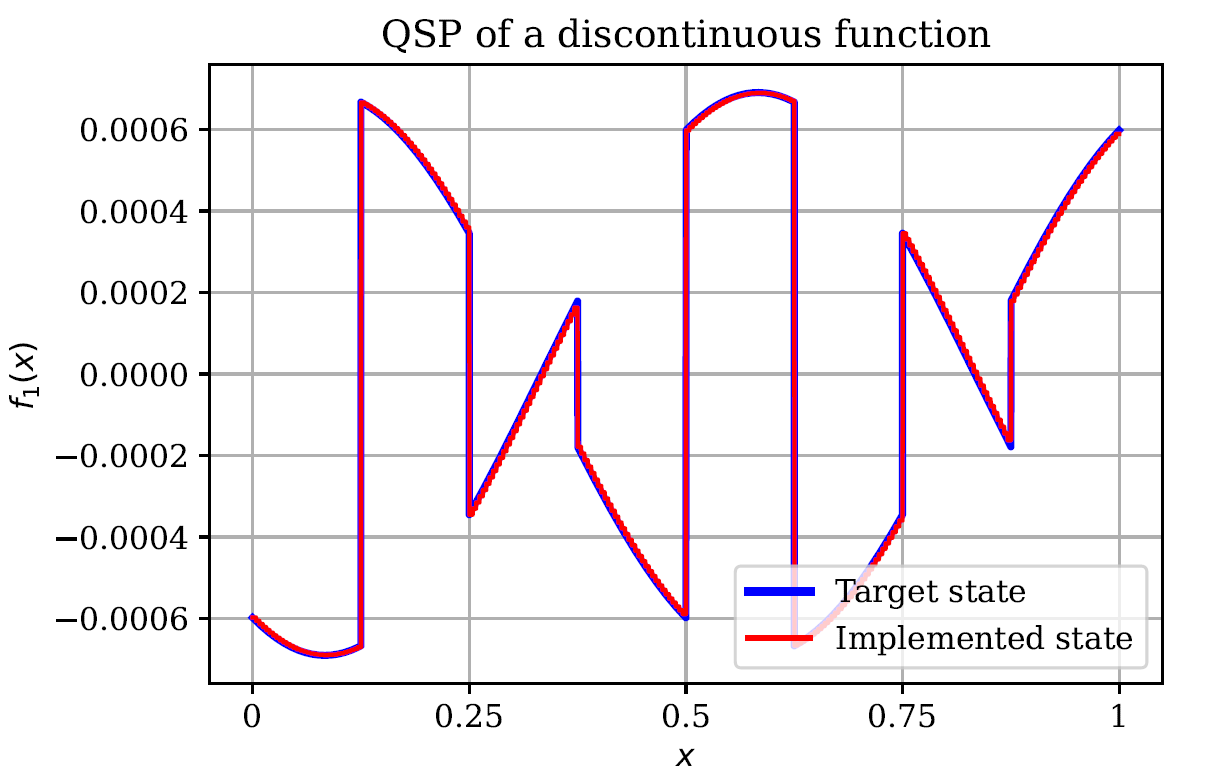}

    \caption{Quantum State Preparation of the non-differentiable function $f_1$ defined Eq.(\ref{"functions test"}) on $n=22$ qubits with infidelity $1-F=2.1\times 10^{-3}$ with parameters $\epsilon_0=0.1$ and $\epsilon_1=1/2^{7}$.}
\label{QSP of discontinuous functions}
\end{figure}

Numerical results Fig. \ref{Infidelity scaling of non-differentiable functions} show scaling laws $1-F\propto \epsilon$ for continuous non-differentiable functions such as $f_2$ and also for non-continuous functions such as $f_1$. The scalings are similar to the ones of differentiable functions Fig.3. The case of $f_3$ is particularly interesting since its first derivative is unbounded but the WSL method still provides an accurate QSP method. It could be explained by the fact that $f_3$ itself is bounded suggesting that the WSL could converge also for some bounded functions alsmost-everywhere differentiable with unbounded first derivative. However, in the particular case of a diverging function with a singularity point such as $f_4$, the WSL fails to prepare accurately the target state due to the diverging values taken by the function in the neighboorhood of the singularity. The quantum state preparation of $f_1$ is presented on Fig. \ref{QSP of discontinuous functions} with fidelity larger than $0.99$.

\section{Theorems and proofs}
\label{Theorems and proofs}
In this Appendix, the main theorems are stated and proofs are given for the Gray-ordered implementation of a quantum state.

\subsection{Definitions : one dimensional case}

In the following, working on $n$ qubits with a $M-$ Walsh series defined on $M$ points associated to a continuous function $f$ defined on $[0,1]$, one has to take into account the discrete space $\mathcal{X}_n=\{0,1/N,...,(N-1)/N\}$, with $N=2^n$, and the continuous space $[0,1]$ with the following norms :

For any vector $\ket{\psi}=\sum_{x\in \mathcal{X}_n} \psi(x)\ket{x}\in \mathcal{H}_2^{\otimes n}$:
\begin{equation}
    ||\ket{\psi}||_{2,N}=\sqrt{\sum_{x\in \mathcal{X}_n}|\psi(x)|^2},
\end{equation}
\begin{equation}
    ||\ket{\psi}||_{\infty,N}=\max_{x\in \mathcal{X}_n}|\psi(x)|.
\end{equation}

For $m<n$ and $M=2^m$:
\begin{equation}
    ||\ket{\psi}||_{2,M}=\sqrt{\sum_{x\in \mathcal{X}_m}|\psi(x)|^2}.
\end{equation}

The fidelity between two states $\ket{\psi_1}\in \mathcal{H}_2^{\otimes n}$ and $\ket{\psi_2}\in \mathcal{H}_2^{\otimes n}$ is
\begin{equation}
    F=|\braket{\psi_1|\psi_2}|^2=|\sum_{x\in \mathcal{X}_n}\psi_1^*(x)\psi_2(x)|^2,
\end{equation}
and the infidelity is defined as $1-F$.

For any function $f$ defined on $[0,1]$ :
\begin{equation}
    ||f||_{\infty,[0,1]}=\max_{x\in [0,1]}|f(x)|,
\end{equation}
\begin{equation}
    ||f||_{\infty,N}=\max_{x\in \mathcal{X}_n}|f(x)|,
\end{equation}
\begin{equation}
    ||f||_{2,N}=\sqrt{\sum_{x\in \mathcal{X}_n}|f(x)|^2},
\end{equation}
\begin{equation}
    ||f||_{2,[0,1]}=\sqrt{\int_0^1|f(x)|^2dx}.
\end{equation}
Let's note the following properties :
\begin{equation}
    ||\ket{\psi}||_{\infty,N} \leq ||\ket{\psi}||_{2,N} \leq \sqrt{N}||\ket{\psi}||_{\infty,N},
\end{equation}
\begin{equation}
    ||f||_{\infty,N} \leq ||f||_{2,N} \leq \sqrt{N}||f||_{\infty,N},
\end{equation}
\begin{equation}
    ||f||_{\infty,N} \leq ||f||_{\infty,[0,1]},
\end{equation}
\begin{equation}
   \forall f \in \mathcal{C}_0([0,1]),\text{   } \frac{1}{\sqrt{N}}||f||_{2,N} \xrightarrow{N\rightarrow +\infty} ||f||_{2,[0,1]},
\end{equation}
\begin{equation}
    ||\ket{\psi_1}-\ket{\psi_2}||_{2,N}\leq \epsilon \implies 1-F\leq \epsilon^2.
\end{equation}

\subsection{Theorems : one dimensional case.}

Let's consider $n$ qubits, a differentiable function $f$ defined on $[0,1]$ such that $||f||_{\infty,[0,1]}\neq 0$, $\epsilon_0 \in ]0,\pi/||f||_{\infty,[0,1]}]$ and $\epsilon_1>0$ such that $||f^{\epsilon_1}||_{\infty,[0,1]}\neq0$ where $f^{\epsilon_1}$ is the Walsh series of $f$ defined Eq(\ref{Walsh serie}).

\paragraph*{\textbf{Theorem1.}}
There is an efficient quantum circuit of size $O(n+1/\epsilon_1)$, depth $O(1/\epsilon_1)$, using one ancillary qubit, to implement a quantum state approximating the target state $\ket{f}$ up to an infidelity $1-F=O((\epsilon_0+\epsilon_1||f'||_\infty)^2)$ and with a  probability of success $P(1)=\Theta(\epsilon_0^2)$.

In the particular case of $\epsilon_1=\epsilon_0$, one can show:

\paragraph*{\textbf{Corollary 1.}}
There is an efficient quantum circuit of size $O(n+1/\sqrt{\epsilon})$, depth $O(1/\sqrt{\epsilon})$, using one ancillary qubit, to implement a quantum state approximating the target state $\ket{f}$ up to an infidelity $1-F\leq \epsilon$ with a  probability of success $P(1)=\Theta(\epsilon)$.

For any function $f$ with values $f(x)$ calculable in time $T_f$, the number of classical computations needed to find the quantum circuit is $O(T_f/\epsilon_1^2)$ (Theorem 1) or $O(T_f/\epsilon)$ (Corollary 1).

\subsection{Proof : one dimensional case}

The proof is based on the six following Lemmas.

\subsubsection{Lemmas}
\textbf{Lemma 1.0.} For any function $f$ continuously defined on $[0,1]$ such that $||f||_{\infty,[0,1]} \neq 0$: $\exists n_0$ such that $ \forall n\ge n_0$, $\forall \epsilon_0 \in ]0,\frac{2\pi}{||f||_{\infty,[0,1]}}[$
, $||\frac{\hat{I}-e^{-i\hat{f}\epsilon_0}}{2}\ket{s}||_{2,N} \neq 0 $ with $N=2^n$.

\textbf{Proof of Lemma 1.0.}

First,  $||f||_{\infty,[0,1]} \neq 0$ implies it exists $x_0\in [0,1]$ such that $f(x)=||f||_{\infty,[0,1]} \neq 0$. The continuity of $f$ implies it exists a neighbourhood $V$ of $x_0$ such that $\forall x \in V, f(x)\neq 0$. Let's note the following equality:

\begin{equation}
\begin{split}
||\frac{\hat{I}-e^{-i\hat{f}\epsilon_0}}{2}\ket{s}||_{2,N}&=\sqrt{\sum_{x\in\mathcal{X}_n}\frac{\sin^2(f(x)\epsilon_0/2)}{N}}.
\end{split}
\end{equation}
The ensemble $\mathcal{X}_n=\{0,\frac{1}{N},...,\frac{N-1}{N} \}$, with $N=2^n$, is a set of dyadic rational, ie rational numbers with a denominator that can be expressed as a power of $2$. Using the fact that dyadic numbers restricted to $[0,1]$ are dense in $[0,1]$, it exists two integers $p,q$ such that $p<2^q$, and $p/2^q \in V$. Futhermore, $p/2^q \in \mathcal{X}_q=\{0,...,\frac{2^q-1}{2^q}\}$ and $\forall q' \ge q, p/2^q \in \mathcal{X}_{q'}$. Let's remark that $\forall q'\ge q, \mathcal{X}_q \subseteq \mathcal{X}_{q'}$. Noting $n_0=q$, one has : $\exists x_1 \in  V$ such that $\forall n\ge n_0, x_1 \in \mathcal{X}_n $ and $f(x_1)\neq 0$.
Futhermore, $0<\epsilon_0<\frac{2\pi}{||f||_{\infty,[0,1]}}$ implies that $0<|f(x_1)|\epsilon_0/2<\pi$ and therefore $\sin^2(f(x_1)\epsilon_0/2) \neq 0$. Finally, as a sum of positive numbers with at least one non-vanishing number,  $\forall n\ge n_0, ||\frac{\hat{I}-e^{-i\hat{f}\epsilon_0}}{2}\ket{s}||_{2,N}\neq 0$, which achieves the proof of Lemma 1.0.

\textbf{Lemma 1.1.}
For any differentiable function $f\in\mathcal{C}^1([0,1])$ and $\epsilon_1>0$: the $M$-Walsh series $f^{\epsilon_1}$ defined Eq.(\ref{Walsh serie}) approximates the function $f$ up to an error $O(\epsilon_1)$:
\begin{equation}
\begin{split}
    ||f-f^{\epsilon_1}||_{\infty,[0,1]} \leq \epsilon_1 ||f'||_{\infty,[0,1]}.
\end{split}
\end{equation}

\textbf{Proof of Lemma 1.1.}
The function $f^{\epsilon_1}$ is a sum of $M$ Walsh functions of order $j \in \{0,...,M-1\}$. The Walsh function of order $j$ is a piecewise function taking values $+1$ and $-1$ on at most $2^p$ different intervals $I_k^p=[k/2^p,(k+1)/2^p[$ with $k\in \{0,...,2^p-1\}$, The $p$ first terms of the dyadic expansion of all $x \in I_k^p$ are equal. Therefore, the function $f^{\epsilon_1}$ is a piecewise function which is constant on each of the $M=2^m$ intervals $I_{k}^m$:

\begin{equation}
\begin{split}
\forall k \in \{0,...,M-1&\},\text{ } \forall x \in I_k^m, \\ f^{\epsilon_1}(x)=f^{\epsilon_1}&(k/M).
\end{split}
\end{equation}
Then, from the definitions of $f^{\epsilon_1}$ and the Walsh coefficients $a_{j}^{f}$ and using the orthonormality of the Walsh functions $\frac{1}{M}\sum_{p=0}^{M-1}w_{j}(p/M)w_{l}(p/M)=\delta_{jl}$, one has 
\begin{equation}
    f^{\epsilon_1}(k/M)=f(k/M).
\end{equation}

Let $x$ be a real number in $I_k^m$, then $f(x)-f^{\epsilon_1}(x)=f(x)-f(k/M)$. The mean value theorem implies it exists $y\in I_k^m$ such that $f(x)-f(k/M)=f'(y)(x-k/M)$. Using $|x-k/M|\leq 1/M <\epsilon_1$ and $|f'(y)|\leq ||f'||_{\infty,[0,1]}$, one has:
\begin{equation}
   \forall x\in[0,1], \text{  } |f(x)-f^{\epsilon_1}(x)|\leq ||f'||_{\infty,[0,1]} \epsilon_1.
\end{equation}

\textbf{Lemma 1.2.}
For any function $f$ continuous on $[0,1]$ such that $||f||_{\infty,[0,1]} \neq 0$, it exists $n_0 \ge0$ and a constant $C_1>0$ such that $\forall n\ge n_0$:

\begin{equation}
    ||f||_{2,N}\ge C_1\sqrt{N},
\end{equation}

with $N=2^n$.

\textbf{Proof of Lemma 1.2.} 
The function $f$ is continuous on $[0,1]$ with a non-zero value. The continuity of $f$ and the fact that the dyadic rational numbers are dense in $[0,1]$ imply  $\exists n_0$ such that $\exists x \in [0,1] \cap \mathcal{X}_{n_0}$ such that $f(x)\neq 0$, ie  $ \forall n \ge n_0, x \in \mathcal{X}_{n}, $ and, therefore, $\forall n\ge n_0, ||f||_{2,N}\neq0$ with $N=2^n$. The sequence $(\frac{1}{\sqrt{N}}||f||_{2,N})_N$ can be rewritten using the Riemann sum of $|f|^2$, $(\frac{1}{\sqrt{N}}||f||_{2,N})^2=S_N(|f|^2)=\sum_{k=1}^{N}(\frac{k}{N}-\frac{k-1}{N})|f(x_k)|^2=\sum_{k=1}^{N}\frac{|f(x_k)|^2}{N}$ which converges toward  $||f||^2_{2,[0,1]}$.
Therefore, $(\frac{1}{\sqrt{N}}||f||_{2,N})_N$ converges toward $l=||f||_{2,[0,1]}>0$ and it exists $n_1$ such that $\forall n \ge n_1, \frac{1}{\sqrt{2^n}}||f||_{2,2^n}> l/2$. By defining $C_1=\min(\min_{n_0\le k \le n_1}(\frac{1}{\sqrt{2^k}}||f||_{2,2^k}),l/2)$,  one has $\forall n \ge n_0,  ||f||_{2,N}\ge C_1\sqrt{N}$.

\textbf{Lemma 1.3.} For any function $f$ defined and continuous on $[0,1]$ with $||f||_{\infty,[0,1]} \neq 0 $, $\exists n_0$ such that $\forall n\ge n_0$, $\forall \epsilon_0 \in ]0,\pi/||f||_{\infty,[0,1]}]$ the normalization factor $\frac{1}{||\frac{\hat{I}-e^{-i\hat{f}\epsilon_0}}{2}\ket{s} ||_{2,N}}$ can be bounded as
\begin{equation}
    \frac{2\sqrt{N}}{\epsilon_0||f||_{2,N}} \leq \frac{1}{||\frac{\hat{I}-e^{-i\hat{f}\epsilon_0}}{2}\ket{s} ||_{2,N}} \leq C_0\frac{2\sqrt{N}}{\epsilon_0||f||_{2,N}},
\end{equation}
which is equivalent to:
\begin{equation}
   1 \leq \frac{\epsilon_0 ||f||_{2,N}}{2||\sin(f\epsilon_0/2)||_{2,N}} \leq C_0,
\end{equation}

with $N=2^n$ and $C_0=\pi/2$.

\textbf{Proof of Lemma 1.3.}
Lemma 1.0. implies it exists $n_0$ such that $\forall n>n_0, \forall \epsilon_0 \in ]0,2\pi/||f||_{\infty,[0,1]}[$, $ ||\frac{\hat{I}-e^{-i\hat{f}\epsilon_0}}{2}\ket{s} ||_{2,N} \neq 0$, with $N=2^n$, ensuring the quantity $1/||\frac{\hat{I}-e^{-i\hat{f}\epsilon_0}}{2}\ket{s} ||_{2,N}$ to be well defined. The left inequality is trivial using the fact that $\forall x\ge0, \sin(x)\le x$. For the right inequality, consider $\alpha \in ]0,\pi]$ and $\epsilon_0\in]0,\frac{2(\pi-\alpha)}{||f||_{\infty,[0,1]}}]$, then, thanks to the fact that the function $(x\mapsto \sin{(x)}/x)$ is decreasing on $[0,\pi]$ :

\begin{equation}
\begin{split}
    &\frac{\epsilon_0 ||f||_{2,N}}{2||\sin(f\epsilon_0/2)||_{2,N}} \\ &\leq \frac{||f||_{2,N}}{\sqrt{\sum_{x\in\mathcal{X}_n}f(x)^2\frac{\sin^2(f(x)(\pi-\alpha)/||f||_{\infty,[0,1]})}{(f(x)(\pi-\alpha)/||f||_{\infty,[0,1]})^2}}} \\&
    \leq \frac{\pi-\alpha}{\sin(\pi-\alpha)}.
\end{split}    
\end{equation}
Therefore, for $\alpha=\pi/2$, one proves Lemma 1.3.

\textbf{Lemma 1.4.} For any function $f$  differentiable on $[0,1]$ with $||f||_{\infty,[0,1]} \neq 0 $,  $\epsilon_1 > 0$ such that   $||f^{\epsilon_1}||_{\infty,[0,1]} \neq 0 $ where $f^{\epsilon_1}$  is the Walsh series defined by Eq.(\ref{Walsh serie}): $\exists n_0$ such that $\forall n\ge n_0$, $\forall \epsilon_0 \in ]0,\pi/||f||_{\infty,[0,1]}]$:
\begin{equation}
\begin{split}
|\frac{1}{||\frac{\hat{I}-e^{-i\hat{f}^{\epsilon_1}\epsilon_0}}{2}\ket{s} ||_{2,N}}-\frac{1}{||\frac{\hat{I}-e^{-i\hat{f}\epsilon_0}}{2}\ket{s} ||_{2,N}}| \\ \leq C_0^2\frac{2N\epsilon_1}{\epsilon_{0}}\frac{||f'||_{\infty,[0,1]}}{||f||_{2,N}||f^{\epsilon_1}||_{2,N}},
\end{split}
\end{equation}
with $N=2^n$, and $C_0=\pi/2$.

\textbf{Proof of Lemma 1.4.}

First, Lemma 1.0. implies it exists $n_1$ such that $\forall n\ge n_1, ||\frac{\hat{I}-e^{-i\hat{f}\epsilon_0}}{2}\ket{s} ||_{2,N} \neq 0$, with $N=2^n$. Then, $f^{\epsilon_1}$ is a function defined on $[0,1]$ taking $2^m$ different values of $f$ on the invervals $I_k^m=[k/2^m,(k+1)/2^m[$ for $k\in \{0,...,2^m-1\}$, with $m=\lfloor \log_2(1/\epsilon_1) \rfloor+1$. Therefore, the fact that $||f^{\epsilon_1}||_{\infty,[0,1]} \neq 0$ implies  $ \exists x \in \mathcal{X}_{m}$ such that $ f^{\epsilon_1}(x)\neq 0$ and by setting $n_0=\max(n_1,m)$, one has $\forall n\ge n_0$, $||\frac{\hat{I}-e^{-i\hat{f}^{\epsilon_1}\epsilon_0}}{2}\ket{s} ||_{2,N}\neq 0$ and $||\frac{\hat{I}-e^{-i\hat{f}\epsilon_0}}{2}\ket{s} ||_{2,N}\neq 0$.

Using the subbadditivity of the $||.||_{2,N}$ norm, one can show that 
\begin{equation}
\begin{split}
&|\frac{1}{||\frac{\hat{I}-e^{-i\hat{f}^{\epsilon_1}\epsilon_0}}{2}\ket{s} ||_{2,N}}-\frac{1}{||\frac{\hat{I}-e^{-i\hat{f}\epsilon_0}}{2}\ket{s} ||_{2,N}}| \\&\leq \frac{||(e^{-i\hat{f}\epsilon_0}-e^{-i\hat{f}^{\epsilon_1}\epsilon_0})\ket{s}||_{2,N}}{2||\frac{\hat{I}-e^{-i\hat{f}^{\epsilon_1}\epsilon_0}}{2}\ket{s} ||_{2,N}\times||\frac{\hat{I}-e^{-i\hat{f}\epsilon_0}}{2}\ket{s} ||_{2,N}}\\
& \leq \sqrt{N}\frac{||\sin((f-f^{\epsilon_1})\epsilon_0/2)||_{2,N}}{||\sin(f\epsilon_0/2)||_{2,N}\times||\sin((f^{\epsilon_1}\epsilon_0/2)||_{2,N}} \\
& \leq N\frac{||f-f^{\epsilon_1}||_{\infty,[0,1]}}{||f||_{2,N}||f^{\epsilon_1}||_{2,N}}\frac{\epsilon_0}{2}(\frac{2C_0}{\epsilon_0})^2,
\end{split}
\end{equation}
where $||f||_{2,N}\leq\sqrt{N}||f||_{\infty,N}\leq\sqrt{N}||f||_{\infty,[0,1]}$ and Lemma 1.3. have been used for the last inequality. Lemma 1.1. achieves the proof:
\begin{equation}
\begin{split}
 &N\frac{||f-f^{\epsilon_1}||_{\infty,[0,1]}}{||f||_{2,N}||f^{\epsilon_1}||_{2,N}}\frac{2C_0^2}{\epsilon_0} \\ &\leq   C_0^2\frac{2N\epsilon_1}{\epsilon_{0}}\frac{||f'||_{\infty,[0,1]}}{||f||_{2,N}||f^{\epsilon_1}||_{2,N}}.
\end{split}
\end{equation}

\textbf{Lemma 1.5}
For any function $f$ defined and continuous on $[0,1]$ with $||f||_\infty \neq 0 $ : $\exists n_0$ integer such that $\forall n\ge n_0$, $\forall \epsilon_0 \in [0,\pi/||f||_{\infty,[0,1]}]$:
\begin{equation}
\begin{split}
    |\frac{\epsilon_0 }{2\sqrt{N}||\frac{\hat{I}-e^{-i\hat{f}\epsilon_0}}{2}\ket{s}||_{2,N}}-\frac{1}{||f||_{2,N}}|\\ \leq \frac{C_0\epsilon_0^3}{24}\frac{||f^3||_{2,N}}{||f||^2_{2,N}},
\end{split}
\end{equation}
with $N=2^n$ and $C_0=\pi/2$.

\textbf{Proof of Lemma 1.5.}
Using the subbadditivity of the $||.||_{2,N}$ norm, the inequality $\forall x$ real, $x-\sin(x)\leq\frac{x^3}{6} $ and Lemma1.2.:
\begin{equation}
\begin{split}
    &|\frac{\epsilon_0}{2\sqrt{N}||\frac{\hat{I}-e^{-i\hat{f} \epsilon_0}}{2}\ket{s}||_{2,N}}-\frac{1}{||f||_{2,N}}|\\&=|\frac{\epsilon_0}{2||\sin(f\epsilon_0/2)||_{2,N}}- \frac{1}{||f||_{2,N}}|\\
    &\leq \frac{||\epsilon_0 f/2-\sin(\epsilon_0 f/2)||_{2,N}}{||\sin(f\epsilon_0/2)||_{2,N}||f||_{2,N}} \\ 
    & \leq \frac{\epsilon_0^3}{48}\frac{||f^3||_{2,N}}{||\sin(f\epsilon_0/2)||_{2,N}||f||_{2,N}}\\
    &\leq \frac{C_0\epsilon_0^2}{24}\frac{||f^3||_{2,N}}{||f||_{2,N}^2}.
 \end{split}
\end{equation}

\textbf{Lemma 1.6.} For any function $f$ continuous on $[0,1]$ such that $||f||_{\infty,[0,1]}\neq 0$ and for any $\epsilon_1>0$ such that $||f^{\epsilon_1}||_{\infty,[0,1]}\neq 0$ where $f^{\epsilon_1}$  is the Walsh series defined by Eq.(\ref{Walsh serie}): $\exists  C_1>0, \exists n_0$ such that $\forall n\ge n_0$:

\begin{equation}
    ||f^{\epsilon_1}||_{2,N}\ge C_1\sqrt{N},
\end{equation}
with $N=2^n$ and $f^{\epsilon_1}$ the Walsh series of $f$ defined Eq.(\ref{Walsh serie}).

\textbf{Proof of Lemma 1.6.}
Let's define  $m=\lfloor \log_2(1/\epsilon_1) \rfloor+1$ and $M=2^m$. $||f^{\epsilon_1}||_{\infty,[0,1]}\neq 0$ implies $\exists x \in \mathcal{X}_m$ such that $f^{\epsilon_1}(x)=f(x)\neq0$. Then,  $||f^{\epsilon_1}||_{2,M}=||f||_{2,M}\neq0$. Let's note $n_0=\min(\{n, ||f||_{2,2^n}\neq0\})\leq m$. Lemma 1.2. on $f$ states $\exists C_1>0$ such that $\forall n\ge n_0, ||f||_{2,N}\ge \sqrt{N} C_1$. Let $n$ be an integer larger than $n_0$, if $m\ge n$, $||f^{\epsilon_1}||_{2,N}=||f||_{2,N}\ge C_1\sqrt{N}$, if $m\le n$, $f^{\epsilon_1}$ takes only $M$ different values of $f$, implying that $\frac{1}{\sqrt{N}}||f^{\epsilon_1}||_{2,N}=\frac{1}{\sqrt{M}}||f||_{2,M}\ge C_1$ since $m\ge n_0$. One concludes $\exists C_1>0, \exists n_0$ such that $\forall n\ge n_0$, $||f^{\epsilon_1}||_{2,N}\ge C_1\sqrt{N}$.

\subsubsection{Proof of Theorem 1}
\label{Proof of Theorem 1}
Let's consider a differentiable function $f$ defined on $[0,1]$ such that $||f||_{\infty,[0,1]}\neq 0$, $\epsilon_0 \in ]0,\pi/||f||_{\infty,[0,1]}]$ and $\epsilon_1>0$ such that $||f^{\epsilon_1}||_{\infty,[0,1]}\neq0$ where $f^{\epsilon_1}$ is the Walsh series of $f$ defined Eq(\ref{Walsh serie}).

The implemented quantum state after measuring $\ket{1}$ for the ancillary qubit $\ket{q_A}$ is
\begin{equation}
\begin{split}
\ket{\psi_{f^{\epsilon_1}}}_{\epsilon_0}=-i\frac{\hat{I}-e^{-i\hat{f}^{\epsilon_1}\epsilon_0}}{2||\frac{\hat{I}-e^{-i\hat{f}^{\epsilon_1}\epsilon_0}}{2}\ket{s} ||_{2,N}}\ket{s},
\end{split}
\end{equation}
where $f^{\epsilon_1}$ is the Walsh series approximating $f$ up to an error $\epsilon_1 ||f'||_\infty$ (Lemma 1.1).

\textbf{Distance between the target state and the implemented state.}

Let's bound the infinite norm of the difference between the implemented quantum state $\ket{\psi_{f^{\epsilon_1}}}_{\epsilon_0}$ and the target quantum state $\ket{f}=\frac{1}{||f||_{2,N}}\sum_{x\in\mathcal{X}_n}f(x)\ket{x}$.

\begin{equation}
\begin{split}
    &||\ket{\psi_{f^{\epsilon_1}}}_{\epsilon_0}-\ket{f}||_{\infty,N} \\ &\leq ||\ket{\psi_{f^{\epsilon_1}}}_{\epsilon_0}-\ket{\psi_{f}}_{\epsilon_0}||_{\infty,N} \\ &+||\ket{\psi_{f}}_{\epsilon_0}-\ket{f}||_{\infty,N}.
\label{inégalité première}
\end{split}
\end{equation}

Let's start to bound the first term by making the difference of the normalization factors appears:

\begin{equation}
\begin{split}
    &||\ket{\psi_{f^{\epsilon_1}}}_{\epsilon_0}-\ket{\psi_{f}}_{\epsilon_0}||_{\infty,N} \\ & = ||(\frac{\hat{I}-e^{-i\hat{f}^{\epsilon_1}\epsilon_0}}{2||\frac{\hat{I}-e^{-i\hat{f}^{\epsilon_1}\epsilon_0}}{2}\ket{s} ||_{2,N}}-\frac{\hat{I}-e^{-i\hat{f}\epsilon_0}}{2||\frac{\hat{I}-e^{-i\hat{f}\epsilon_0}}{2}\ket{s} ||_{2,N}})\ket{s}||_{\infty,N} \\
    & \leq|\frac{1}{||\frac{\hat{I}-e^{-i\hat{f}^{\epsilon_1}\epsilon_0}}{2}\ket{s} ||_{2,N}}-\frac{1}{||\frac{\hat{I}-e^{-i\hat{f}\epsilon_0}}{2}\ket{s} ||_{2,N}}| \\ &\times  ||\frac{\hat{I}-e^{-i\hat{f}^{\epsilon_1}\epsilon_0}}{2}\ket{s}||_{\infty,N} \\
    &+||\frac{e^{-i\hat{f}^{\epsilon_1}\epsilon_0}-e^{-i\hat{f}\epsilon_0}}{2||\frac{\hat{I}-e^{-i\hat{f}\epsilon_0}}{2}\ket{s}||_{2,N}}\ket{s}||_{\infty,N} \\
    &\leq|\frac{1}{||\frac{\hat{I}-e^{-i\hat{f}^{\epsilon_1}\epsilon_0}}{2}\ket{s} ||_{2,N}}-\frac{1}{||\frac{\hat{I}-e^{-i\hat{f}\epsilon_0}}{2}\ket{s} ||_{2,N}}| \\&\times  \frac{1}{\sqrt{N}}\max_{x\in\mathcal{X}_n}|\sin(f^{\epsilon_1}(x) \epsilon_0/2))| \\
    &+\frac{1}{||\frac{\hat{I}-e^{-i\hat{f}\epsilon_0}}{2}\ket{s}||_{2,N}}\frac{1}{\sqrt{N}}\max_{x\in\mathcal{X}_n}|\sin(\frac{(f^{\epsilon_1}(x)-f(x))\epsilon_0}{2}))|\\
    &\leq C_0^2\sqrt{N}\frac{||f'_0||_{\infty,[0,1]}||f||_{\infty,[0,1]}}{||f||_{2,N} ||f^{\epsilon_1}||_{2,N}}\epsilon_1 +C_0\frac{||f'||_{\infty,[0,1]}}{||f||_{2,N}}\epsilon_1,
 \end{split}
\end{equation}

where in the last inequality, one uses $|\sin(x)|\leq|x|$, $||f^{\epsilon_1}||_{\infty,N} \leq ||f^{\epsilon_1}||_{\infty,[0,1]} \leq ||f||_{\infty,N}$, Lemma1.1., Lemma 1.3. and Lemma 1.4.

Lemma 1.2 and 1.6 and implies that there is a constant $B$ depending only on $f$ such that 
\begin{equation}
C_0^2\sqrt{N}\frac{||f||_{\infty,[0,1]}}{||f||_{2,N} ||f^{\epsilon_1}||_{2,N}} +C_0\frac{1}{||f||_{2,N}} \leq \frac{B}{\sqrt{N}},
\end{equation}

leading to the following bound on the first term
\begin{equation}
 ||\ket{\psi_{f^{\epsilon_1}}}_{\epsilon_0}-\ket{\psi_{f}}_{\epsilon_0}||_{\infty,N}\leq \frac{B}{\sqrt{N}}||f'||_{\infty,[0,1]}\epsilon_1.
\end{equation}

The second term in inequality (\ref{inégalité première}) can also be bounded by using the Taylor expansion of the exponential term $e^{-i\hat{f}\epsilon_0}=\hat{I}-i\hat{f}\epsilon_0+R_1(-i\hat{f}\epsilon_0)$ with $R_1(x)=\sum_{k=2}^{+\infty}\frac{x^k}{k!}$ and by making the difference of the norms appears:

\begin{equation}
\begin{split}
    &||\ket{\psi_{f}}_{\epsilon_0}-\ket{f}||_{\infty,N} \\ & =||(-i \frac{\hat{I}-(\hat{I}-i\hat{f}\epsilon_0+R_1(-i\hat{f}\epsilon_0))}{2||\frac{\hat{I}-e^{-i\hat{f}\epsilon_0}}{2}\ket{s}||_{2,N}}-\frac{\sqrt{N}}{||f||_{2,N}}\hat{f})\ket{s}||_{\infty,N} \\
    & \leq |\frac{\epsilon_0}{2||\frac{\hat{I}-e^{-i\hat{f}\epsilon_0}}{2}\ket{s}||_{2,N}}-\frac{\sqrt{N}}{||f||_{2,N}}|\times||\hat{f}\ket{s}||_{\infty,N}\\ &+\frac{||R_1(-i\hat{f}\epsilon_0)\ket{s}||_{\infty,N}}{2||\frac{\hat{I}-e^{-i\hat{f}\epsilon_0}}{2}\ket{s}||_{2,N}}.
\end{split}
\end{equation}

The Taylor inequality applied on the remainders of the cosinus and sinus functions implies $||R_1(-i\hat{f}\epsilon_0)\ket{s}||_{\infty,N}\leq\frac{\epsilon_0^2}{2\sqrt{N}}||f^2||_{\infty,[0,1]}+\frac{\epsilon_0^3}{6\sqrt{N}}||f^3||_{\infty,[0,1]}$ and using Lemma1.2. and Lemma1.5.:
\begin{equation}
\begin{split}
    &||\ket{\psi_{f}}_{\epsilon_0}-\ket{f}||_{\infty,N}
    \\& \leq \epsilon_0\frac{C_0||f^2||_{\infty,[0,1]}}{2||f||_{2,N}}+ \\&\epsilon_0^2(\frac{C_0}{48}\frac{||f^3||_{2,N}}{||f||^2_{2,N}}||f||_{\infty,[0,1]}+\frac{C_0||f^3||_{\infty,[0,1]}}{6||f||_{2,N}}).
\end{split}
\end{equation}

Using Lemma1.2 and the fact that $\epsilon_0\leq \pi/||f||_{\infty,[0,1]}$, there is a constant $A$, depending only $f$, such that

\begin{equation}
\begin{split}
    ||\ket{\psi_{f}}_{\epsilon_0}-\ket{f}||_{\infty,N}
    & \leq A\frac{\epsilon_0}{\sqrt{N}}.
\end{split}
\end{equation}

One can rewritte the inequality in term of the L2 norm, using $||.||_{2,N}\leq\sqrt{N}||.||_{\infty,N}$, one gets:
\begin{equation}
    ||\ket{\psi_{f^{\epsilon_1}}}_{\epsilon_0}-\ket{f}||_{2,N} \leq A\epsilon_0+B||f'||_{\infty,[0,1]}\epsilon_1.
\label{inequality L2}
\end{equation}

Let's define the fidelity between the target state and the implemented state $F=|\braket{f|\psi_{f^{\epsilon_1}}}_{\epsilon_0}|^2$ and the infidelity $1-F$. One can show that Eq.(\ref{inequality L2}) implies
\begin{equation}
    1-F \leq (A\epsilon_0+B||f'||_{\infty,[0,1]}\epsilon_1)^2,
\end{equation}
which concludes that $1-F=O((\epsilon_0+||f'||_{\infty,[0,1]}\epsilon_1)^2)$.

\textbf{Bounds on the probability of success.}

The probability of measuring the ancilla qubit $\ket{q_A}$ in state $\ket{1}$ with $\epsilon_0 \in ]0,\pi/||f||_{\infty,[0,1]}]$ and $\epsilon_1 \ge 0 $ is:
\begin{equation}
\begin{split}
P(1)&=||\frac{\hat{I}-e^{-i\hat{f}^{\epsilon_1}\epsilon_0}}{2}\ket{s}||_{2,N}^2\\&=\frac{1}{N}|| \sin{(\frac{f^{\epsilon_1}\epsilon_0}{2})} ||_{2,N}^2.
\end{split}
\end{equation}

The upper bound is trivial and comes from the inequality $\forall x \ge 0$ ,  $\sin(x)\leq x$:
\begin{equation}
    P(1)\leq \frac{||f^{\epsilon_1}||_{2,N}^2\epsilon_0^2}{4N} \leq \frac{||f||_{\infty,[0,1]}^2\epsilon_0^2}{4},
\end{equation}
where one has to use $||f^{\epsilon_1}||_{2,N}\leq \sqrt{N} ||f^{\epsilon_1}||_{\infty,N}\leq \sqrt{N} ||f||_{\infty,N} \leq \sqrt{N} ||f||_{\infty,[0,1]}$.

The lower bound comes from the fact the function $(x\mapsto \text{sinc}(x)=\sin(x)/x)$ decreases on $[0,\pi/2]$:
\begin{equation}
\begin{split}
P(1)&=\frac{1}{N}\sum_{x\in\mathcal{X}_n}\sin^2(\frac{f^{\epsilon_1}(x)\epsilon_0}{2})\\
& =\frac{1}{N}\sum_{x\in\mathcal{X}_n}(\frac{f^{\epsilon_1}(x)\epsilon_0}{2})^2\text{sinc}^2(\frac{f^{\epsilon_1}(x)\epsilon_0}{2})\\
& \ge \frac{1}{N}\sum_{x\in\mathcal{X}_n}(\frac{f^{\epsilon_1}(x)\epsilon_0}{2})^2\text{sinc}^2(\pi/2) \\
& \ge \frac{\epsilon_0^2}{\pi^2 N}||f^{\epsilon_1}||_{2,N}^2.
\end{split}
\end{equation}
Finally, using Lemma 1.6., it exists a constant $D'$, independent of $\epsilon_1$, such that $||f^\epsilon_1||_{2,N}^2 \ge N D'$. Therefore, it exists a constant $D$, independent of $\epsilon_0, \epsilon_1$ and $N$, such that $P(1)\ge D\epsilon_0^2$ concluding on $P(1)=\Theta(\epsilon_0^2)$.

\textbf{Complexities.}

The protocol starts with $n+1$ Hadamard gates to prepare the state $ \ket{s}=\frac{1}{\sqrt{N}}\sum_{x \in \mathcal{X}_n}\ket{x}$. Then, the controlled-$\hat{U}_{f^{\epsilon_1},\epsilon_0}$ has asymptotically the same size and depth than the unitary $\hat{U}_{f^{\epsilon_1},\epsilon_0}$ since the controlled operation only changes CNOT into Toffoli gates and single qubit rotations into controlled-rotations. The number of single-qubit gates and CNOT gates is $O(M)$, which is also $O(1/\epsilon_1)$. Depth for $\hat{U}_{f^{\epsilon_1},\epsilon_0}$ is also $O(1/\epsilon_1)$ \cite{welch2014efficient}. Finally, a Hadamard gate and Phase gate are applied on $\ket{q_A}$ to perform the right interference giving an approximation of the target state up to an infidelity $O((\epsilon_0+||f'||\epsilon_1)^2)$. Therefore, the total size is $O(n+1/\epsilon_1)$ while the depth is $O(1/\epsilon_1)$.

For any function $f$ with values $f(x)$ calculable in time $T_f$, the number of classical computations to compute the $M$-Walsh coefficients associated to the Walsh Series of $f$ is $O(T_fM^2)$ which is also $O(T_f/\epsilon_1^2)$.

\subsection{Multi-dimensional quantum state preparation}

\begin{figure}

     \centering
     \includegraphics[width=0.4\textwidth]{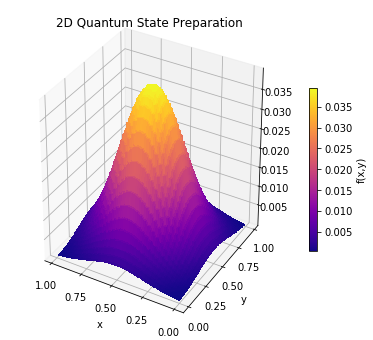}
     

    \caption{2D Quantum State Preparation of a 2D Gaussian function $f_2(x,y)=e^{-10((x-0.5)^2+(y-0.5)^2)}$ on $n=20$ qubits ($n_x=n_y=10)$ and infidelity $1-F=6.2\times10^{-3}$ with parameters $\epsilon_0=0.1$, $\epsilon_{1,x}=\epsilon_{1,y}=1/2^{5}$.}
\label{QSP2D}
\end{figure}

Multidimensional QSP is crucial for many PDEs modelling phenomena appearing in several dimensions. For instance, magnetic fields exists only in spaces with more than two dimensions or Kinetic plasma simulations consider the Vlasov-Maxwell equations in 2 or 4 or 6 dimensions and are particularly challenging to solve even on supercomputers  \cite{filbet2003numerical}. Quantum algorithms could play a crucial role to overcome the large computational cost of solving multidimensional PDE problems \cite{engel2019quantum}. In FIG.\ref{QSP2D}, a bivariate gaussian is implemented on $20$ qubits with a fidelity larger than $0.99$.

\paragraph*{Definitions.} Let's consider the problem of preparing a d-dimensional initial state on $n=n_1+...+n_d$ qubits where $n_i$ is the number of qubits associated to i-th axis. 

Let's note $\mathcal{X}_{n_i}$ the ensemble of the $N_i=2^{n_i}$ positions in $[0,1]$ represented by $n_i$ qubits $\mathcal{X}_{n_i}=\{\sum_{j=0}^{n_i-1} q_j/ 2^{j+1} \text{ such as }\forall j  \text{ }q_j\in\{0,1\}  \}=\{0,1/N_i,2/N_i,3/N_i,...,(N_i-1)/N_i\}$. Let's define $\vec{n}=(n_1,...,n_d)$ and $\mathcal{X}_{\vec{n}}=\{(r_1,...,r_d)\in \mathcal{X}_{n_1}\times...\times \mathcal{X}_{n_d}\}$.
We denote by $\ket{\vec{r}}$ the state $\ket{\vec{r}}=\ket{r_1}...\ket{r_d}=\ket{r_1}\otimes...\otimes \ket{r_d} \in \mathcal{H}_2^{\otimes n}$.

Let's consider a differentiable real-valued function $f$ on $[0,1]^d$ and $\hat{f}$ the associated diagonal operator in the position basis $\{\ket{\vec{r}}\}$ such that $\hat{f}\ket{\vec{r}}=f(\vec{r})\ket{\vec{r}}$. $\hat{f}$ encodes the amplitude of the target state $\ket{f}=\sum_{\vec{r}\in\mathcal{X}_{\vec{n}}} f(\vec{r})\ket{r}$.

In the multidimensional case, the function $f$ is developed into a Walsh series with respect to each axis $i$ which gives the equality on the discrete position space $\forall \vec{r} \in \mathcal{X}_{\vec{n}}$:
\begin{equation}
\begin{split}
    f(\vec{r})&=\sum_{j_1=0}^{N_1-1}...\sum_{j_d=0}^{N_d-1} a_{j_1,...,j_d,N_1,...,N_d}^fw_{j_1,...,j_d}(\vec{r})\\ &=\sum_{\vec{j}=\vec{0}}^{\vec{N}-1}a_{\vec{j},\vec{N}}^f w_{\vec{j}}(\vec{r}),
\end{split}
\end{equation}
with $w_{j_1,...,j_d}(\vec{r})=w_{\vec{j}}(\vec{r})=w_{j_1}(r_1)\times ...\times w_{j_d}(r_d)$ and $a_{\vec{j},\vec{N}}^f$ the multi-dimensional Walsh coefficient :
\begin{equation}
    a_{\vec{j},\vec{N}}^f=\frac{1}{N}\sum_{\vec{r} \in \mathcal{X}_{\vec{n}}}f(\vec{r})w_{\vec{j}}(\vec{r}),
\end{equation}
with $N=N_1\times ... \times N_d=2^{n_1}\times...\times2^{n_d}$

Let's note $f^{\vec{\epsilon}}$ the approximation of $f$ up to an error $\epsilon= \sum_{i=1}^d ||\partial_i f||_{\infty,[0,1]^d} \times \epsilon_i$, with $\epsilon_i>0$ the error associated to each spatial axis $i \in \{0,...,d\}$, defined by

\begin{equation}
f^{\vec{\epsilon}}=\sum_{\vec{j}=\vec{0}}^{\vec{M}-1} a_{\vec{j},\vec{M}}^f w_{\vec{j}},
\label{Walsh serie multidim}
\end{equation}

where $\forall i, m_i=\lfloor \log_2(1/\epsilon_i) \rfloor+1$ and $M_i=2^{m_i}$ such that $\frac{1}{M_i} < \epsilon_i$.

The implemented quantum state after measuring the ancilla qubit $\ket{q_A}=\ket{1}$ is 
\begin{equation}
    \ket{\psi_{f^{\vec{\epsilon}}}}_{\epsilon_0}=-i\frac{\hat{I}-e^{-i\hat{f}^{\vec{\epsilon}}\epsilon_0}}{2||\frac{\hat{I}-e^{-i\hat{f}^{\vec{\epsilon}}\epsilon_0}}{2}\ket{s} ||_{2,\vec{N}}}\ket{s},
\end{equation}

with $\epsilon_0>0$ and $ \ket{s}=\hat{H}^{\otimes n}\ket{0,...,0}=\frac{1}{\sqrt{N}}\sum_{\vec{r} \in \mathcal{X}_{\vec{n}}}\ket{\vec{r}}$ with $N=N_1\times...\times N_d$.

Let's define the different multi-dimensional norms used in this paper :

For any vector $\ket{\psi}=\sum_{\vec{r}\in \mathcal{X}_{\vec{n}}} \psi(\vec{r})\ket{\vec{r}}\in \mathcal{H}_2^{\otimes n}$:
\begin{equation}
    ||\ket{\psi}||_{2,\vec{N}}=\sqrt{\sum_{\vec{r}\in \mathcal{X}_{\vec{n}}}|\psi(\vec{r})|^2},
\end{equation}
\begin{equation}
    ||\ket{\psi}||_{\infty,\vec{N}}=\max_{\vec{r}\in \mathcal{X}_{\vec{n}}}|\psi(\vec{r})|,
\end{equation}
with $\vec{N}=(N_1,...,N_d)=(2^{n_1},...,2^{n_d})$.

 For $m_i<n_i$ $\forall i$ and $M_i=2^{m_i}$:
\begin{equation}
    ||\ket{\psi}||_{2,\vec{M}}=\sqrt{\sum_{\vec{r}\in \mathcal{X}_{\vec{m}}}|\psi(\vec{r})|^2}.
\end{equation}

For any function $f$ defined on $[0,1]^d$ :
\begin{equation}
    ||f||_{\infty,[0,1]^d}=\max_{\vec{r}\in [0,1]^d}|f(\vec{r})|,
\end{equation}
\begin{equation}
    ||f||_{\infty,\vec{N}}=\max_{\vec{r}\in \mathcal{X}_{\vec{n}}}|f(\vec{r})|,
\end{equation}
\begin{equation}
    ||f||_{2,\vec{N}}=\sqrt{\sum_{\vec{r}\in \mathcal{X}_{\vec{n}}}|f(\vec{r})|^2},
\end{equation}
\begin{equation}
    ||f||_{2,[0,1]^d}=\sqrt{\int_{[0,1]^d}|f(\vec{r})|^2dV}.
\end{equation}
Let's note the following properties :
\begin{equation}
\begin{split}
&||\ket{\psi}||_{\infty,\vec{N}} \leq ||\ket{\psi}||_{2,\vec{N}} \\& \leq \sqrt{N_1+...+N_d}||\ket{\psi}||_{\infty,\vec{N}},
\end{split}
\end{equation}
\begin{equation}
    ||f||_{\infty,\vec{N}} \leq ||f||_{2,\vec{N}} \leq \sqrt{N_1+...+N_d}||f||_{\infty,\vec{N}},
\end{equation}
\begin{equation}
    ||f||_{\infty,\vec{N}} \leq ||f||_{\infty,[0,1]^d},
\end{equation}

\begin{equation}
\begin{split}
   &\forall f \in \mathcal{C}_0([0,1]^d),\\ \frac{1}{\sqrt{N_1\times...\times N_d}}&||f||_{2,\vec{N}} \xrightarrow{\forall i, N_i\rightarrow +\infty} ||f||_{2,[0,1]^d}.
\end{split}
\end{equation}

\subsection{Theorems : multi-dimensional case.}

Let's consider $n_1+...+n_d$ qubits, a differentiable function $f$ defined on $[0,1]^d$ such that $||f||_{\infty,[0,1]^d}\neq 0$, $\epsilon_0 \in ]0,\pi/||f||_{\infty,[0,1]^d}]$ and $\epsilon_1>0,...,\epsilon_d>0$ such that $||f^{\vec{\epsilon}}||_{\infty,[0,1]^d}\neq0$ where $f^{\vec{\epsilon}}$ is the Walsh series of $f$ defined Eq. (\ref{Walsh serie multidim}).

\paragraph*{\textbf{Theorem 2.}}
There is an efficient quantum circuit of size $O(n_1+...+n_d+1/(\epsilon_1...\epsilon_d))$ and depth $O(1/(\epsilon_1...\epsilon_d))$, which, using one ancillary qubit, implements the quantum state $\ket{f}$ with a  probability of success $P(1)=\Theta(\epsilon_0^2)$ and infidelity $1-F=O((\epsilon_0+\sum_{i=1}^d\epsilon_i ||\partial_i f||_{\infty,[0,1]^d})^2)$.

In the particular case $n_1=...=n_d=n$ and $\epsilon_0=\epsilon_1=...=\epsilon_d$, one can show the following corollary.
\paragraph*{\textbf{Corollary 2.}}

There is an efficient quantum circuit of size $O(nd+1/\epsilon^{d/2})$, depth $O(1/\epsilon^{d/2})$, using one ancillary qubit, to implement a quantum state approximating the target state $\ket{f}$ up to an infidelity $1-F\leq \epsilon$ with a probability of success $P(1)=\Theta(\epsilon)$.

For any function $f$ with values $f(\vec{r})$ calculable in time $T_f$, the number of classical computations needed to find the quantum circuit is $O(T_f/(\epsilon_1...\epsilon_d)^2)$ (Theorem 2) or $O(T_f/\epsilon^d)$ (Corollary 2).

\subsection{Proof : multi-dimensional case.}
\label{Proof : multi-dimensional case.}
The proof of Theorem 2 is similar to the proof of Theorem 1. It starts with the following six lemmas.

\subsubsection{Lemmas}

\textbf{Lemma 2.0.} For any function $f$ continuously defined on $[0,1]^d$ such that $||f||_{\infty,[0,1]^d} \neq 0$ : $ \exists \vec{n}_0=(n_{0,1},...,n_{0,d})$ such that $  \forall n_1 \ge n_{0,1}$,...,$\forall n_d\ge n_{0,d}$, $ \forall \epsilon_0 \in ]0,\frac{2\pi}{||f||_{\infty,[0,1]^d}}[$, $||\frac{\hat{I}-e^{-i\hat{f}\epsilon_0}}{2}\ket{s}||_{2,\vec{N}} \neq 0 $ with $\vec{N}=(N_1,...,N_d)=(2^{n_1},...,2^{n_d})$.

\textbf{Proof of Lemma 2.0.}

Proof of Lemma 2.0 is very similar to proof of Lemma 1.0.

First, the fact that $||f||_{\infty,[0,1]^d} \neq 0$ implies it exists $\vec{r}_0\in [0,1]^d$ such that $f(\vec{r})=||f||_{\infty,[0,1]^d} \neq 0$. The continuity of $f$ implies it exists a neighbourhood $V$ of $\vec{r}_0$ such that $\forall \vec{r} \in V, f(\vec{r})\neq 0$. Let's note the following equality:

\begin{equation}
\begin{split}
||\frac{\hat{I}-e^{-i\hat{f}\epsilon_0}}{2}\ket{s}||_{2,\vec{N}}&=\sqrt{\sum_{\vec{r}\in\mathcal{X}_{\vec{n}}}\frac{\sin^2(f(\vec{r})\epsilon_0/2)}{N}},
\end{split}
\end{equation}
with $N=N_1\times ... \times N_d$.

The ensemble $\mathcal{X}_{\vec{n}}=\mathcal{X}_{n_1} \times ... \times \mathcal{X}_{n_d}$, is a set of vectors of dyadic rational. Using the same argument as in the proof of Lemma 1.0. for each axis, that dyadic rationals are dense in the $[0,1]$ : it exists a vector $\vec{n}_0=(n_{0,1},...,n_{0,d})$ and $\exists \vec{r}_1\in V$ such that $\forall n_1\ge n_{1,0},...,n_d \ge n_{d,0}$ $ \vec{r} \in  \mathcal{X}_{\vec{n}}$, with $\vec{n}=(n_1,...,n_d)$, and $f(\vec{r}_1 )\neq 0$. 

Futhermore, $0<\epsilon_0<\frac{2\pi}{||f||_{\infty,[0,1]^d}}$ implies that $0<|f(\vec{r}_1)|\epsilon_0/2<\pi$ and therefore $\sin^2(f(\vec{r}_1)\epsilon_0/2) \neq 0$. Finally, as a sum of positive numbers with at least one non-vanishing number,  $\forall n_1\ge n_{0,1},...\forall n_d \ge n_{0,d}, ||\frac{\hat{I}-e^{-i\hat{f}\epsilon_0}}{2}\ket{s}||_{2,\vec{N}}\neq 0$, with $\vec{N}=(N_1,...,N_d)=(2^{n_1},...,2^{n_d})$, which achieves the proof of Lemma 2.0.

\textbf{Lemma 2.1.}
For any differentiable function $f\in\mathcal{C}^1([0,1]^d)$ and $\epsilon_1>0$,...,$\epsilon_d>0$: the truncated Walsh series $f^{\vec{\epsilon}}$ defined Eq.(\ref{Walsh serie multidim}) approximates the function $f$ up to an error $O(\epsilon_1...+ \epsilon_d)$ : 
\begin{equation}
\begin{split}
    ||f-f^{\vec{\epsilon}}||_{\infty,[0,1]^d} \le \sum_{i=1}^d\epsilon_i ||\partial_i f||_{\infty,[0,1]^d}.
\end{split}
\end{equation}

\textbf{Proof of Lemma 2.1.}
The proof is similar to the one of Lemma 1.1. 

The function $f^{\vec{\epsilon}}$ is a sum of $M=M_1 \times ... \times M_d$ products of Walsh functions $w_{\vec{j}}(\vec{r})=w_{j_1}(r_1)\times ...\times w_{j_d}(r_d)$ with $j_i \in \{0,...,M_i-1\}$. 

Each Walsh function of order $j_i\le M_i$ is a piecewise function taking values $+1$ and $-1$ on at most $M_i$ different intervals $I_{k_i}^{m_i}=[k_i/M_i,(k_i+1)/M_i[$ for $k_i\in \{0,...,M_i\}$, related to the i-th axis. Therefore,  $w_{\vec{j}}=w_{j_1}\times ...\times w_{j_d}$ and the function $f^{\vec{\epsilon}}$ are piecewise functions constant on the $M$ different volumes $I_{\vec{k}}^{\vec{m}}=I_{k_1}^{m_1}\times...\times I_{k_d}^{m_d}$ with $\forall i, k_i\in \{0,...,M_i\}$:

\begin{equation}
\forall \vec{k}, \text{  } \forall \vec{r} \in I_{\vec{k}}^{\vec{m}}, \text{ } f^{\vec{\epsilon}}(\vec{r})=f^{\vec{\epsilon}}(k_1/M_1,...,k_d/M_d).
\end{equation}
Let's note $\vec{r}_{\vec{k},\vec{m}}=(k_1/M_1,...,k_d/M_d)$.
Then, from the definitions of $f^{\vec{\epsilon}}$, $a_{\vec{j},\vec{M}}^{f}$ one has:
\begin{equation}
    f^{\vec{\epsilon}}(\vec{r}_{\vec{k},\vec{m}})=f(\vec{r}_{\vec{k},\vec{m}}).
\end{equation}

Now, let's consider $\vec{r} \in I_{\vec{k}}^{\vec{m}}$  then $f(\vec{r})-f^{\epsilon_1}(\vec{r})=f(\vec{r})-f(\vec{r}_{\vec{k},\vec{m}})$. Let's consider the curve $ \gamma : \begin{pmatrix} [0,1] &\rightarrow& [0,1]^d  \\ t &\mapsto& t\vec{r}+(1-t)\vec{r}_{\vec{k},\vec{m}}\end{pmatrix}$. The mean value theorem on the function $g(t)=f(\gamma(t))$ implies it exists $t_1\in [0,1]$ such that $g(1)-g(0)=g'(t_1)(1-0)$ with $\forall t \in [0,1], g'(t)=(\partial_i f(\gamma(t))).(\gamma'(t))=\sum_{i=1}^d \partial_if(\gamma(t)) \times (r_i-k_i/M_i)$. Finally, noting that  $\forall i,  |r_i-k_i/M_i|\leq 1/M_i < \epsilon_i $ and $|\partial_if(\gamma(t))|\leq ||\partial_i f||_{\infty,[0,1]^d}$, one has :
\begin{equation}
   \forall \vec{r}\in[0,1]^d, \text{  } |f(\vec{r})-f^{\vec{\epsilon}}(\vec{r})| \le \sum_{i=1}^d  \epsilon_i ||\partial_if||_{\infty,[0,1]^d}.
\end{equation}
In particular : $||f-f^{\vec{\epsilon}}||_{\infty,[0,1]^d} \le \sum_{i=1}^d\epsilon_i ||\partial_i f||_{\infty,[0,1]^d}$.

\textbf{Lemma 2.2.}
For any function $f$ continuously defined on $[0,1]^d$ such that $||f||_{\infty,[0,1]^d} \neq 0$ : $ \exists \vec{n}_0=(n_{0,1},...,n_{0,d})$ such that $  \forall n_1 \ge n_{0,1}$,...,$\forall n_d\ge n_{0,d}$:

\begin{equation}
    ||f||_{2,\vec{N}}\ge C_1\sqrt{N},
\end{equation}
with $N=N_1\times...\times N_d=2^{n_1}\times ... \times 2^{n_d} $ and $\vec{N}=(N_1,...,N_d)$.

\textbf{Proof of Lemma 2.2.} The function $f$ is continuous on $[0,1]^d$ with a non-zero value. The continuity of $f$ and the fact that the dyadic rational numbers are dense in $[0,1]$ implies  $\exists \vec{n}_0=(n_{0,1},...,n_{0,d})$ such that $\exists \vec{r} \in [0,1]^d \cap \mathcal{X}_{\vec{n}_0}$ such that $f(\vec{r})\neq \vec{0}$,ie  $ \forall n_1 \ge n_{0,1},..., \forall n_d\ge n_{0,d}, \vec{r} \in \mathcal{X}_{\vec{n}}$ with $\vec{n}=(n_{1},...,n_{d})$ implying that $||f||_{2,\vec{N}}\neq0$ with $\vec{N}=(2^{n_0},...2^{n_d})$ and $||f||_{2,[0,1]^d}\ne 0$. The quantity $\frac{1}{\sqrt{N}}||f||_{2,\vec{N}}$ can be seen as a Riemman sum over a partition of $[0,1]^d$ composed of subrectangles of volume $1/N$. Therefore, $\frac{1}{\sqrt{N}}||f||_{2,\vec{N}}\xrightarrow{\forall i, N_i \rightarrow+ \infty} ||f||_{2,[0,1]^d}=l\ne 0$, meaning that $\forall \epsilon >0, \exists \vec{n}_1=(n_{1,1},...,n_{1,d})$ such that $\forall n_1\ge n_{1,1},...,\forall n_d \ge n_{1,d}$, $-\epsilon \le \frac{1}{\sqrt{N}}||f||_{2,\vec{N}}-l \le \epsilon$. One concludes the proof by setting $\epsilon=\frac{l}{2}$ such that $\exists \vec{n}_2=(n_{2,1},...,n_{2,d})$ such that $\forall n_1\ge n_{2,1},...,\forall n_d \ge n_{2,d}$ such that $||f||_{2,\vec{N}}\ge \frac{l}{2}\sqrt{N}$ with $C_1=l/2$.

\textbf{Lemma 2.3.} For any function $f$ defined and continuous on $[0,1]^d$ with $||f||_{\infty,[0,1]^d} \neq 0 $, $\exists \vec{n}_0=(n_{0,1},...,n_{0,d})$ such that $\forall n_1\ge n_{0,1},...,\forall n_d\ge n_{0,d}$, $\forall \epsilon_0 \in ]0,\pi/||f||_{\infty,[0,1]^d}]$ the normalization factor $\frac{1}{||\frac{\hat{I}-e^{-i\hat{f}\epsilon_0}}{2}\ket{s} ||_{2,\vec{N}}}$ can be bounded as
\begin{equation}
    \frac{2\sqrt{N}}{\epsilon_0||f||_{2,\vec{N}}} \leq \frac{1}{||\frac{\hat{I}-e^{-i\hat{f}\epsilon_0}}{2}\ket{s} ||_{2,\vec{N}}} \leq C_0\frac{2\sqrt{N}}{\epsilon_0||f||_{2,\vec{N}}}, 
\end{equation}
which is equivalent to:
\begin{equation}
   1 \leq \frac{\epsilon_0 ||f||_{2,\vec{N}}}{2||\sin(f\epsilon_0/2)||_{2,\vec{N}}} \leq C_0,
\end{equation}

with $N=N_1\times...\times N_d=2^{n_1}\times...\times 2^{n_d}$, $\vec{N}=(N_0,...,N_d)$ and $C_0=\pi/2$.

\textbf{Proof of Lemma 2.3.}
Lemma 2.0. implies it exists $\vec{n}_0=(n_{0,1},...,n_{0,d})$ such that $\forall n_1\ge n_{0,1},...,\forall n_d\ge n_{0,d}, \forall\epsilon_0 \in ]0,2\pi/||f||_{\infty,[0,1]^d}[$, $ ||\frac{\hat{I}-e^{-i\hat{f}\epsilon_0}}{2}\ket{s} ||_{2,\vec{N}} \neq 0$, with $\vec{N}=(N_0,...,N_d)=(2^{n_0},...,2^{n_d})$, ensuring the quantity $1/||\frac{\hat{I}-e^{-i\hat{f}\epsilon_0}}{2}\ket{s} ||_{2,\vec{N}}$ to be well defined. Left inequality is trivial using the fact that $\forall x\ge0, \sin(x)\le x$. For the right inequality, consider $\alpha \in ]0,\pi]$ and $\epsilon_0\in]0,\frac{2(\pi-\alpha)}{||f||_{\infty,[0,1]^d}}]$, then, thanks to the fact that the function $(\text{sinc}:x\mapsto \sin(x)/x)$ decreases on $[0,\pi]$:

\begin{equation}
\begin{split}
    &\frac{\epsilon_0 ||f||_{2,\vec{N}}}{2||\sin(f\epsilon_0/2)||_{2,\vec{N}}} \\& \leq \frac{||f||_{2,\vec{N}}}{\sqrt{\sum_{\vec{r}\in\mathcal{X}_{\vec{n}}}f(\vec{r})^2\frac{\sin^2(f(\vec{r})(\pi-\alpha)/||f||_{\infty,[0,1]^d})}{(f(\vec{r})(\pi-\alpha)/||f||_{\infty,[0,1]^d})^2}}} \\&
    \leq \frac{\pi-\alpha}{\sin(\pi-\alpha)}.
\end{split}    
\end{equation}
 Therefore, for $\alpha=\pi/2$, one prooves Lemma 1.3.

\textbf{Lemma 2.4.} For any function $f$  differentiable on $[0,1]^d$ with $||f||_{\infty,[0,1]^d} \neq 0 $,  $\epsilon_1\ge0,...,\epsilon_d\ge0$ such that   $||f^{\vec{\epsilon}}||_{\infty,[0,1]^d} \neq 0 $ where $f^{\vec{\epsilon}}$  is the Walsh series defined by Eq.(\ref{Walsh serie multidim}): $\exists \vec{n}_0=(n_{0,1},...,n_{0,d})$ such that $\forall n_1\ge n_{0,1},...,n_d\ge n_{0,d}$, $\forall \epsilon_0 \in ]0,\pi/||f||_{\infty,[0,1]^d}]$:
\begin{equation}
\begin{split}
&|\frac{1}{||\frac{\hat{I}-e^{-i\hat{f}^{\vec{\epsilon}}\epsilon_0}}{2}\ket{s} ||_{2,\vec{N}}}-\frac{1}{||\frac{\hat{I}-e^{-i\hat{f}\epsilon_0}}{2}\ket{s} ||_{2,\vec{N}}}| \\ &\leq C_0^2\frac{2N}{\epsilon_{0}||f||_{2,\vec{N}}||f^{\vec{\epsilon}}||_{2,\vec{N}}}\sum_{i=1}^d\epsilon_i ||\partial_i f||_{\infty,[0,1]^d},
\end{split}
\end{equation}
with $N=N_1\times...\times N_d=2^{n_1}\times...\times 2^{n_d}$, and $C_0=\pi/2$.

\textbf{Proof of Lemma 2.4}

First, Lemma 2.0. implies it exists $\vec{n}_1=(n_{1,1},...,n_{1,d})$ such that $\forall n_1\ge n_{1,1},...,\forall n_d\ge n_{1,d}, \forall \epsilon_0 \in ]0,2\pi/||f||_{\infty,[0,1]^d}[$, $ ||\frac{\hat{I}-e^{-i\hat{f}\epsilon_0}}{2}\ket{s} ||_{2,\vec{N}} \neq 0$, with $\vec{N}=(N_1,...,N_d)=(2^{n_1},...,2^{n_d})$. Then, $f^{\vec{\epsilon}}$ is a function defined on $[0,1]^d$ taking $M=M_1\times...\times M_d=2^{m_1}\times...\times2^{m_d}$ different values of $f$ on the invervals $I_{\vec{k}}^{\vec{m}}=I_{k_1}^{m_1}\times...\times I_{k_d}^{m_d}=[k_1/2^{m_1},(k_1+1)/2^{m_1}[\times...\times[k_d/2^{m_d},(k_d+1)/2^{m_d}[ $ with $\forall i, k_i\in \{0,...,2^{m_i}-1\}$ and $m_i=\lfloor \log_2(1/\epsilon_i) \rfloor+1$. Therefore, the fact that $||f^{\vec{\epsilon}}||_{\infty,[0,1]^d} \neq 0$ implies  $ \exists \vec{r} \in \mathcal{X}_{\vec{m}}$ such that $ f^{\vec{\epsilon}}(\vec{r})\neq 0$ and by setting $\vec{n}_0=(n_{0,1},...,n_{0,d})=(\max(n_{1,1},m_1),...,\max(n_{1,d},m_d))$, one has $\forall n_1\ge n_{0,1},...,\forall n_d \ge n_{0,d}$, $||\frac{\hat{I}-e^{-i\hat{f}^{\vec{\epsilon}}\epsilon_0}}{2}\ket{s} ||_{2,\vec{N}}\neq 0$ and $||\frac{\hat{I}-e^{-i\hat{f}\epsilon_0}}{2}\ket{s} ||_{2,\vec{N}}\neq 0$ with $\vec{N}=(N_0,...,N_d)=(2^{n_1},...,2^{n_d})$.

Using the subbadditivity of the $||.||_{2,\vec{N}}$ norm, one can show that 
\begin{equation}
\begin{split}
&|\frac{1}{||\frac{\hat{I}-e^{-i\hat{f}^{\vec{\epsilon}}\epsilon_0}}{2}\ket{s} ||_{2,\vec{N}}}-\frac{1}{||\frac{\hat{I}-e^{-i\hat{f}\epsilon_0}}{2}\ket{s} ||_{2,\vec{N}}}| \\&\leq \frac{||(e^{-i\hat{f}\epsilon_0}-e^{-i\hat{f}^{\vec{\epsilon}}\epsilon_0}\ket{s})||_{2,\vec{N}}}{2||\frac{\hat{I}-e^{-i\hat{f}^{\vec{\epsilon}}\epsilon_0}}{2}\ket{s} ||_{2,\vec{N}}\times||\frac{\hat{I}-e^{-i\hat{f}\epsilon_0}}{2}\ket{s} ||_{2,\vec{N}}}\\
& \leq \sqrt{N}\frac{||\sin((f-f^{\vec{\epsilon}})\epsilon_0/2)||_{2,\vec{N}}}{||\sin(f\epsilon_0/2)||_{2,\vec{N}}\times||\sin((f^{\vec{\epsilon}}\epsilon_0/2)||_{2,\vec{N}}} \\
& \leq N\frac{||f-f^{\vec{\epsilon}}||_{\infty,[0,1]^d}}{||f||_{2,\vec{N}}||f^{\vec{\epsilon}}||_{2,\vec{N}}}\frac{\epsilon_0}{2}(\frac{2C_0}{\epsilon_0})^2,
\end{split}
\end{equation}
where, for the last inequality, $||f||_{2,\vec{N}}\leq\sqrt{N}||f||_{\infty,\vec{N}}\leq\sqrt{N}||f||_{\infty,[0,1]^d}$ and Lemma 2.3. have been used. Lemma 2.1. achieves the proof:
\begin{equation}
\begin{split}
 &N\frac{||f-f^{\epsilon_1}||_{\infty,[0,1]^d}}{||f||_{2,\vec{N}}||f^{\vec{\epsilon}}||_{2,\vec{N}}}\frac{2C_0^2}{\epsilon_0}\\& \leq   C_0^2\frac{2N}{\epsilon_{0}||f||_{2,\vec{N}}||f^{\vec{\epsilon}}||_{2,\vec{N}}}\sum_{i=1}^d\epsilon_i ||\partial_i f||_{\infty,[0,1]^d},
\end{split}
\end{equation}
with $N=N_1\times...\times N_d=2^{n_1}\times...\times 2^{n_d}$, and $C_0=\pi/2$.

\textbf{Lemma 2.5}
For any function $f$ defined and continuous on $[0,1]^d$ with $||f||_{\infty,[0,1]^d} \neq 0 $ : it exists $\vec{n}_0=(n_{0,1},...,n_{0,d})$ such that $\forall n_1\ge n_{0,1},...,\forall n_d\ge n_{0,d}, \forall \epsilon_0 \in ]0,\pi/||f||_{\infty,[0,1]^d}]$:
\begin{equation}
\begin{split}    &|\frac{\epsilon_0 }{2\sqrt{N}||\frac{\hat{I}-e^{-i\hat{f}\epsilon_0}}{2}\ket{s}||_{2,\vec{N}}}-\frac{1}{||f||_{2,\vec{N}}}| \\ &\leq \frac{C_0\epsilon_0^2}{24}\frac{||f^3||_{2,\vec{N}}}{||f||^2_{2,\vec{N}}},
\end{split}
\end{equation}
with $N=N_1\times...\times N_d=2^{n_1}\times...\times 2^{n_d}$ and $C_0=\pi/2$.

\textbf{Proof of Lemma 2.5.}
First, Lemma 2.0. implies it exists $\vec{n}_0=(n_{0,1},...,n_{0,d})$ such that $\forall n_1\ge n_{0,1},...,\forall n_d\ge n_{0,d}, \forall \epsilon_0 \in ]0,\pi/||f||_{\infty,[0,1]^d}]$, $ ||\frac{\hat{I}-e^{-i\hat{f}\epsilon_0}}{2}\ket{s} ||_{2,\vec{N}} \neq 0$ and $||f||_{2,\vec{N}}$, with $\vec{N}=(N_0,...,N_d)=(2^{n_0},...,2^{n_d})$, ensuring the quantity $1/||\frac{\hat{I}-e^{-i\hat{f}\epsilon_0}}{2}\ket{s} ||_{2,\vec{N}}$ and $1/||f||_{2,\vec{N}}$ to be well defined. Using the subbadditivity of the $||.||_{2,\vec{N}}$ norm, the inequality $\forall x$ real, $x-\sin(x)\leq\frac{x^3}{6} $ and Lemma 2.3.:
\begin{equation}
\begin{split}
    &|\frac{\epsilon_0 }{2\sqrt{N}||\frac{\hat{I}-e^{-i\hat{f}\epsilon_0}}{2}\ket{s}||_{2,\vec{N}}}-\frac{1}{||f||_{2,\vec{N}}}|\\&=|\frac{\epsilon_0}{2||\sin(f\epsilon_0/2)||_{2,\vec{N}}}- \frac{1}{||f||_{2,\vec{N}}}|\\
    &\leq \frac{||\epsilon_0 f/2-\sin(\epsilon_0 f/2)||_{2,\vec{N}}}{||\sin(f\epsilon_0/2)||_{2,\vec{N}}||f||_{2,\vec{N}}} \\ 
    & \leq \frac{\epsilon_0^3}{48}\frac{||f^3||_{2,\vec{N}}}{||\sin(f\epsilon_0/2)||_{2,\vec{N}}||f||_{2,\vec{N}}}\\
    &\leq \frac{C_0\epsilon_0^2}{24}\frac{||f^3||_{2,\vec{N}}}{||f||_{2,\vec{N}}^2}.
 \end{split}
\end{equation}

\textbf{Lemma 2.6.} For any function $f$ continuous on $[0,1]^d$ such that $||f||_{\infty,[0,1]^d}\neq 0$ and $\epsilon_1>0,...,\epsilon_d>0$ such that $||f^{\vec{\epsilon}}||_{\infty,[0,1]^d} \neq 0 $: $\exists C_1>0, \exists \vec{n}_0=(n_{0,1},...,n_{0,d})$ such that $\forall n_1\ge n_{0,1},...,n_d\ge n_{0,d}$:

\begin{equation}
    ||f^{\vec{\epsilon}}||_{2,\vec{ N}}\ge C_1\sqrt{N},
\end{equation}
with $\vec{N}=(2^{n_1},...,2^{n_d})$, $N=N_1\times...\times N_d$, $\vec{\epsilon}=(\epsilon_1,...,\epsilon_d)$ and $f^{\vec{\epsilon}}$ the Walsh series of $f$ defined Eq.(\ref{Walsh serie multidim}).

\textbf{Proof of Lemma 2.6}
Let's define  $\forall i \in \{1,...,d\}, m_i=\lfloor \log_2(1/\epsilon_i) \rfloor+1$ and $M_i=2^{m_i}$.  $||f^{\vec{\epsilon}}||_{\infty,[0,1]^d}\neq 0$ implies $\exists \vec{r} \in \mathcal{X}_{\vec{m}}$ such that $f^{\vec{\epsilon}}(\vec{r})=f(\vec{r})\neq0$. Then,  $||f^{\vec{\epsilon}}||_{2,\vec{M}}=||f||_{2,\vec{M}}\neq0$. As shown in the proof of Lemma2.0., $\exists \vec{n}_0=(n_{0,1},...,n_{0,d})$ such that $\forall n_1\ge n_{0,1},...,\forall n_d \ge n_{0,d}, ||f||_{2,\vec{N}}\neq0 $ with $\vec{N}=(2^{n_1},...,2^{n_d})$. Lemma 2.2. on $f$ states $\exists C_1>0$ such that $\forall n_1\ge n_{0,1},...,\forall n_d\ge n_{0,d}, ||f||_{2,\vec{N}}\ge \sqrt{N} C_1$, with $N=N_1\times...\times N_d$. Let $\vec{n}=(n_1,...,n_d)$ be a vector such that $\forall i, n_i\ge n_{0,i}$. Let's note that if $\forall i, m_i\ge n_i$, $||f^{\epsilon_1}||_{2,\vec{N}}=||f||_{2,\vec{N}}\ge C_1\sqrt{N}$. Otherwise, one has $\prod_{i=1}^d \frac{1}{\sqrt{N_i}}||f^{\vec{\epsilon}}||_{2,\vec{N}}=\prod_{i=1}^d \frac{1}{\sqrt{N_i'}}||f||_{2,\vec{N'}} \ge C_1 $ where $\vec{N'}=(N_0',...,N_d')$ is defined as $\forall i : N_i'=N_i$ if $m_i \ge n_i$, $N_i'=M_i$ if $m_i \le n_i$.  One concludes $\exists C_1>0, \exists \vec{n}_0$ such that $\forall n_1\ge n_{0,1},...,\forall n_d \ge n_{0,d}$, $||f^{\vec{\epsilon}}||_{2,\vec{N}}\ge C_1\sqrt{N}$.

\subsubsection{Proof of Theorem 2}

Let's consider $n_1+...+n_d$ qubits, a differentiable function $f$ defined on $[0,1]^d$ such that $||f||_{\infty,[0,1]^d}\neq 0$, $\epsilon_0 \in ]0,\pi/||f||_{\infty,[0,1]^d}]$ and $\epsilon_1>0,...,\epsilon_d>0$ such that $||f^{\vec{\epsilon}}||_{\infty,[0,1]^d}\neq0$ where $f^{\vec{\epsilon}}$ is the Walsh series of $f$ defined Eq(\ref{Walsh serie multidim}).

The implemented quantum state after measuring $\ket{1}$ for the ancillary qubit $\ket{q_A}$ is
\begin{equation}
    \ket{\psi_{f^{\vec{\epsilon}}_0}}_{\epsilon_0}=-i\frac{\hat{I}-e^{-i\hat{f}^{\vec{\epsilon}}_0\epsilon_0}}{2||\frac{\hat{I}-e^{-i\hat{f}^{\vec{\epsilon}}_0\epsilon_0}}{2}\ket{s} ||_{2,\vec{N}}}\ket{s},
\end{equation}
where $f^{\vec{\epsilon}}_0$ is the Walsh series approximating $f$ up to an error $\sum_{i=1}^d\epsilon_i ||\partial_i f||_{\infty,[0,1]^d}$ (Lemma 2.1).

\textbf{Distance between the target state and the implemented state.}

Let's bound the infinite norm of the difference between the implemented quantum state $\ket{\psi_{f^{\vec{\epsilon}}}}_{\epsilon_0}$ and the target quantum state $\ket{f}=\frac{1}{||f||_{2,\vec{N}}}\sum_{\vec{r}\in\mathcal{X}_{\vec{n}}}f(\vec{r})\ket{\vec{r}}$.

\begin{equation}
\begin{split}
    &||\ket{\psi_{f^{\vec{\epsilon}}_0}}_{\epsilon_0}-\ket{f}||_{\infty,\vec{N}} \\ &\leq ||\ket{\psi_{f^{\vec{\epsilon}}_0}}_{\epsilon_0}-\ket{\psi_{f}}_{\epsilon_0}||_{\infty,\vec{N}} \\&+||\ket{\psi_{f}}_{\epsilon_0}-\ket{f}||_{\infty,\vec{N}}.
\label{inégalité première 2}
\end{split}
\end{equation}

Let's start to bound the first term by making the difference of the normalization factors appears:

\begin{equation}
\begin{split}
    &||\ket{\psi_{f^{\vec{\epsilon}}_0}}_{\epsilon_0}-\ket{\psi_{f}}_{\epsilon_0}||_{\infty,\vec{N}} \\&= ||\frac{\hat{I}-e^{-i\hat{f}^{\vec{\epsilon}}_0\epsilon_0}}{2||\frac{\hat{I}-e^{-i\hat{f}^{\vec{\epsilon}}_0\epsilon_0}}{2}\ket{s} ||_{2,\vec{N}}}\ket{s}-\frac{\hat{I}-e^{-i\hat{f}\epsilon_0}}{2||\frac{\hat{I}-e^{-i\hat{f}\epsilon_0}}{2}\ket{s} ||_{2,\vec{N}}}\ket{s}||_{\infty,\vec{N}} \\
    & \leq|\frac{1}{||\frac{\hat{I}-e^{-i\hat{f}^{\vec{\epsilon}}\epsilon_0}}{2}\ket{s} ||_{2,\vec{N}}}-\frac{1}{||\frac{\hat{I}-e^{-i\hat{f}\epsilon_0}}{2}\ket{s} ||_{2,\vec{N}}}|\\&\times  ||\frac{\hat{I}-e^{-i\hat{f}^{\vec{\epsilon}}_0\epsilon_0}}{2}\ket{s}||_{\infty,\vec{N}} \\
    &+||\frac{e^{-i\hat{f}^{\vec{\epsilon}}\epsilon_0}-e^{-i\hat{f}\epsilon_0}}{2||\frac{\hat{I}-e^{-i\hat{f}\epsilon_0}}{2}\ket{s}||_{2,\vec{N}}}\ket{s}||_{\infty,\vec{N}} \\
    &\leq|\frac{1}{||\frac{\hat{I}-e^{-i\hat{f}^{\vec{\epsilon}}\epsilon_0}}{2}\ket{s} ||_{2,\vec{N}}}-\frac{1}{||\frac{\hat{I}-e^{-i\hat{f}\epsilon_0}}{2}\ket{s} ||_{2,\vec{N}}}|\\&\times  \frac{1}{\sqrt{N}}\max_{\vec{r}\in\mathcal{X}_{\vec{n}}}|\sin(f^{\vec{\epsilon}}_0(\vec{r}) \epsilon_0/2))| \\
    &+\frac{1}{||\frac{\hat{I}-e^{-i\hat{f}\epsilon_0}}{2}\ket{s}||_{2,\vec{N}}}\frac{1}{\sqrt{N}}\max_{\vec{r}\in\mathcal{X}_{\vec{n}}}|\sin((f^{\vec{\epsilon}}_0(\vec{r})-f(\vec{r}))\epsilon_0/2))|\\
    &\leq (C_0^2\sqrt{N}\frac{||f||_{\infty,[0,1]^d}}{||f||_{2,\vec{N}} ||f^{\vec{\epsilon}}_0||_{2,\vec{N}}} +\frac{C_0}{||f||_{2,\vec{N}}})\sum_{i=1}^d\epsilon_i ||\partial_i f||_{\infty,[0,1]^d},
\end{split}
\end{equation}

where, in the last inequality, one uses $|\sin(x)|\leq|x|$, $||f^{\vec{\epsilon}}||_{\infty,\vec{N}} \leq ||f^{\vec{\epsilon}}||_{\infty,[0,1]^d} \leq ||f||_{\infty,\vec{N}}$, Lemma 2.1., Lemma 2.3. and Lemma 2.4.

Lemma 2.2 and 2.6 imply there is a constant $B>0$ depending only on $f$ such that 
\begin{equation}
C_0^2\sqrt{N}\frac{||f||_{\infty,[0,1]^d}}{||f||_{2,\vec{N}} ||f^{\vec{\epsilon}}_0||_{2,\vec{N}}} +C_0\frac{1}{||f||_{2,\vec{N}}} \leq \frac{B}{\sqrt{N}}.
\end{equation}

Leading to the bound on the first term
\begin{equation} \begin{split}
&||\ket{\psi_{f^{\vec{\epsilon}}_0}}_{\epsilon_0}-\ket{\psi_{f}}_{\epsilon_0}||_{\infty,\vec{N}}\\& \leq \frac{B}{\sqrt{N}}\sum_{i=1}^d\epsilon_i ||\partial_i f||_{\infty,[0,1]^d}.
\end{split}
\end{equation}

The second term in inequality (\ref{inégalité première 2}) is bounded by using the Taylor expansion of the exponential term $e^{-i\hat{f}\epsilon_0}=\hat{I}-i\hat{f}\epsilon_0+R_1(-i\hat{f}\epsilon_0)$ with $R_1(x)=\sum_{k=2}^{+\infty}\frac{x^k}{k!}$ and by making the difference of the norms appear:

\begin{equation}
\begin{split}
    &||\ket{\psi_{f}}_{\epsilon_0}-\ket{f}||_{\infty,\vec{N}} \\& =|| -i \frac{\hat{I}-(\hat{I}-i\hat{f}\epsilon_0+R_1(-i\hat{f}\epsilon_0))}{2||\frac{\hat{I}-e^{-i\hat{f}\epsilon_0}}{2}\ket{s}||_{2,\vec{N}}}\ket{s} \\&-\frac{\sqrt{N}}{||f||_{2,\vec{N}}}\hat{f}\ket{s}||_{\infty,\vec{N}} \\
    & \leq |\frac{\epsilon_0}{2||\frac{\hat{I}-e^{-i\hat{f}\epsilon_0}}{2}\ket{s}||_{2,\vec{N}}}-\frac{\sqrt{N}}{||f||_{2,\vec{N}}}|\times||\hat{f}\ket{s}||_{\infty,\vec{N}}\\&+\frac{||R_1(-i\hat{f}\epsilon_0)\ket{s}||_{\infty,\vec{N}}}{2||\frac{\hat{I}-e^{-i\hat{f}\epsilon_0}}{2}\ket{s}||_{2,\vec{N}}}.
\end{split}
\end{equation}

The Taylor inequality applied on the remainders of the cosinus and sinus functions implies $||R_1(-i\hat{f}\epsilon_0)\ket{s}||_{\infty,\vec{N}}\leq\frac{\epsilon_0^2}{2\sqrt{N}}||f^2||_{\infty,[0,1]^d}+\frac{\epsilon_0^3}{6\sqrt{N}}||f^3||_{\infty,[0,1]^d}$ and using Lemma 2.3. and Lemma 2.5.:
\begin{equation}
\begin{split}
    &||\ket{\psi_{f}}_{\epsilon_0}-\ket{f}||_{\infty,\vec{N}}
    \\& \leq \epsilon_0\frac{C_0||f^2||_{\infty,[0,1]^d}}{2||f||_{2,\vec{N}}}\\&+\epsilon_0^2(\frac{C_0}{24}\frac{||f^3||_{2,\vec{N}}}{||f||^2_{2,\vec{N}}}||f||_{\infty,[0,1]^d}+\frac{C_0||f^3||_{\infty,[0,1]^d}}{6||f||_{2,\vec{N}}}).
\end{split}
\end{equation}

Using Lemma2.2 and the fact that $\epsilon_0^2 \leq \pi \epsilon_0/||f||_{\infty,[0,1]^d}$, there is a constant $A>0$ depending only on $f$ such that

\begin{equation}
\begin{split}
    ||\ket{\psi_{f}}_{\epsilon_0}-\ket{f}||_{\infty,\vec{N}}
    & \leq A\frac{\epsilon_0}{\sqrt{N}}.
\end{split}
\end{equation}

Finally, using the norm inequality : $||.||_{2,\vec{N}}\leq\sqrt{N}||.||_{\infty,\vec{N}}$, one gets :
\begin{equation}
    ||\ket{\psi_{f^{\vec{\epsilon}}}}_{\epsilon_0}-\ket{f}||_{2,\vec{N}} \leq A\epsilon_0+B\sum_{i=1}^d\epsilon_i ||\partial_i f||_{\infty,[0,1]^d}.
\end{equation}

This result implies in terms of infidelity $1-F=1-|\braket{f|\psi_{f^{\vec{\epsilon}}}}_{\epsilon_0}|^2$:
\begin{equation}
    1-F \leq (A\epsilon_0+B\sum_{i=1}^d\epsilon_i ||\partial_i f||_{\infty,[0,1]^d})^2,
\end{equation}
which concludes that $1-F=O((\epsilon_0+\sum_{i=1}^d\epsilon_i ||\partial_i f||_{\infty,[0,1]^d})^2)$.

\textbf{Bounds on the probability of success.}

The probability of measuring the ancilla qubit $\ket{q_A}$ in state $\ket{1}$ with $\epsilon_0 \in ]0,\pi/||f||_{\infty,[0,1]^d}]$ is:
\begin{equation}
\begin{split}
&P(1)=||\frac{\hat{I}-e^{-i\hat{f}^{\vec{\epsilon}}\epsilon_0}}{2}\ket{s}||_{2,\vec{N}}^2= \\ &\frac{1}{N}|| \sin{(\frac{f^{\vec{\epsilon}}\epsilon_0}{2})} ||_{2,\vec{N}}^2.
\end{split}
\end{equation}

The upper bound comes from the inequality $\forall x \ge 0$,  $\sin(x)\leq x$:
\begin{equation}
    P(1)\leq \frac{||f^{\vec{\epsilon}}||_{2,\vec{N}}^2\epsilon_0^2}{4N} \leq \frac{||f||_{\infty,[0,1]^d}^2\epsilon_0^2}{4},
\end{equation}
where one has to use $||f^{\vec{\epsilon}}||_{2,\vec{N}}\leq \sqrt{N} ||f^{\vec{\epsilon}}||_{\infty,\vec{N}}\leq \sqrt{N} ||f||_{\infty,\vec{N}} \leq \sqrt{N} ||f||_{\infty,[0,1]^d}$.

The lower bound comes from the fact the function $(x\mapsto \text{sinc}(x)=\sin(x)/x)$ decreases on $[0,\pi/2]$:
\begin{equation}
\begin{split}
P(1)&=\frac{1}{N}\sum_{x\in\mathcal{X}_n}\sin^2(\frac{f^{\vec{\epsilon}}(x)\epsilon_0}{2})\\
& =\frac{1}{N}\sum_{x\in\mathcal{X}_n}(\frac{f^{\vec{\epsilon}}(x)\epsilon_0}{2})^2\text{sinc}^2(\frac{f^{\vec{\epsilon}}(x)\epsilon_0}{2})\\
& \ge \frac{1}{N}\sum_{x\in\mathcal{X}_n}(\frac{f^{\vec{\epsilon}}(x)\epsilon_0}{2})^2\text{sinc}^2(\pi/2) \\
& \ge \frac{\epsilon_0^2}{\pi^2 N}||f^{\vec{\epsilon}}||_{2,\vec{N}}^2.
\end{split}
\end{equation}

Lemma 1.6. implies it exists a constant $D'$, independent of $\vec{\epsilon}$, such that $||f^{\vec{\epsilon}}||_{2,\vec{N}}^2 \ge N D'$. Therefore, it exists a constant $D$, independent of $\epsilon_0, \vec{\epsilon}$ and $N$, such that $P(1)\ge D\epsilon_0^2$. This concludes that $P(1)=\Theta(\epsilon_0^2)$.

\textbf{Complexities.}

The protocol starts with $n_1+...+n_d+1$ Hadamard gates  to prepare the state $ \ket{s}=\frac{1}{\sqrt{N}}\sum_{\vec{r} \in \mathcal{X}_{\vec{n}}\ket{r}}$. The number of single-qubit gates and CNOT gates to implement the diagonal unitary $\hat{U}_{f^{\vec{\epsilon}},\epsilon_0}$, and so the controlled-$\hat{U}_{f^{\vec{\epsilon}},\epsilon_0}$, is $O( \frac{1}{\epsilon_1...\epsilon_d})$. Finally, a Hadamard gate and Phase gate are applied on $\ket{q_A}$ to perform the right interference giving the target state up to an infidelity $1-F=O((\epsilon_0+\sum_{i=1}^d\epsilon_i ||\partial_i f||_{\infty,[0,1]^d})^2)$. The size is $O(n_1+...+n_d+\frac{1}{\epsilon_1...\epsilon_d})$ while the depth is $\frac{1}{\epsilon_1...\epsilon_d})$ independent with the number of qubits.

In the particular case $n_1=...=n_d=n$ and $\epsilon_1=...=\epsilon_d=\epsilon$, size  becomes $O(nd+\frac{1}{\epsilon^d})$ and depth $O(\frac{1}{\epsilon^d})$.

For any function $f$ with values $f(\vec{r})$ calculable in time $T_f$, the number of classical computations to compute the Walsh coefficients of the $\vec{M}$-Walsh Series of $f$, with $\vec{M}=(M_1,...,M_d)$ is $O(T_f(M_1...M_d)^2)$ which is also $O(\frac{T_f}{(\epsilon_1...\epsilon_d)^2})$.

\subsection{Sparse Walsh Series Loader}

Sparse Walsh Series is an efficient tool to design quantum circuits for quantum state preparation with particularly small depth and size.  For some real-valued functions of particular interests, Fig. 3 shows that one can obtain quantum circuits with smaller depth than the "dense" Walsh Series loader (see Theorem 1) in order to reach the same accuracy. The problem of finding the "best" sparse Walsh Series which approximates the target real-valued function is called the Minimax series problem \cite{yuen1975function}. One efficient way to get a sparse Walsh Series approximating a function $f$ is to compute the Walsh Series of $f$ on $M$ points as written in Eq.(\ref{Walsh serie}). Then, one can sort the absolute value of the $M$ coefficients in decreasing order and implement the largest ones once a target infidelity is reached. Other possible methods include threshold
sampling, data compression \cite{yuen1975function} or efficient estimation of the number M of best Walsh coefficients \cite{kushilevitz1991learning}.

\paragraph*{Definitions.} Let's consider the problem of loading unto $n$ qubits a sparse Walsh Series:
\begin{equation}
f_s=\sum_{j\in S} a_j w_j
\label{sparse Ws}
\end{equation}
with real coefficients $a_j$, $ S\subseteq \{0,...,2^{n_0}-1\}$, $n_0$ an integer such that $n\geq n_0$ and. The sparsity of the Walsh series is defined as the number of terms in the series, i.e. the cardinal of $S$ : $s=|S|$. The complexity of the quantum circuit implementing a sparse Walsh series depends directly on a parameter $k$  defined as the maximum Hamming weight of the binary decomposition of the Walsh coefficient indices: $k=\max_{j\in S}(\sum_{i=0}^{l_j}j_i)$ with $j=\sum_{i=0}^{l_j}j_i2^i$. and $j_i\in\{0,1\}$. The parameter $k$ is also the maximum number of qubits on which one Walsh operator is implemented implying that $k\leq n_0$.

\paragraph*{\textbf{Theorem 3.}} Let $f_s$ be a sparse Walsh series with sparsity $s$ and parameters $n_0$ and $k$ as defined in Eq.(\ref{sparse Ws}) such that $||f_s||_{\infty,[0,1]}\neq 0$. Then, $\forall n\geq n_0$ and $\epsilon_0 \in ]0,\pi/||f_s||_{\infty,[0,1]}]$, there is a quantum circuit of size $O(n+sk)$, depth $O(sk)$ which, using one ancillary qubit, implements the quantum state $\ket{f_s}=\frac{1}{||f_s||_{2,N}}\sum_{x\in \mathcal{X}_n}f_s(x)\ket{x}$ with a probability of success $P(1)=\Theta(\epsilon_0^2)$ and infidelity $1-F\leq \epsilon_0^2$.

Let's now consider a function $f$ approximated by a sparse Walsh series $f_s$.
\paragraph*{\textbf{Corollary 3.}} Let $f$ be a real valued function defined on $[0,1]$ such that $||f||_{\infty,[0,1]}\neq 0$ and $f_s$ be a sparse Walsh series of sparsity $s$ and parameters $n_0$ and $k$ as defined in Eq.(\ref{sparse Ws}) such that $||f_s||_{\infty,[0,1]} \neq 0$ and $||f-f_s||_{\infty,[0,1]}\leq \epsilon_1$ Then, $\forall n\ge n_0$ and $\forall \epsilon_0 \in ]0,\pi/||f_s||_\infty]$, there is a quantum circuit of size $O(n+sk)$, depth $O(sk)$ which, using one ancillary qubit, implements the quantum state $\ket{f}=\frac{1}{||f||_{2,N}}\sum_{x\in \mathcal{X}_n}f(x)\ket{x}$ with a probability of success $P(1)=\Theta(\epsilon_0^2)$ and infidelity $1-F\leq (\epsilon_0+\epsilon_1)^2$.

\textbf{Proof.} 

The proof of this theorem and corollary is very similar than the proof of theorem 1. It is based on the following Lemmas :

\textbf{Lemma 3.0.} For any sparse Walsh series $f_s$ of parameter $n_0$ as defined in Eq.(\ref{sparse Ws}) such that $||f_s||_{\infty,[0,1]} \neq 0$: $\forall n\ge n_0$, $\forall \epsilon_0 \in ]0,\frac{2\pi}{||f_s||_{\infty,[0,1]}}[$
, $||\frac{\hat{I}-e^{-i\hat{f_s}\epsilon_0}}{2}\ket{s}||_{2,N} \neq 0 $ with $N=2^n$.

\textbf{Proof of Lemma 3.0.}

The function $f_s$ is a sum of $s$ Walsh functions of order $j \in \{0,...,2^{n_0}-1\}$. The Walsh function of order $j$ is a piecewise function taking values $+1$ and $-1$ on at most $2^p$ different intervals $I_k^p=[k/2^p,(k+1)/2^p[$ with $p\leq n_0$ and $k\in \{0,...,2^p-1\}$. Therefore, the function $f_s$ is a piecewise function which is constant on each of the $N_0=2^{n_0}$ intervals $I_{k}^{n_0}$:

\begin{equation}
\begin{split}
\forall k \in \{0,...,N_0-1&\},\text{ } \forall x \in I_k^{n_0}, \\ f_s(x)&=f_s(k/N_0).
\end{split}
\end{equation}
The fact that $||f_s||_{\infty,[0,1]} \neq 0$ implies it exists $k_0\in \{0,...,2^{n_0}-1\}$ such that $\forall x\in I_{k_0}^{n_0}$, $f_s(x)=||f_s||_{\infty,[0,1]} \neq 0$. Let's note that $\forall n\geq n_0$, $\mathcal{X}_{n_0}\subseteq \mathcal{X}_{n}$ and $I_{k_0}^{n_0} \cap \mathcal{X}_{n_0} \neq 0$. Therefore :

\begin{equation}
\begin{split}
||\frac{\hat{I}-e^{-i\hat{f}_s\epsilon_0}}{2}\ket{s}||_{2,N}&=\sqrt{\sum_{x\in\mathcal{X}_n}\frac{\sin^2(f_s(x)\epsilon_0/2)}{N}}. \\
& \geq \sqrt{\sum_{x\in I_{k_0}^{n_0} \cap \mathcal{X}_{n_0} }\frac{\sin^2(f_s(x)\epsilon_0/2)}{N}} \\
& \geq \frac{\sin(||f_s||_{\infty,[0,1]}\epsilon_0/2)}{\sqrt{N}}
\end{split}
\end{equation}
 Then,  $0<\epsilon_0<\frac{2\pi}{||f_s||_{\infty,[0,1]}}$ implies that $0<||f_s||_{\infty,[0,1]}\epsilon_0/2<\pi$ and therefore $\sin(||f_s||_{\infty,[0,1]}\epsilon_0/2) > 0$. Finally, $\forall \epsilon_0 \in ]0,\frac{2\pi}{||f_s||_{\infty,[0,1]}}[$,  $\forall n\ge n_0, ||\frac{\hat{I}-e^{-i\hat{f}_s\epsilon_0}}{2}\ket{s}||_{2,N}\neq 0$, which achieves the proof of Lemma 3.0.

\textbf{Lemma 3.1.}
For any sparse Walsh series $f_s$ of parameter $n_0$ as defined in Eq.(\ref{sparse Ws}) such that $||f_s||_{\infty,[0,1]} \neq 0$, it exists a constant $C_1>0$ depending only on $f_s$ and $n_0$ such that $\forall n\ge n_0$:

\begin{equation}
    ||f_s||_{2,N}= C_1\sqrt{N},
\end{equation}

with $N=2^n$.

\textbf{Proof of Lemma 3.1.} 
Let's note $I_k^{n_0}=[k/N_0,(k+1)/N_0[$ with $k\in \{0,...,N_0-1\}$ and $N_0=2^{n_0}$ such that $\bigcup_{k=0}^{N_0-1}I_k^{n_0}= [0,1[$ and $\forall n\ge n_0$, $|I_k^{n_0}\cap \mathcal{X}_n|=N/N_0$ with $N=2^n$. Then:

\begin{equation}
\begin{split}
    ||f_s||_{2,N}&=\sqrt{\sum_{x\in \mathcal{X}_n}|f_s(x)|^2}\\&=\sqrt{\sum_{k=0}^{N_0-1}\sum_{x\in I_k^{n_0}\cap \mathcal{X}_n}|f_s(x)|^2}
    \\& =\sqrt{\sum_{k=0}^{N_0-1}|f_s(k/N_0)|^2 \times \sum_{x\in I_k^{n_0}\cap \mathcal{X}_n}}1
    \\&=\sqrt{\sum_{k=0}^{N_0-1}|f_s(k/N_0)|^2 \times N/N_0}
    \\&=C_1\sqrt{N}
\end{split}
\end{equation}
with $C_1=\sqrt{\frac{1}{N_0}\sum_{k=0}^{N_0-1}|f_s(k/N_0)|^2}=||f_s||_{2,N_0}/\sqrt{N_0} >\frac{1}{\sqrt{N_0}}||f_s||_{\infty,[0,1]} > 0$.

\textbf{Lemma 3.3.} For any sparse Walsh series $f_s$ of parameter $n_0$ as defined in Eq.(\ref{sparse Ws}) such that $||f_s||_{\infty,[0,1]} \neq 0$, $\forall n\ge n_0$, $\forall \epsilon_0 \in ]0,\pi/||f_s||_{\infty,[0,1]}]$ the normalization factor $\frac{1}{||\frac{\hat{I}-e^{-i\hat{f}_s\epsilon_0}}{2}\ket{s} ||_{2,N}}$ can be bounded as
\begin{equation}
    \frac{2}{\epsilon_0 C_1} \leq \frac{1}{||\frac{\hat{I}-e^{-i\hat{f}_s\epsilon_0}}{2}\ket{s} ||_{2,N}} \leq \frac{\pi}{\epsilon_0 C_1},
\end{equation}
which is equivalent to:
\begin{equation}
   1 \leq \frac{\epsilon_0 C_1 \sqrt{N}}{2||\sin(f_s\epsilon_0/2)||_{2,N}} \leq C_0,
\end{equation}

with $N=2^n$, $C_0=\pi/2$ and $C_1=||f_s||_{2,N_0}/\sqrt{N_0}$.

\textbf{Proof of Lemma 3.3.}
Lemma 3.0. implies $\forall n>n_0, \forall \epsilon_0 \in ]0,2\pi/||f_s||_{\infty,[0,1]}[$, $ ||\frac{\hat{I}-e^{-i\hat{f}_s\epsilon_0}}{2}\ket{s} ||_{2,N} \neq 0$, with $N=2^n$, ensuring the quantity $1/||\frac{\hat{I}-e^{-i\hat{f}_s\epsilon_0}}{2}\ket{s} ||_{2,N}$ to be well defined. The left inequality is trivial using the fact that $\forall x\ge0, \sin(x)\le x$ and Lemma 3.2. For the right inequality, consider $\alpha \in ]0,\pi]$ and $\epsilon_0\in]0,\frac{2(\pi-\alpha)}{||f_s||_{\infty,[0,1]}}]$, then, thanks to the fact that the function $(x\mapsto \sin{(x)}/x)$ is decreasing on $[0,\pi]$ :

\begin{equation}
\begin{split}
    &\frac{\epsilon_0 ||f_s||_{2,N}}{2||\sin(f_s\epsilon_0/2)||_{2,N}} \\ &\leq \frac{||f_s||_{2,N}}{\sqrt{\sum_{x\in\mathcal{X}_n}f_s(x)^2\frac{\sin^2(f_s(x)(\pi-\alpha)/||f_s||_{\infty,[0,1]})}{(f_s(x)(\pi-\alpha)/||f_s||_{\infty,[0,1]})^2}}} \\&
    \leq \frac{\pi-\alpha}{\sin(\pi-\alpha)}.
\end{split}    
\end{equation}
Therefore, for $\alpha=\pi/2$ and using Lemma 3.2, one proves Lemma 3.3.

\textbf{Lemma 3.4}
For any sparse Walsh series $f_s$ of parameter $n_0$ as defined in Eq.(\ref{sparse Ws}) such that $||f_s||_{\infty,[0,1]} \neq 0$, $\forall n\ge n_0$, $\forall \epsilon_0 \in [0,\pi/||f_s||_{\infty,[0,1]}]$:
\begin{equation}
\begin{split}
    |\frac{\epsilon_0 }{2\sqrt{N}||\frac{\hat{I}-e^{-i\hat{f}_s\epsilon_0}}{2}\ket{s}||_{2,N}}-\frac{1}{||f_s||_{2,N}}|\\ \leq C_2\epsilon_0^2/\sqrt{N},
\end{split}
\end{equation}
with $N=2^n$,  $C_2=\frac{\pi}{48}\frac{||f_s^3||_{2,N_0}}{||f_s||_{2,N_0}^2}$.

\textbf{Proof of Lemma 3.4.}
Using the subbadditivity of the $||.||_{2,N}$ norm, the inequality $\forall x$ real, $x-\sin(x)\leq\frac{x^3}{6} $ and Lemma 3.2.:
\begin{equation}
\begin{split}
    &|\frac{\epsilon_0}{2\sqrt{N}||\frac{\hat{I}-e^{-i\hat{f_s} \epsilon_0}}{2}\ket{s}||_{2,N}}-\frac{1}{||f_s||_{2,N}}|\\&=|\frac{\epsilon_0}{2||\sin(f_s\epsilon_0/2)||_{2,N}}- \frac{1}{||f_s||_{2,N}}|\\
    &\leq \frac{||\epsilon_0 f_s/2-\sin(\epsilon_0 f_s/2)||_{2,N}}{||\sin(f_s\epsilon_0/2)||_{2,N}||f_s||_{2,N}} \\ 
    & \leq \frac{\epsilon_0^3}{48}\frac{||f_s^3||_{2,N}}{||\sin(f_s\epsilon_0/2)||_{2,N}||f_s||_{2,N}}\\
    &\leq \frac{C_0\epsilon_0^2}{24}\frac{||f_s^3||_{2,N}}{||f_s||_{2,N}^2}=C_2\epsilon_0^2/\sqrt{N}.
 \end{split}
\end{equation}
with $C_2=\frac{C_0}{24}\frac{||f_s^3||_{2,N_0}}{||f_s||_{2,N_0}^2}$ and $C_0=\pi/2$.

\subsubsection{Proof of Theorem 3}

Let's consider $f_s=\sum_{j\in S}a_jw_j$ a sparse Walsh series of parameters $s, n_0$ and  $k$. The quantum state implemented by the sparse WSL is \begin{equation}
\begin{split}
  \ket{\psi_{f_s}}=-i\frac{\hat{I}-e^{-i\hat{f}_s\epsilon_0}}{2||\frac{\hat{I}-e^{-i\hat{f}_s\epsilon_0}}{2}\ket{s} ||_2}\ket{s} ,
\end{split}
\end{equation}
with $\hat{f}_s=\sum_{x\in \mathcal{X}_n}f_s(x)\ket{x}\bra{x}$ and $\ket{f_s}=\frac{1}{||f_s||_{2,N}}\sum_{x\in \mathcal{X}_n}f_s(x)\ket{x}$.

First, one can expand the following term $e^{-i\hat{f}_s\epsilon_0}=\hat{I}-i\hat{f}_s\epsilon_0+R_1(-i\hat{f}_s\epsilon_0)$ where $R_1$ is the remainder of the Taylor series of the exponential function. Then:

\begin{equation}
\begin{split}
    &||\ket{\psi_{f_s}}_{\epsilon_0}-\ket{f_s}||_{\infty,N}\\ &\leq \frac{||R_1(-i\hat{f}_s \epsilon_0)\ket{s}||_{\infty,[0,1]}}{2||\frac{\hat{I}-e^{-i\hat{f}_s\epsilon_0}}{2}\ket{s} ||_2} \\&+ |\frac{\epsilon_0}{2||\frac{\hat{I}-e^{-i\hat{f}_s\epsilon_0}}{2}\ket{s} ||_2}-\frac{\sqrt{N}}{||f_s||_{2,N}}|\times ||\hat{f}_s\ket{s}||_{\infty,N}
\label{inégalité première 3}
\end{split}
\end{equation}
The first term can be bounded using the remainders of the cosinus and sinus functions: $||R_1(-i\hat{f}_s \epsilon_0)\ket{s}||_{\infty,[0,1]}\epsilon_0)||\leq ||R_{\cos}(\hat{f}_s \epsilon_0)\ket{s}||_{\infty,[0,1]}+||R_{\sin}(\hat{f}_s \epsilon_0)\ket{s}||_{\infty,[0,1]}$ and using the Taylor inequality, one has:
\begin{equation}
   ||R_{\cos}(\hat{f}_s \epsilon_0)\ket{s}||_{\infty,[0,1]}\leq \frac{\epsilon_0^2 ||f_s^2||_{\infty,[0,1]}}{2\sqrt{N}}  
\end{equation}
and
\begin{equation}
   ||R_{\sin}(\hat{f}_s \epsilon_0)\ket{s}||_{\infty,[0,1]}\leq \frac{\epsilon_0^3 ||f_s^3||_{\infty,[0,1]}}{6\sqrt{N}}  
\end{equation}
Therefore, using Lemma 3.3, $\exists C>0$ such that:
\begin{equation}
\begin{split}
    &\frac{||R_1(-i\hat{f}_s \epsilon_0)\ket{s}||_{\infty,[0,1]}}{2||\frac{\hat{I}-e^{-i\hat{f}_s\epsilon_0}}{2}\ket{s} ||_2} \leq C\frac{\epsilon_0}{\sqrt{N}} 
\end{split}
\end{equation}
Then, using Lemma 3.4, the second term is bounded such as:
\begin{equation}
\begin{split}
    & |\frac{\epsilon_0}{2||\frac{\hat{I}-e^{-i\hat{f}_s\epsilon_0}}{2}\ket{s} ||_2}-\frac{\sqrt{N}}{||f_s||_{2,N}}|\times ||\hat{f}_s\ket{s}||_{\infty,N} \\ & \leq C_2 \epsilon_0^2 ||f_s||_{\infty,[0,1]}/\sqrt{N}\leq C'\frac{\epsilon_0}{\sqrt{N}} 
\end{split}
\end{equation}
with $C'>0$. Therefore, $\exists C''>0$ such that $\forall n\ge n_0$, $\forall \epsilon_0 \in [0,\pi/||f_s||_{\infty,[0,1]}]$ 

\begin{equation}
    ||\ket{\psi_{f_s}}_{\epsilon_0}-\ket{f_s}||_{2,N} \leq C''\epsilon_0
\end{equation}
Finally, one has $1-F=O(\epsilon_0^2)$.

In the case where the sparse Walsh series $f_s$ approximates a given function $f$ up to an accuracy $\epsilon_1$ such that:
\begin{equation}
    ||f-f_s||_{\infty,[0,1]}\leq \epsilon_1,
\end{equation}
the quantum state $\ket{\psi_{f_s}}_{\epsilon_0}$ approximates the target quantum state $\ket{f}$ up to an error $O((\epsilon_0+\epsilon_1)^2)$ in terms of infidelity :

\begin{equation}
\begin{split}
    ||&\ket{f_s}-\ket{f}||_{\infty,N}=\max_{x\in \mathcal{X}_n}|\frac{f_s(x)}{||f_s||_{2,N}}-\frac{f(x)}{||f||_{2,N}}| \\
    &\leq \frac{||f_s-f||_{\infty,[0,1]}}{||f_s||_{2,N}}+|\frac{1}{||f||_{2,N}}-\frac{1}{||f_s||_{2,N}}|\times ||f||_{\infty,[0,1]} \\
    & \leq \frac{\epsilon_1}{||f_s||_{2,N}}+\frac{||f-f_s||_{2,N}}{||f||_{2,N}||f_s||_{2,N}} ||f||_{\infty,[0,1]} \\& \leq C_3 \epsilon_1/\sqrt{N}
\end{split}
\end{equation}
with $C_3>0$. Therefore, $||\ket{\psi_{f_s}}_{\epsilon_0}-\ket{f}||_{2,N} =O(\epsilon_0+\epsilon_1)$ and the infidelity between $\ket{\psi_{f_s}}_{\epsilon_0}$and $\ket{f}$ is $1-F=O((\epsilon_0+\epsilon_1)^2)$.
The probability of success is bounded as in theorem 1 : $P(1)=\Theta(\epsilon_0^2)$. The complexity to load an $(s,k,n_0)$-sparse Walsh series is given by the implementation of $s$ controlled Walsh operators. Each of them is composed of at worst $2k$ Toffoli gates and one control-$R_z$ gates making the depth $O(sk)$, the size $O(n+sk)$ with the $n$ dependency due to the $n+1$ initial Hadamard gates.

\section{Amplitude Amplification}
\label{Amplitude Amplification}
The Walsh Series Loader is based on a repeat-until-success scheme with a probability of success $P(1)=\Theta(\epsilon_0^2)$. It is possible to reach $P(1)=O(1)$ by performing an amplitude amplification scheme \cite{brassard2002quantum} at the cost of modifying the size to $O(n^2/\epsilon_0+1/(\epsilon_0\epsilon_1))$ and the depth to $O(n/\epsilon_0+1/(\epsilon_0\epsilon_1))$ or, using one additional ancilla qubit, to a size $O(n/\epsilon_0+1/(\epsilon_0\epsilon_1))$ and a depth $O(n/\epsilon_0+1/(\epsilon_0\epsilon_1))$. The total time is reduced  by a quadratic factor with respect to the parameter $\epsilon_0$, but the size and depth of the associated quantum circuits are larger. In particular, it makes the depth of the WSL dependent with the number of qubits $n$.

The amplitude amplification scheme on $n+1$ qubits consists to implement $k$ times the operator $\hat{U}_{tot}=-\hat{U}_{\psi}\hat{U}_P$ with the two following unitaries:
\begin{equation}
    \hat{U}_{\psi}=\hat{I}-2\ket{\psi_3}\bra{\psi_3}
\end{equation}

\begin{equation}
    \hat{U}_P=\hat{I}-2\hat{P}
\end{equation}

where $\ket{\psi_3}=\frac{\hat{I}+e^{-i\hat{f}^{\epsilon_1}\epsilon_0}}{2}\ket{s}\ket{0}-i\frac{\hat{I}-e^{-i\hat{f}^{\epsilon_1}\epsilon_0}}{2}\ket{s}\ket{1}$, $\hat{P}=\hat{I}_{\text{position}}\otimes\ket{q_A=1}\bra{q_A=1}$ is the projector on the target subspace and $\hat{I}=\hat{I}_{\text{position}}\otimes \hat{I}_A$ is the identity on the $n+1$ qubits.

The operator  $\hat{U}_\psi$ can be simply rewritten as a product of the operator $\hat{U}_3$ defined as $\hat{U}_3=(\hat{P_1}\hat{H}\otimes \hat{I}_2^{\otimes n})(\text{control-}\hat{U}_{f^{\epsilon_1},\epsilon_0})(\hat{H}^{\otimes(n+1)})$, with $\hat{P_1}=\begin{pmatrix} 1 &0 \\ 0 & -i\end{pmatrix}$ such that $\hat{U}_3\ket{0}=\ket{\psi_3}$ , and a $n-$anti-controlled-$(Z)-$Pauli gate $\Lambda_n(-Z)$ which applies a phase $-1$ only if all qubits are in state $\ket{0}$:
\begin{equation}
    \hat{U}_\psi=\hat{U}_3 \Lambda_n(Z) \hat{U}_3^\dagger
\end{equation}

$\hat{U}_3$ is composed of $n+1$ Hadamard gates, a Phase gates and the controlled-diagonal unitary controlled-$\hat{U}_{f^{\epsilon_1},\epsilon_0}$,  which is implemented with a quantum circuit of size and depth $O(1/\epsilon_1)$. The $\Lambda_n(Z)$ unitary can be implemented without ancilla qubits using a quantum circuit of quadratic size $O(n^2)$ and linear depth $O(n)$ \cite{saeedi2013linear}, or, with one ancillary qubit, using a quantum circuit of linear size $O(n)$ and linear depth $O(n)$ \cite{barenco1995elementary}. The second unitary $\hat{U}_P$ simply corresponds to a $Z-$Pauli gate applied on the ancilla qubit $\ket{q_A}$. Let's define the positive parameter $\theta$ such that $\sin(\theta)=||\frac{\hat{I}-e^{-i\hat{f}^{\epsilon_1}\epsilon_0}}{2}\ket{s}||_{2,N}=\Theta(\epsilon_0)$ (Appendix (\ref{Proof of Theorem 1})). One needs to apply the operator $\hat{U}_{tot}^k$ with $k=\lfloor \pi/4\theta\rfloor$ in order to get $P(1)=O(1)$ \cite{brassard2002quantum}. Finally, one can show $\theta=\Theta(\epsilon_0)$ from $\sin(\theta)=\Theta(\epsilon_0)$, implying that $k=\Theta(1/\epsilon_0)$. Therefore, the WSL scheme using amplitude amplification has an overall size $O((n+1/\epsilon_1)/\epsilon_0)$ and depth $O((n+1/\epsilon_1)/\epsilon_0)$  using one additional ancilla qubit.

\end{document}